\RequirePackage[2020-02-02]{latexrelease}
\documentclass[10pt, letter]{article}
\usepackage[letterpaper, margin=0.75in]{geometry}
\usepackage{bm}
\usepackage{amsmath,mathtools}
\usepackage{amsthm}
\usepackage{amssymb}
\usepackage{mathrsfs}
\usepackage[dvipsnames]{xcolor}
\usepackage{graphicx} 
\usepackage{subcaption}
\usepackage[ruled, vlined, linesnumbered]{algorithm2e}
\usepackage[textsize=tiny,textwidth=0.6in,disable]{todonotes}
\usepackage{authblk}
\graphicspath{{figures/}}

\newcommand{\rl}{r_L}
\newcommand{\rh}{r_H}

\newcommand{\rs}{r_s}

\newcommand{\bxi}{\bm{\xi}}
\newcommand{\ubxi}{\underline{\bm{\xi}}}
\newcommand{\bxiL}{\bm{\xi}_L}
\newcommand{\bxiLr}{\bm{\xi}_L^{\rl}}
\newcommand{\bxiLri}{\bm{\xi}_L^{\rl,(i)}}
\newcommand{\ubxiL}{\underline{\bm{\xi}}_L}
\newcommand{\bxiH}{\bm{\xi}_H}
\newcommand{\bxiHr}{\bm{\xi}_H^{\rh}}
\newcommand{\bxiHri}{\bm{\xi}_H^{\rh,(i)}}
\newcommand{\ubxiH}{\underline{\bm{\xi}}_H}
\newcommand{\bbeta}{\bm{\eta}}

\newcommand{\bbetas}{\bm{\eta}_s}
\newcommand{\bbetah}{\bm{\eta}_H}
\newcommand{\bbetahr}{\bm{\eta}_{H,\rh}}
\newcommand{\bbetal}{\bm{\eta}_L}
\newcommand{\bbetalr}{\bm{\eta}_{L,\rl}}

\newcommand{\ubeta}{\underline{\bm{\eta}}}
\newcommand{\ubetah}{\underline{\bm{\eta}}_H}

\newcommand{\ubetal}{\underline{\bm{\eta}}_{L}}

\newcommand{\ubbetas}{\underline{\bm{\eta}}_s}

\newcommand{\hQh}{\widehat{Q}_H}
\newcommand{\hQl}{\widehat{Q}_L}
\newcommand{\hmul}{\widehat{\mu}_L}

\newcommand{\hQmfab}{\widehat{Q}^{MFAB}}

\newcommand{\AmatH}{\bm{A}_H}
\newcommand{\AmatHt}{\bm{A}_H^\mathrm{T}}
\newcommand{\AmatHr}{\bm{A}_{H,\rh}}
\newcommand{\AmatHrt}{\bm{A}_{H,\rh}^\mathrm{T}}

\newcommand{\AmatL}{\bm{A}_L}
\newcommand{\AmatLt}{\bm{A}_L^\mathrm{T}}
\newcommand{\AmatLr}{\bm{A}_{L,\rl}}
\newcommand{\AmatLrt}{\bm{A}_{L,\rl}^\mathrm{T}}

\newcommand{\EE}[1]{\mathbb{E}\left[#1\right]}
\newcommand{\EExi}[1]{\mathbb{E}_{\xi}\left[#1\right]}
\newcommand{\Var}[1]{\mathbb{V}ar\left[#1\right]}
\newcommand{\StDev}[1]{\mathbb{V}ar^{1/2}\left[#1\right]}
\newcommand{\Cov}[2]{\mathrm{Cov}\!\!\left[#1,#2\right]}
\newcommand{\Covsq}[2]{\mathrm{Cov}^{2}\!\!\left[#1,#2\right]}

\newcommand{\ratio}{\upsilon}

\DeclareMathOperator*{\argmax}{arg\,max}
\newtheorem{proposition}{Proposition}[section]
\newtheorem{theorem}{Theorem}[section]
\newtheorem{corollary}[theorem]{Corollary}

\usepackage[title]{appendix}

\usepackage[T1]{fontenc}
\usepackage{times}
\usepackage{url}

\usepackage[printwatermark]{xwatermark}
\newwatermark[allpages,color=red!30,angle=0,scale=0.5,xpos=58,ypos=130]{SAND2023-01382O}

\title{\textbf{Multifidelity uncertainty quantification with models based on dissimilar parameters}}

\author[1]{Xiaoshu Zeng}
\author[2]{Gianluca Geraci}
\author[2]{Michael S. Eldred}
\author[2]{John D. Jakeman}
\author[3]{Alex A. Gorodetsky}
\author[1]{Roger Ghanem}

\affil[1]{University of Southern California, Los Angeles, CA}
\affil[2]{Sandia National Laboratories, Albuquerque, NM}
\affil[3]{University of Michigan, Ann Arbor, MI}

\begin{document}

\maketitle

\begin{abstract}
	Multifidelity uncertainty quantification (MF UQ) sampling approaches have been shown to significantly reduce the variance of statistical estimators while preserving the bias of the highest-fidelity model, provided that the low-fidelity models are well correlated. However, maintaining a high level of correlation can be challenging, especially when models depend on different input uncertain parameters, which drastically reduces the correlation. Existing MF UQ approaches do not adequately address this issue. 
	In this work, we propose a new sampling strategy that exploits a shared space to improve the correlation among models with dissimilar parametrization. We achieve this by transforming the original coordinates onto an auxiliary manifold using the adaptive basis (AB) method~\cite{Tipireddy2014}. The AB method has two main benefits: (1) it provides an effective tool to identify the low-dimensional manifold on which each model can be represented, and (2) it enables easy transformation of polynomial chaos representations from high- to low-dimensional spaces. This latter feature is used to identify a shared manifold among models without requiring additional evaluations.
	We present two algorithmic flavors of the new estimator to cover different analysis scenarios, including those with legacy and non-legacy high-fidelity data. We provide numerical results for analytical examples, a direct field acoustic test, and a finite element model of a nuclear fuel assembly. For all examples, we compare the proposed strategy against both single-fidelity and MF estimators based on the original model parametrization.  
\end{abstract}

\section{Introduction}
\label{SEC:intro}
Model-based predictions of complex phenomena are inherently subject to uncertainty. To address this, a wide range of uncertainty quantification (UQ) algorithms have been developed and successfully applied to various applications~\cite{zeng2017serviceability, chen2017probabilistic, najm2009uncertainty, li2016quantifying, eldred2011design}. However, most of these methods rely on evaluations of a single highly accurate numerical model, which can become computationally intractable for models that are expensive to evaluate. To mitigate this computational cost, multifidelity (MF) methods have been developed~\cite{Giles2008,Peherstorfer2016b,Gorodetsky_GEJ_JCP_2020}. These methods combine limited high-fidelity numerical simulations with less accurate, but cheaper to evaluate, lower-fidelity models. Low-fidelity models typically arise from coarse spatial/temporal discretizations and/or simplifying physics assumptions. Despite their reduced accuracy, low-fidelity models are often sufficiently correlated with the highest-fidelity model, allowing for extensive sampling to reduce the cost of estimating high-fidelity statistics.

MF methods can be classified into two main categories: sampling-based methods, which are derived from Monte Carlo (MC)~\cite{Giles2008,Cliffe_GST_CVS_2011,Nobile2015,Peherstorfer2016b,Haji_NT_NM_2016,Fairbanks2017,geraci_multifidelity_2017,Gorodetsky_GEJ_JCP_2020,Schaden_U_SISC_2020}, and surrogate-based approaches~\cite{Teckentrup_JWG_SIAMUQ_2015,HajiAli_NTT_CMAME_2016,Jakeman_EGG_2019,Kennedy_O_B_2000,Rumpfkeil_B_AIAAJ_2020,LeGratiet_G_IJUQ_2014,Gorodetsky_JGE_IJUQ_2020,Gorodetsky_2021}. The main idea introduced in this paper applies to both categories; however, the focus is on sampling-based methods. Sampling methods produce estimates of statistics with an error that decays at a rate independent of the number of parameters and the smoothness of the model's output.

Sampling MF methods use one or more low-fidelity models to reduce the error in statistics of the quantity of interest (QoI), e.g., central moments. Their mean squared error (MSE) is the sum of two contributions: variance and bias. If the highest-fidelity is unbiased, the MSE can only be reduced by decreasing the estimator variance. Control variate approaches have been introduced to decrease the estimator variance by leveraging lower-fidelity models with known means; for instance, \cite {Lavenberg_W_MS_1981}. However, for low-fidelity models used in science and engineering, the knowledge of their means is an unrealistic assumption. To address this limitation, MF approaches have been introduced that evaluate the means of the low-fidelity models along with the high-fidelity statistics. A non-exhaustive list of references includes~\cite{Giles2008,Cliffe_GST_CVS_2011,Peherstorfer2016b,Haji_NT_NM_2016,Peherstorfer_WG_SIAMR_2018,Schaden_U_SISC_2020,Nobile2015,Fairbanks2017,geraci_multifidelity_2017,Gorodetsky_GEJ_JCP_2020}. For all these methods, the variance reduction directly depends on the correlation between the high- and low-fidelity models and the relative computational expense. Coarsening the spatial and/or temporal discretizations is the most straightforward technique to generate correlated models, particularly for partial differential equations, and has been relied upon extensively in literature, such as~\cite{Giles2008,Cliffe_GST_CVS_2011,Haji_NT_NM_2016}. Simplified modeling assumptions are often necessary in physical models, making it difficult to preserve the original parametrization. This can lead to more parameters being introduced in low-fidelity models or \textit{vice versa}.
In the current MF practice, non-shared parameters are generally ignored, although engineers and scientists usually embed physical knowledge into the low-fidelity models' construction to avoid this situation. However, this is impossible to do for complex multi-physics and/or multi-discipline applications.

The work presented in this paper addresses a crucial gap in the field by introducing a novel contribution. Our pivotal idea is to abandon the original parametrization of models and instead rely on a shared manifold among models, which can be lower-dimensional. This shared manifold is obtained by transforming the original coordinates of the models.
For each model, we propose a mapping from the original input parameters to a more convenient \emph{adapted space} where variables can be ordered according to their importance in representing the model's output. This allows the adapted space to be truncated to include only the important variables while quantifying the error committed during this truncation. The truncated adapted space is referred to as the \emph{auxiliary space}, with the adapted coordinates that span it referred to as \emph{important directions}. By defining a common \emph{shared space} among all models based on their auxiliary spaces, we can demonstrate that the performance and generality of MF UQ estimators can be enhanced by sampling the models over this shared space rather than their original coordinates.

Our approach is inspired by the conjecture that the system response often depends on a limited number of variables that can be obtained as a function of the original input parameters. Our method is agnostic of the problem's physics and relies only on data, making it applicable to a large class of numerical problems, even those whose parameters may not have a strict physical meaning, such as the coefficients in a Reynold-Averaged Navier Stokes turbulence model.
Dimension reduction strategies have previously been used to identify dimensionless groups responsible for the most variability in a system's response, e.g.,~\cite{Jofre2020}.
We postulate that if a shared manifold exists, it captures the common underlying causes of a model's response, thus maximizing the correlation among models.           
Any dimension reduction strategy can be used, but we initially designed our approach by exploiting the Active Subspace (AS) method proposed by Constantine and collaborators~\cite{Constantine2015}. We have previously presented our original idea in a series of contributions~\cite{Geraci2018,Geraci2018b,Geraci2019} but only considered a simplified context where each model was fully captured by a one-dimensional auxiliary space and, as a consequence, the shared space was also one-dimensional. Despite this strong assumption, the method was still effective in realistic application scenarios, as demonstrated in previous works~\cite{Constantine2015,Geraci2019}.

The present contribution addresses the theoretical and algorithmic challenges presented by situations where more than one important direction is required to represent a model effectively. We integrate the Adaptive Basis (AB) method~\cite{Tipireddy2014, zeng2021accelerated, zeng2022projection}, which has useful features for its integration within a MF UQ approach.
AB can identify the auxiliary space by relying on a set of random realizations for the model's QoI, and a first-order polynomial regression can be built to provide the necessary information, avoiding the need for derivative information. This inaccurate polynomial chaos expansion is only needed to build the transformation among variables and is never required to be accurate. Our approach can be applied to problems that are usually not amenable to surrogate-based methods, such as high-dimensional problems and problems with poor solution regularity. Moreover, a crude PCE is sufficient to guide the construction of the best shared manifold without requiring additional models' evaluations, taking into account the cost limitations associated with realistic problems.

The original contributions of this work are:
\begin{itemize}
    \item[I.] A framework for designing MF sampling estimators with models based on different parametrizations is introduced and analyzed;
    \item[II.] The AB dimension reduction method is integrated in the MF sampling algorithm and we derive estimators for the statistics that enable this coupling without extensive re-sampling of the models;  
    \item[III.] Two practical algorithms are designed to enable the application of this framework in contexts with different computational requirements, e.g., legacy or non-legacy high-fidelity data, various degrees of relative computational expense between high- and low-fidelity models, \textit{etc.}
    \item[IV.] Several numerical examples verify the methodology and gauge its effectiveness against both single- and MF approaches based on the original models' parametrizations.
\end{itemize}

The remainder of the paper is organized as follows. In Section~\ref{SEC:motivation}, we provide an example to illustrate the relevance of dissimilar parametrization in the practice of MF UQ. In Section~\ref{SEC:Background}, we introduce the mathematical background for our new framework. We discuss the MF sampling strategy in Section~\ref{SSEC:MFUQ} and the AB approach in Section~\ref{SSEC:AB}. Our novel framework, which integrates AB within MF UQ, is discussed in Section~\ref{SEC:ABforMF}. In Section~\ref{SEC:MFAB_practical}, we present the practical implementation of the approach, where we offer two different algorithmic variants. Numerical results are presented in Section~\ref{SEC:numres}, where we explore the properties and performance of our new estimators with respect to both single- and multi-fidelity approaches based on the models' original parametrizations. Finally, conclusions and perspectives are provided in Section~\ref{SEC:conclusions}.

\section{Dissimilar parametrization in MF UQ: a constructive example}
\label{SEC:motivation}
\begin{figure}[htb]
	\centering
	\begin{subfigure}{0.5\textwidth}
		\centering
		\includegraphics[width=0.75\textwidth]{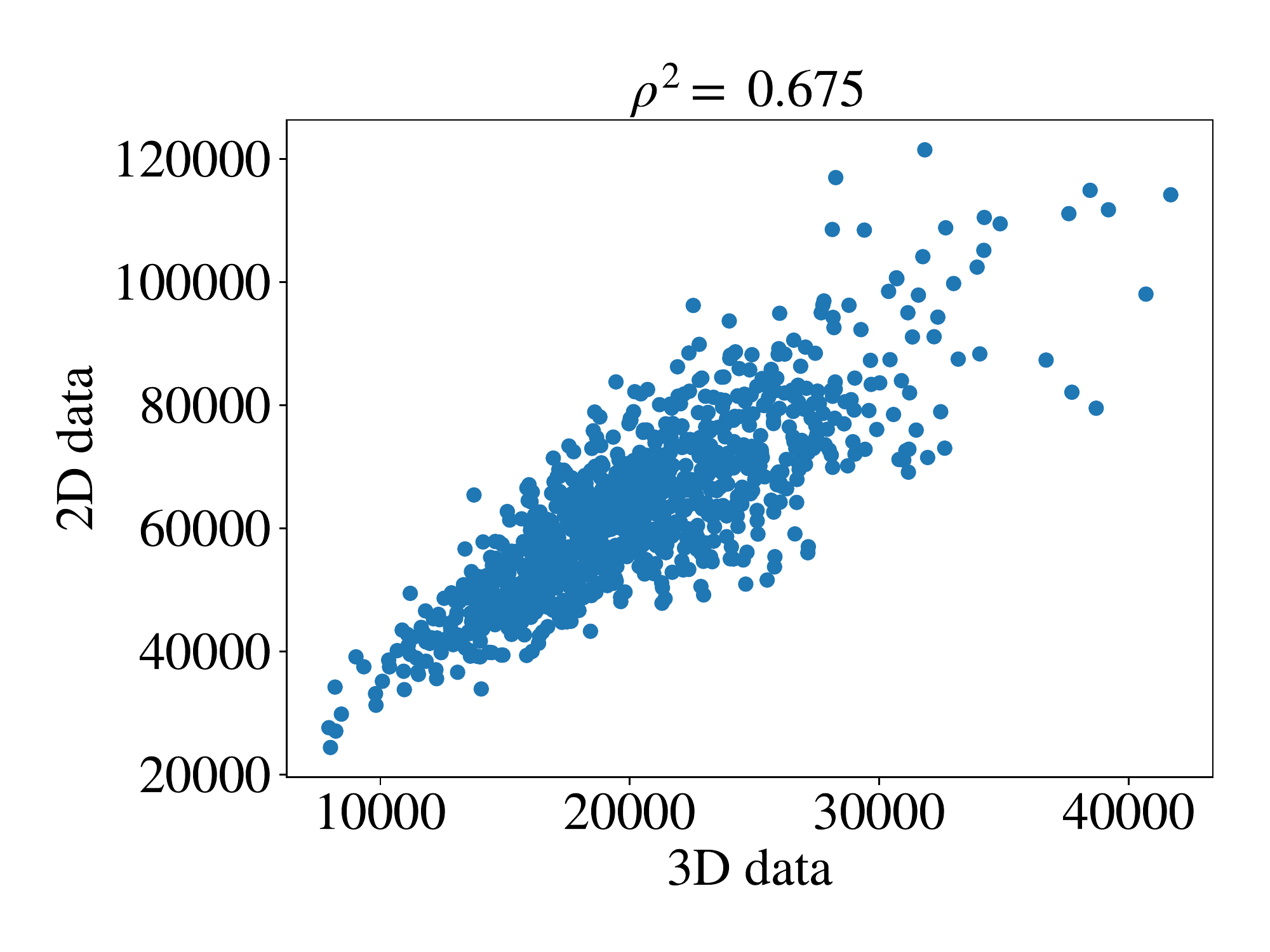}
		\caption{Exit pressure $P_e$ [Pa] for the elliptical geometry case}
		\label{fig:scatterplot_3D_2D_ellipse}
	\end{subfigure}%
	~ 
	\begin{subfigure}{0.5\textwidth}
		\centering
		\includegraphics[width=0.75\textwidth]{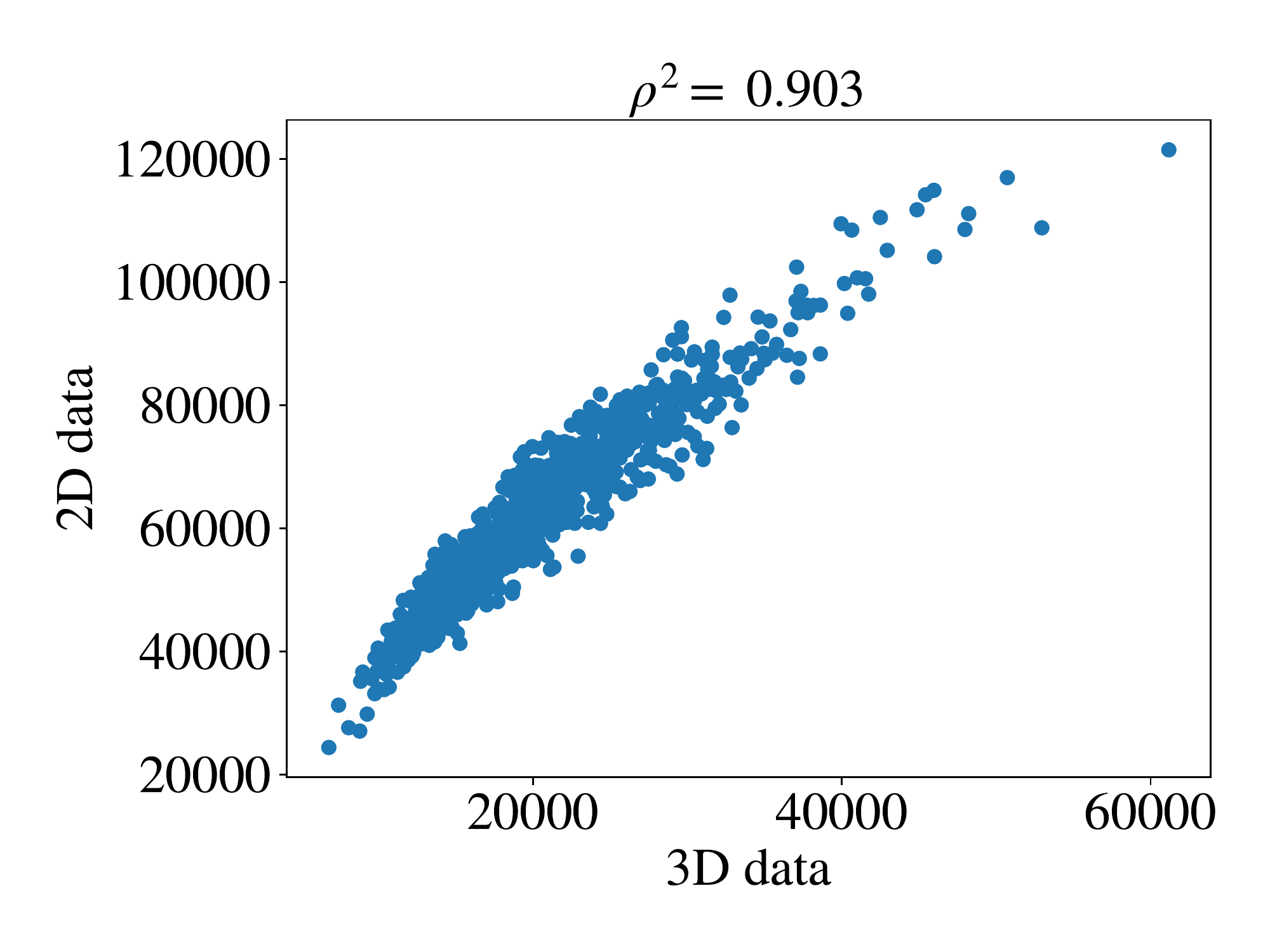}
		\caption{Exit pressure $P_e$ [Pa] for the circular geometry case}
		\label{fig:scatterplot_3D_2D_circular}
	\end{subfigure}
	\caption{Scatter plot for the nozzle flow in the elliptical (\subref{fig:scatterplot_3D_2D_ellipse}) and circular (\subref{fig:scatterplot_3D_2D_circular}) geometry cases. The data represent the exit pressure $P_e$ [Pa] with uncertain stagnation pressure $P_0\sim \mathcal{N}(658612.5, 118212.5)$ and exit area geometry. In the elliptical case, the 3D model has uncertain parameters $r_a$ and $r_b$, while the 2D is only described by $r_a$. In the circular section case, the geometrical parameter is $r_a$ for both models. We consider the geometrical parameters distributed as: $r_a/r_t, r_b/r_t \sim \mathcal{N}(2, 1/6)$.}
	\label{fig:nozzle_scatter_original}
\end{figure}

In this section, we illustrate how dissimilar parametrizations can naturally arise in MF UQ analyses. We show that this situation can arise either because the number of parameters is different or because different parameters are used across models. Our strategy addresses both scenarios. For simplicity, we resort to a textbook compressible fluid dynamics problem to illustrate the idea. Consider an isentropic supersonic flow through the divergent portion of a converging-diverging de Laval nozzle \cite{Anderson1990} that results in supersonic exit flow and an exit pressure greater than the ambient pressure (under-expanded flow). Isentropic relationships describe the flow quantities and the exit pressure $P_e$ can be expressed as
\begin{equation}
\label{eq:nozzle_Pe}
 P_e = P_0 \left( 1 + \dfrac{\gamma-1}{2} M_e^2 \right)^{-\dfrac{\gamma}{\gamma-1}},
\end{equation}
in which $P_0$ indicates the stagnation (reservoir) pressure, $\gamma$ the ratio of polytropic coefficients (equal to 1.4 for air modeled as diatomic gas), and $M_e$ the Mach number of the fluid velocity at the exit of the duct. The Mach number $M_e$ can be obtained by solving the non-linear equation for the isentropic flows with a section area variation such that
\begin{equation}
 \label{eq:nozzle_Me}
 \frac{1}{M_e} \left[ \frac{2}{\gamma+1} \left( 1 + \frac{\gamma-1}{2} M_e^2 \right) 
               \right]^{ \frac{\gamma+1}{2(\gamma-1)} }
  - \frac{A_e}{A^\star} = 0, 
\end{equation}
where $A_e$ and $A^\star$ indicate the area of the exit section and the throat (at which sonic conditions exist), respectively. We consider uncertainty in the reservoir pressure $P_0$ and in the geometry of the duct, e.g., a nozzle with a circular throat section with radius $r_t$ and an elliptical exit section with axes $r_a$ and $r_b$. The exit pressure is only affected by the area ratio
\begin{equation}
 \frac{A_e}{A^\star} = \frac{\pi r_a r_b}{\pi r_t^2} = \frac{r_a r_b}{r_t^2}.
\end{equation}
The simplest low-fidelity model can be obtained by considering (planar) two-dimensional flow, i.e., all derivatives in the transversal direction are zero. The model is agnostic of the area variation due to the transversal dimension $r_b$; therefore the area ratio is linear in $r_a/r_t$
\begin{equation}
 \left. \frac{A_e}{A^\star} \right|_{2D}= \frac{r_a}{r_t}.
\end{equation}
Even in a simple problem like this, preserving the models' parametrization does not follow naturally. In Figure~\ref{fig:scatterplot_3D_2D_ellipse}, we show the scatter plot for the exit pressure $P_e$ as function of the uncertain stagnation pressure $P_0\sim \mathcal{N}(658612.5, 118212.5)$ Pa and the nozzle geometry with parameters $r_a/r_t$ and $r_b/r_t$ for the 3D model and $r_a/r_t$ for the 2D model, which we assume to be distributed as $ \mathcal{N}(2, 1/6)$. 
The scatter plot is a graphical representation of the linear correlation between the models and the spread of the data in Figure~\ref{fig:scatterplot_3D_2D_ellipse} shows how dissimilar parametrization negatively impacts the correlation; the effect of the extra parameter in the 3D model de-correlates the models, even if models' variability along the other (common) variable is correlated.

A more subtle case is observed in the presence of a common shared parameter with a different physical meaning in distinct models. For instance, if the above target model has a circular exit section, rather than an elliptical section, we would have
$r_a = r_b$ and the area ratio would become
\begin{equation}
 \frac{A_e}{A^\star} = \frac{\pi r_a^2}{\pi r_t^2} = \frac{r_a^2}{r_t^2}.
\end{equation}
Although this parameter is nominally the same in both models, in the 3D configuration the area would depend on its squared value; on the other hand, the low-fidelity model depends linearly on it.  
In Figure~\ref{fig:scatterplot_3D_2D_circular}, we show how the scatter plot is impacted in this scenario; although the spread of data is reduced compared to the previous case, the models are still not perfectly correlated.

Understanding the role played by the models' parameters is easy in simple situations and it can be used to define well-correlated low-fidelity models. However, in complex multi-physics and/or multi-discipline applications this would be impossible. The method presented in this paper has been designed to aid in both the situations described above; in particular, it leverages an optimally constructed mapping among models (and their variables) without requiring the definition of new low-fidelity based on a physics-informed parametrization.

\section{Mathematical background}
\label{SEC:Background}

Let $(\Omega, \mathcal{F}, \mathcal{P})$ be a probability space and $\bm \xi=(\xi_1,...,\xi_d)^T \in \mathbb{R}^d$ be a vector of random variables defined on this space. The focus of this paper is the accurate estimation of moments, e.g., the expected value of a QoI $Q \coloneqq Q(\bm\xi) \in \mathbb{R}$. Various numerical quadrature methods can be used to compute such expectations, e.g., sparse grids~\cite{Gerstner_G_NA_1998,Narayan_J_SISC_2014}, cubature rules~\cite{Jakeman_N_CMAME_2018,Keshavarzzadeh_SISC_2018}, PCE~\cite{Xiu_K_SISC_2002,Jakeman_FNEP_CMAME_2019}, and Gaussian processes~\cite{Ohagan_JSPI_1991}. However, these methods often cannot be applied to high-dimensional and/or non-smooth QoIs. In these settings MC-based sampling methods are appealing since their rate of convergence, while slow, is independent of both dimension and the smoothness of the QoI.  

\subsection{Monte Carlo-based MF estimation of statistics}
\label{SSEC:MFUQ}

Given $N$ samples $\{\bm{\xi}^{(i)}\}_{i=1}^N$ drawn from the joint-distribution $p(\bxi)$, the MC estimator of the mean $\mathbb{E}_{\xi}[Q]$ is
\begin{equation}
\widehat{Q} = \frac{1}{N}\sum_{i=1}^{N} Q^{(i)}, \quad \mathrm{where} \quad Q^{(i)} \coloneqq Q\left( \bm{\xi}^{(i)} \right).
\end{equation}
The MC estimator is itself a random variable, i.e., different realizations corresponds to distinct set of $N$ samples for $\bm{\xi}$. However, MC is unbiased, i.e., $\EE{\widehat{Q}} = \mathbb{E}_{\xi}[Q]$, and converges almost surely as $N \rightarrow \infty$ with an error $\lvert \widehat{Q} - \EExi{Q} \lvert$ that follows (by the law of large numbers) a normal distribution with zero mean and standard deviation $\sqrt{\Var{Q}/N}$.

Several MF strategies have been proposed in the literature to reduce the error of MC estimators for a fixed computational budget, for example, multilevel Monte Carlo (MLMC) \cite{Giles2008}, Multi index Monte Carlo (MIMC) \cite{Haji_NT_NM_2016}, MF Monte Carlo (MFMC) \cite{Peherstorfer2016b} and Multilevel MF (MLMF) \cite{Nobile2015,Fairbanks2017,geraci_multifidelity_2017}. These methods leverage the correlation between the high-fidelity (HF) model and one or several less expensive and less accurate lower-fidelity (LF) models to reduce the MSE of the MC estimator. 
In the following we focus on Approximate Control Variate (ACV) MC estimation~\cite{Gorodetsky_GEJ_JCP_2020}, which is a generalization of many existing MF sampling techniques like MLMC and MFMC.

The ACV framework,
\begin{equation} \label{eq:ACVgeneric}
 \widehat{Q}^{ACV} = \widehat{Q} + \sum_{i=1}^M \alpha_i \left( \widehat{Q}_i - \widehat{\mu}_i \right), 
\end{equation}
where $M$ is the number of LF models and $\widehat{Q}_i$ and $\widehat{\mu}_i$ are two approximations of the expected value of the $i$th LF model obtained using two different sets of samples. Different ACV estimators can be obtained from this framework by varying the structure of the sets of samples used by $\widehat{Q}_i$ and $\widehat{\mu}_i$ as explained in \cite{Gorodetsky_GEJ_JCP_2020}. The best sampling scheme is generally problem-dependent; in this work we focus on the use of a single LF model such that $M=1$ and
\begin{equation} \label{eq:ACV1}
 \widehat{Q}^{MF}\left(\alpha,\ubxi\right) = \widehat{Q}_{H}(\ubxiH) + \alpha \left( \widehat{Q}_{L}(\ubxiL^1) - \widehat{\mu}_{L}(\ubxiL^{2}) \right),
\end{equation}
where the subscripts $H$ and $L$ denote the HF and LF models and the different sample sets used to evaluate each model. We indicate the collection of samples for the HF model as $\ubxiH = \left\{ \bm{\xi}^{(i)} \right\}_{i=1}^{N}$, the collection of samples for the LF model as $\ubxiL^1 = \left\{ \bm{\xi}^{1,(i)} \right\}_{i=1}^{N}$ and $\ubxiL^2 = \left\{ \bm{\xi}^{2,(i)} \right\}_{i=1}^{N_L}$, respectively, and the ordered collection of all samples as $\ubxi = \left( \ubxiH, \ubxiL^1, \ubxiL^2 \right) = \left( \ubxiH, \ubxiL \right)$, where $\ubxiL = \left( \ubxiL^1, \ubxiL^2 \right)$. 
For ease of discussion and without loss of generality, we focus on the sample configuration that results in the MFMC estimator introduced in~\cite{Ng2014,Peherstorfer2016b}. This estimator uses $N$ samples to evaluate $\widehat{Q}_H$ and $\ratio N = N_L$ total samples for the LF model, for some scalar $\ratio>1$, where the first $N$ samples are used for $\widehat{Q}_{L}$ ($\ubxiL \supset \ubxiL^1=\ubxiH$), while the full set $\ubxiL$ is used for $\widehat{\mu}_{L}$ ($\ubxiL^2=\ubxiL$) with $\bm{\xi} \subset \bm{\xi}_{L}^2$.
The variance of the MFMC estimator is~\cite{Ng2014}
\begin{equation}
	\label{eq:var_MF}
 \Var{ \widehat{Q}^{MF} } = \frac{\Var{ Q }}{N} \left( 1 - \frac{\ratio-1}{\ratio} \rho^2 \right), 
\end{equation}
where $\rho$ indicates the Pearson's correlation coefficient between the HF and LF model. This variance is minimized when the coefficient $\alpha$ and the ratio $\ratio$ are equal to
\begin{equation}
	\label{eq:MF_samp_ratio}
  \alpha = - \rho \frac{\StDev{Q_{H}}}{\StDev{Q_{L}}} \quad \mathrm{and} \quad \ratio = \sqrt{ \frac{\rho^2}{1-\rho^2}
  \frac{\mathcal{C}_{H}}{\mathcal{C}_{L}} },
\end{equation}
where $\mathcal{C}_{H}$ and $\mathcal{C}_{L}$ represent the computational cost of obtaining a single HF and LF model, respectively. From Eq~\eqref{eq:var_MF} it follows that, if a specified estimator variance $\varepsilon^2$ is given, the number of HF model evaluations is chosen as 
\begin{equation}
	\label{eq:required_N_HF}
 	N = \frac{\Var{Q}}{\varepsilon^2} \left( 1 - \frac{\ratio-1}{\ratio} \rho^2 \right).
\end{equation}
Similarly, one could equivalently minimize the variance for a prescribed computational cost.

In this framework, the HF model is considered the truth and not an approximation of a HF process, i.e., the mean squared error of the estimator is equal to the estimator's variance. The variance reduction depends on the correlation between the HF and LF model and their relative cost. Therefore, it is necessary to increase the correlation between the HF and LF models to improve the efficiency of a MF estimator for a fixed computational cost of the LF model. In the following sections we will show that this can be accomplished by sampling the models in a shared space, which alleviates the degradation of the correlation due to the presence of non-shared parameters (as described in the previous section).

\subsection{Identifying important input features using adaptive basis}
\label{SSEC:AB}

The AB method was introduced as a dimension reduction strategy for UQ in~\cite{Tipireddy2014, zeng2021accelerated}. Our goal in dimension reduction is to identify the most significant input parameters of a model for a certain QoI. Two relevant classes of methods are those that identify important parameters among the original coordinates (e.g., ANOVA~\cite{Saltellibook,Crestaux2009}) and those that seek transformations of the original coordinates to express the model with a different set of parameters (e.g., Principal Component Analysis~\cite{jolliffe2016principal}, AS~\cite{Constantine2015,constantine2015book}, and AB). We will focus on linear approaches as they seek a linear combination of the original parameters, defining an \emph{adapted space}. If the number of important directions is lower than the original model's dimensionality, these variables effectively describe a low-dimensional manifold (the adapted space). We will use the adapted space of each available model to define a shared manifold that enhances the correlation among models.

In this work, we focus on AB for two reasons. Firstly, AB, like the AS method, can be implemented from a random set of model evaluations. By solving a linear regression from this set of realizations, we can obtain the first-order polynomial coefficients. Unlike the straightforward version of the AS method, this knowledge is sufficient to identify more than one important direction. Secondly, the PCE construction within AB can be used to estimate several statistical properties for designing the estimator, without requiring additional model runs.

\subsubsection{Fundamentals of polynomial chaos expansions and adaptive basis}
Given a set of independent Gaussian variables $\bm{\xi}$ and the Hilbert space $\mathcal{H}$ generated by the linear span of $\bm{\xi}$, we can approximate any QoI $Q \in L^2(\Omega, \mathcal{F}(\mathcal{H}), \mathcal{P})$ by the PCE \cite{ghanem2003stochastic}, which can be truncated to a total degree $p$ 
\begin{equation}
\label{eq:3.1}
Q(\bm \xi) = \sum_{\bm{\alpha}\in\mathcal{J}}c_{\bm{\alpha}}\psi_{\bm \alpha}(\bm \xi) \approx \sum_{\bm{\alpha}\in\mathcal{J}_p} c_{\bm{\alpha}}\psi_{\bm \alpha}(\bm \xi) .
\end{equation}
Here $\bm{\alpha} = (\alpha_1,...,\alpha_d) \in \mathcal{J}\coloneqq(\mathbb{N}_0)^d$ is a $d$-dimensional multi-index, $\{\psi_{ \bm{\alpha}}(\bm{\xi}): \; \bm{\alpha}\in \mathcal{J}\} $ are orthonormal Hermite polynomials, which are a complete basis for $\mathcal{H}$, $\mathcal{J}_p = \{\bm{\alpha} \in \mathcal{J}: \; |\bm{\alpha}| \leq p \}$, and $|\bm{\alpha}| = \sum_{i=1}^{d} \alpha_i$. 
The PCE coefficients can be expressed by projection as
\begin{equation}
\label{eq:3.2}
c_{\bm{\alpha}} = \frac{\langle Q, \psi_{\bm \alpha} \rangle}{\langle \psi_{\bm \alpha}^2 \rangle} 
=\langle Q, \psi_{\bm \alpha} \rangle,  \quad \bm{\alpha} \in \mathcal{J}_p\, ,
\end{equation}
where $\langle \cdot, \cdot \rangle$ is the $L^2$ inner product. The PCE converges to $Q$ in the mean squared sense as the number of terms in the expansion tends to infinity, i.e., $\mathrm{card}(\mathcal{J}_p)\to\infty$. Without loss of generality, we assume a Gaussian distribution for the uncertain parameters $\bm{\xi}$; however, one could rely on generalized PCE~\cite{Xiu_K_SISC_2002} for distributions belonging to the Askey-Wiener scheme or on transformations, e.g., the Nataf transformation~\cite{Liu_K_PEM_1986} for arbitrary distributions. Several approaches have been proposed to obtain the PCE coefficients; see, e.g.,~\cite{le2002,reagana2003,Hampton_D_CMAME_2015,Narayan_JZ_MC_2017,Blatman_S_JCP_2011,Jakeman_ES_JCP_2015}. In this work we rely on regression-based strategies to leverage available samples generated via random sampling, which is natural in the MF UQ sampling strategy. Therefore, we will work with Ordinary Least-Square (OLS) regression to obtain the coefficients $c_{\bm{\alpha}}$.

The AB method efficiently identifies the important directions in the parameter space by constructing an isometry (rotation matrix) that transforms the Gaussian inputs into a new basis. 
Specifically, the AB method constructs a rotation matrix  $\bm A$ on $\mathbb{R}^{d\times d}$ such that $\bm{AA}^T=\bm I$. This matrix defines the so-called adapted variables $\bbeta$ as
\begin{equation}
\label{eq:3.5}
\bm \eta = \bm{\eta}(\bm{\xi}) = \bm{A\xi}.
\end{equation}
Because $\bm{\xi}$ are independent Gaussian variables and \eqref{eq:3.5} is a linear transformation, $\bm{\eta}$ are also independent Gaussian variables. The PCE expansion of $Q$ can be expressed in the adapted variables as
\begin{equation}
\label{eq:3.6}
{Q}^{\bm A}(\bm{\eta}) = \sum_{\bm{\beta}\in\mathcal{J}_{p}}c_{\bm{\beta}}^{\bm A}\psi_{\bm \beta}(\bm \eta) \,,
\end{equation}
where the superscript $\bm{A}$ on $Q$ and $c_{\bm{\beta}}$ denotes that the expansion is in terms of new variables generated by the rotation matrix $\bm{A}$. 
Since $\bm{\xi}$ and $\bm{\eta}$ are both sets of independent Gaussian variables, $\{\psi_{ \bm{\alpha}} (\bm{\xi}) : \; \bm{\alpha} \in \mathcal{J}_p\} $ and $\{\psi_{ \bm{\beta}} (\bm{\eta}) : \; \bm{\beta} \in \mathcal{J}_p\}$ span the same space; thus, we have two equivalent expansions ${Q}^{\bm{A}}(\bm{\eta}(\bm{\xi})) \equiv {Q}(\bm{\xi})$ with ``$\equiv$'' denoting equivalence. Furthermore, by defining $\psi_{\bm{\beta}} ^{\bm{A}}(\bm{\xi})\coloneqq \psi_{ \bm{\beta} } (\bm{\eta}) = \psi_{ \bm{\beta} } (\bm{A \xi})$, the new PCE coefficients can be computed from the knowledge of the expansion in the original coordinates
\begin{equation}
\label{eq:3.7}
c_{\bm{\beta}}^{\bm A} = \sum_{\bm{\alpha}\in\mathcal{J}_{p}} c_{\bm{\alpha}} \langle \psi_{\bm \alpha}, \psi_{\bm \beta}^{\bm A} \rangle, \quad \bm{\beta}\in \mathcal{J}_{p}\,.
\end{equation}
In summary, the AB method requires the evaluation of a rotation matrix $\bm{A}$, which can be used to derive a PCE that is distinct from the PCE expressed in the original coordinates. The flexibility of this approach arises from the fact that the rotation matrix can be determined using only the first-order coefficients, as outlined in Algorithm~\ref{algo1}. To obtain the first-order PCE coefficients, we can use OLS to fit a model to the available samples of the QoI (Step 1). We obtain the first row of $\bm{A}$ using these coefficients (Step 2), while all subsequent rows are constructed in Step 4 after rearranging the coefficients in descending order of importance (Step 3). The algorithm is concluded by applying the Gram-Schmidt procedure (Step 5) to ensure that $\bm{A} \bm{A}^{\mathrm{T}} = \bm{I}$.
\begin{algorithm}[htb]
	\caption{Construction of rotation matrix by Gaussian adaptation \cite{zeng2022projection}}\label{algo1}
	
	Use least squares regression to estimate the coefficients $c_{\bm{e}_0}$ and $\{c_{\bm e_i}\} _{i=1}^d$ of the first-order PCE
	\begin{equation}
		\label{eq:3.7_2}
		Q(\bm \xi) = \sum_{\bm{\alpha}\in\mathcal{J}_1} c_{\bm{\alpha}}\psi_{\bm \alpha}(\bm \xi) = c_{\bm{e}_0} + \sum_{i=1}^{d} c_{\bm e_i} \xi_i\,,
	\end{equation}
	with $\bm{e}_0$ the multi-index of all zeros, and $\bm e_i$ the multi-index with one at $i$-th location and zeros elsewhere.
	\\
	
	Construct the first row of the rotation matrix $\bm{A} \in \mathbb{R} ^{d \times d}$ such that
	\begin{equation}
		\label{eq:3.8}
		\eta_1 =  \sum_{i=1}^{d} A_{1i} \xi_i = \sum_{i=1}^{d} c_{\bm e_i} \xi_i \quad \Longrightarrow \quad A_{1i} = c_{\bm e_i} \quad \mathrm{for} \quad i=1,\dots,d.
	\end{equation}\\

	Rank the first-order coefficients $\{c_{\bm e_i}\} _{i=1}^d$ by absolute value in descending order and record their indices in the original coordinates as $\{ \kappa_j \}_{j=1}^{d}$;
	
	For $j = 2,\ldots,d$ construct the $j^{\text{th}}$ row of $\bm{A}$ such that
	\begin{equation}
		\label{eq:3.8_2}
		\eta_j  =  \xi_{\kappa_{j-1}}.
	\end{equation}
	
	Perform Gram-Schmidt procedure on $\bm{A}$ to make it an isometry (rotation matrix).
\end{algorithm}

The algorithm requires only a few QoI evaluations because only the coefficients of a first-order PCE are necessary to define the rotation matrix. In regression-based strategies, the required number of random samples is in the order of $N \sim \mathcal{O}(P \ln P)$ or, in some cases, $N \sim \mathcal{O}(P^2 \ln P)$~\cite{cohen2017optimal}, where $P$ is the total number of terms in the PCE. AB enables the recovery of all $d$ dimensions of $\bm{\eta}$ without incurring additional computational expense. Note that in the scenario where the first-order derivatives to the original coordinates are zero, the aforementioned algorithm may be unable to identify the important directions. However, this issue has been addressed in literature~\cite{Tipireddy2014} by devising a high-order adaptation, which can be used without modifications in our context.

The importance of the variables $\bm{\eta}$ is captured in the order of the rows in the matrix $\bm{A}$. Then, a dimension reduction can be performed by partitioning the adapted variables $\bm{\eta}$ as
\begin{align}
\label{eq:Amat-split}
\bm \eta = \begin{pmatrix} \bm{A}_r\bm{\xi}\\ \bm{A}_{\neg r}\bm{\xi}\end{pmatrix}=\begin{pmatrix} \bm{\eta}_r\\ \bm{\eta }_{\neg r}\end{pmatrix} \,,
\end{align}
where $\bm{\eta}_r$ are defined as the first $r$ important directions. The PCE $Q^{\bm{A}_r}$, defined on the first $r$ important directions, satisfies
\begin{align}
	\label{eq:3.9}
	Q^{\bm{A}_r}(\bm{\eta_r})= \sum_{\bm{\gamma}\in\mathcal{J}_p^r} c_{\bm{\gamma}}^{\bm{A}_r} \psi_{\bm{\gamma}}(\bm{\eta}_r)\approx Q(\bm{\xi}),
\end{align}
where $\bm{\gamma} = (\gamma_1,...,\gamma_r)$ is a $r$-dimensional multi-index and $\mathcal{J}^r_p \coloneqq \{\bm{\alpha} \in \mathcal{J}: \; |\bm{\alpha}| \leq p \}$ such that $\{ \psi_{\bm{\gamma}}(\bm{\eta}_r) : \; \bm{\gamma} \in \mathcal{J}_p^r \}$ is a basis of order up to $p$ defined on $\bm{\eta}_r$. Then, by defining $\psi_{\bm{\gamma}} ^{\bm{A}_r}(\bm{\xi})\coloneqq \psi_{ \bm{\gamma} } (\bm{\eta}_r) = \psi_{ \bm{\gamma} } (\bm{A}_r \bm{\xi})$ and assuming that ${Q}^{\bm{A}_r}(\bm{\eta}_r(\bm{\xi}))$ and ${Q}(\bm{\xi})$ are equivalent expansions of the QoI, that is, ${Q}^{\bm{A}_r}(\bm{\eta}_r(\bm{\xi})) \equiv {Q}(\bm{\xi})$, we can obtain the PCE coefficients by projection as
\begin{equation}
\label{eq:3.10}
c_{\bm{\gamma}}^{\bm{A}_r} = \langle Q, \psi_{ \bm{\gamma}}^{\bm{A}_r} \rangle \,, \quad \bm{\gamma} \in \mathcal{J}^r_p \,.
\end{equation}
The estimation of the PCE coefficients in Eq~\eqref{eq:3.9} (with $r \leq d$) requires function evaluations as training data; samples generated in the $\bm{\eta}_r$ space can be mapped to $\bm{\xi}$ to allow for the evaluation of the model . We introduce $\bm{\eta}_{\neg r} = \bm{0}$ to regularize this transformation 
\footnote{This is required to preserve a bijective mapping} 
\begin{equation}
\label{eq:3.11}
\bm{\xi}^r = \bm{\xi}^r(\bm{\eta}_r) = \bm{A}^T \begin{pmatrix}
\bm{\eta}_r \\ \bm{0}
\end{pmatrix} = \bm{A}_r^T\bm{\eta}_r\,.
\end{equation}
The superscript $r$ on $\bm{\xi}$ signifies that the full space coordinates $\bm{\xi}^r \in \mathbb{R}^d$ are approximated using a transformation of the $r$-dimensional adapted coordinates $\bm{\eta}_r \in \mathbb{R}^r$. The accuracy of the adapted PCE in Eq~\eqref{eq:3.9} can be affected by the transformation in~\eqref{eq:3.11}, especially when the variation of $Q$ along the directions $\bm{\eta}_{\neg r}$ is significant. If $Q$ does not change significantly in these directions, the error introduced by the transformation is negligible when computing the adapted PCE.

In the following section, we introduce estimators to assess and manage the accuracy of the adapted PCE in Eq~\eqref{eq:3.9}. These estimators will help guide the integration of AB into the MF UQ estimator.

\section{MF estimation via embedded adaptive basis}
\label{SEC:ABforMF}

The first step for obtaining an MF estimator with embedded AB, which we denote with MFAB, is to reformulate the MF estimator (Eq~\eqref{eq:ACV1}) 
by introducing the change of coordinates explained in the previous section
\begin{equation}
 \begin{split}
  \bxiH   = \AmatHt \bbeta_H\,, \qquad
  \bxiL^1 = \AmatLt \bbeta_L^1\,, \qquad \textrm{and} \qquad
  \bxiL^2 = \AmatLt \bbeta_L^2\,,
 \end{split}
\end{equation}
which leads to 
\begin{equation} \label{eq:genericMFAB}
 \hQmfab(\ubeta; \alpha) = \hQh(\AmatHt \ubetah) + \alpha \left( \hQl(\AmatLt \ubetal^1) - \hmul(\AmatLt \ubetal^2) \right),
\end{equation}
where we defined the collection of samples as $\ubetah = \left\{ \bbeta^{(i)} \right\}_{i=1}^N$, $\ubetal^1 = \left\{ \bbetal^{(i),1} \right\}_{i=1}^N$, $\ubetal^2 = \left\{ \bbetal^{(i),2} \right\}_{i=1}^{\ratio N}$, and $\ubeta = \left( \ubetah, \ubetal^1, \ubetal^2 \right)$, as done in Section~\ref{SSEC:MFUQ} The estimators in Eq.~\eqref{eq:genericMFAB} are defined as
\begin{equation} \label{eq:genericMFAB_est}
 \begin{split}
  \hQh(\AmatHt \ubetah) &= \frac{1}{N} \sum_{i=1}^N Q_H( \bxi_H^{(i)} ) = \frac{1}{N} \sum_{i=1}^N Q_H( \AmatHt \bbeta_H^{(i)} ) \\
  \hQl(\AmatLt \ubetal^1) &= \frac{1}{N} \sum_{i=1}^N Q_L( \bxi_L^{1,(i)} ) = \frac{1}{N} \sum_{i=1}^N Q_L( \AmatLt \bbeta_L^{1,(i)} ) \\
  \hmul(\AmatLt \ubetal^2) &= \frac{1}{\ratio N} \sum_{i=1}^{\ratio N} Q_L( \bxi_L^{2,(i)} ) = \frac{1}{\ratio N} \sum_{i=1}^{\ratio N} Q_L( \AmatLt \bbeta_L^{2,(i)} ), \\
 \end{split}
\end{equation}
where, as discussed in Section~\ref{SSEC:MFUQ}, we have $\ubxiH = \ubxiL^1$ and $(\ubxiL^2 \setminus \ubxi_L^1) \cap \ubxi_L^1 = \emptyset$.

Up to this point, we have only changed the coordinates for each model. However, it is not guaranteed that $\bbeta_H$ and $\bbeta_L$ (as well as $\bxiH$ and $\bxiL$) will have the same cardinality, since $d_H \coloneqq \mathrm{card}(\bbeta_H) = \mathrm{card}(\bxi_H)$ is generally different from $d_L \coloneqq \mathrm{card}(\bbeta_L) = \mathrm{card}(\bxi_L)$. Moreover, we want to take advantage of the dimension reduction offered by AB. In the previous section, we demonstrated how to truncate the model by selecting the first $r$ important directions. Here, we consider a truncation independent for each model, i.e., $\rh \leq d_H$ and $\rl \leq d_L$. The values of $\rh$ and $\rl$ can be chosen to control the MFAB performance, as described later. Assuming that the HF and LF adapted coordinates $\bbetah$ and $\bbetal$ are arranged in decreasing order of importance (see Algorithm~\ref{algo1}), we can define a vector of shared (important) directions as
\begin{equation}
 \label{eq:beta_share}
 \bbetas = \left[ \eta_{s,1}, \dots, \eta_{s,r_s} \right]^{\mathrm{T}} \in \mathbb{R}^{r_s},
\end{equation}
where $\rs = \mathrm{max}(\rh,\rl)$ and 
\begin{equation}
 \eta_{s,i} = \left\{
 \begin{split}
  \eta_{H,i} = \eta_{L,i} \quad &\mathrm{if} \quad i \leq \mathrm{min}(\rh,\rl) \\
  \eta_{H,i}              \quad &\mathrm{if} \quad \rl < i \leq \rh \\
  \eta_{L,i}              \quad &\mathrm{if} \quad \rh < i \leq \rl. \\
 \end{split}
 \right.
\end{equation}
To satisfy this definition, the HF and LF realizations must be associated with the same samples of $\bbetas$ in the adapted space. We achieve this by proposing two novel sampling algorithms, which we will discuss later in Section~\ref{SEC:MFAB_practical}. The current section is devoted to how a MF estimator is constructed with embedded AB.

Starting from Eq~\eqref{eq:3.11} we can determine the map from the shared variables $\bbetas$ to original coordinates corresponding to the truncation, either $\rh$ or $\rl$, for the high- and low-fidelity model, respectively
\begin{equation}
	\label{eq:adp2org}
 \begin{split}
 \bxiHr &= \AmatHrt \bbetahr = \left[ \AmatHrt \quad \bm{0}_{d_H \times (\rs-\rh) } \right] \bbeta_s 
                             = \mathcal{A}_{H,\rh}^\mathrm{T} \bbeta_s\\
 \bxiLr &= \AmatLrt \bbetalr = \left[ \AmatLrt \quad \bm{0}_{d_L \times (\rs-\rl) } \right] \bbeta_s 
                             = \mathcal{A}_{L,\rl}^\mathrm{T} \bbeta_s.
 \end{split}                            
\end{equation}
These transformations allow for re-writing the estimators in Eqs.~\eqref{eq:genericMFAB_est} as function of the shared coordinates\footnote{The symbol ``$\eqqcolon$'' is the reverse of ``$\coloneqq$'', denoting that the right-hand side is defined as the left-hand side.}
\begin{equation} \label{eq:genericMFAB_est_share}
  \begin{split}
  \hQh(\AmatHt \ubetah)  &=\frac{1}{N} \sum_{i=1}^N Q_H\left( \bxi_H^{(i)} \right) 
                         \approx \frac{1}{N} \sum_{i=1}^N Q_H\left( \bxiHri \right) 
                         \approx\frac{1}{N} \sum_{i=1}^N Q_H \left( \mathcal{A}_{H,\rh}^\mathrm{T} \bbeta_s^{(i)} \right) 
                        \eqqcolon \hQh^{\rh}(\rh,\ubbetas) \\
  \hQl(\AmatLt \ubetal^1) &=\frac{1}{N} \sum_{i=1}^N Q_L\left( \bxi_L^{1,(i)} \right)
                           \approx \frac{1}{N} \sum_{i=1}^N Q_L\left( \bxiLri \right) 
                           \approx \frac{1}{N} \sum_{i=1}^N Q_L \left( \mathcal{A}_{L,\rl}^\mathrm{T} \bbeta_s^{(i)} \right) \eqqcolon \hQl^{\rl}(\rl,\ubbetas) \\
  \hmul(\AmatLt \ubetal^2) &=\frac{1}{\ratio N} \sum_{i=1}^{\ratio N} Q_L\left( \bxi_L^{2,(i)} \right)
                          \approx \frac{1}{\ratio N} \sum_{i=1}^{\ratio N} Q_L\left( \bxi_{L}^{\rl,(i)} \right) 
                          \approx \frac{1}{\ratio N} \sum_{i=1}^{\ratio N} Q_L \left( \mathcal{A}_{L,\rl}^\mathrm{T} \bbeta_s^{2,(i)} \right) \eqqcolon \hmul^{\rl}(\rl,\ubbetas^2), 
 \end{split}
\end{equation}
where $\ubbetas = \left\{ \bbeta_s^{(i)} \right\}_{i=1}^N$ and $\ubbetas^2 = \left\{ \bbetas^{2,(i)} \right\}_{i=1}^N$.

Finally, the MFAB estimator~\eqref{eq:genericMFAB} can be written as a function of the samples in the shared coordinates as 
\begin{equation}
 \label{eq:MFAB_share}
 \hQmfab\left( \ubbetas, \ubbetas^2; \alpha, \rh, \rl \right) = \hQh^{\rh}(\rh,\ubbetas) 
                                                              + \alpha \left( 
                                                                \hQl^{\rl}(\rl,\ubbetas)
                                                              - \hmul^{\rl}(\rl,\ubbetas^2)
                                                              \right).
\end{equation}
We reiterate that even though the MFAB estimator necessitates samples solely from the shared coordinates $\ubbetas$ and $\ubbetas^2$, each model employs a separate rotation matrix (refer to Eq~\eqref{eq:adp2org}). The following proposition illustrates the correlation between the samples for the HF and LF models.
\begin{proposition}[Original coordinate mapping through common shared space]
\label{prop:coor}
Assume $\rh = \rl$ and that the samples are generated on the common shared space $\bbetas$, then the original coordinates, obtained as $\bxiHr = \AmatHrt \bbetahr$ and $\bxiLr = \AmatLrt \bbetalr$ (see Eq~\eqref{eq:adp2org}), are related through 
\begin{equation}
	\label{eq:mapH2L}
	\begin{split}
		\bxiHr = \AmatHrt \AmatLr \bxiLr\,, \qquad
		\bxiLr = \AmatLrt \AmatHr \bxiHr \,.
	\end{split}
\end{equation}
\begin{proof}
 The proof is provided in Appendix~\ref{prop:coor_proof}.
\end{proof}
\end{proposition}
The properties of the MFAB estimator~\eqref{eq:MFAB_share} depend on the numbers of important directions $\rh$ and $\rl$, which affect both correlation between the truncated models and the estimator's MSE. The following proposition capture the effect of $\rh$ and $\rl$ in the MSE.

\begin{proposition}[MSE error for an MFAB estimator as function of $\rh$ and $\rl$]
\label{Prop:MFAB_MSE}
\begin{equation}
 \mathrm{MSE}\left[ \widehat{Q}^{MFAB} \right] = \Var{ \widehat{Q}^{MFAB} } 
      + \left( \EE{Q_H \left( \mathcal{A}_{H,\rh}^\mathrm{T} \bbeta_s \right)} - \EE{ Q_{H}(\bxiH) } \right)^2 = \sigma^2\left( \rh, \rl \right) + \delta^2 \left( \rh \right), 
\end{equation}
where
\begin{equation}
 \sigma^2\left( \rh, \rl \right) 
 = \Var{ \hQh^{\rh}(\rh,\ubbetas) } \left( 1 - \frac{\ratio-1}{\ratio} \rho^2_{r_L, r_H} \right),
\end{equation}
$\rho^2_{r_L, r_H}$ is the correlation between the models' samples in the adapted coordinates $Q_L( \mathcal{A}_{L,\rl}^\mathrm{T} \bbeta_s )$ and $Q_H( \mathcal{A}_{H,\rh}^\mathrm{T} \bbeta_s )$, i.e.,
\begin{equation}
 \rho^2_{r_L, r_H} = \frac{ \Covsq{Q_H \left( \mathcal{A}_{H,\rh}^\mathrm{T} \bbetas \right)}{Q_L \left( \mathcal{A}_{L,\rl}^\mathrm{T} \bbetas \right)} }{ \Var{Q_H \left( \mathcal{A}_{H,\rh}^\mathrm{T} \bbetas \right)} \Var{\mathcal{A}_{L,\rl}^\mathrm{T} \bbetas} },
\end{equation}
and the estimator's squared bias is
\begin{equation}
 \delta^2 \left( \rh \right) = \left( \EE{Q_H \left( \mathcal{A}_{H,\rh}^\mathrm{T} \bbeta_s \right)} - \EE{ Q_{H}(\bxiH) } \right)^2.
\end{equation} 
\begin{proof}
 The proof is provided in Appendix~\ref{Prop:MFAB_MSE_proof}.
\end{proof}
\end{proposition}
We see that the estimator's bias is introduced due to the truncation of HF in the AB method, while it is not present in the MF estimator obtained by sampling the models in the original coordinates. It is easy to show that for $\rh \rightarrow d_H$ the term $\delta^2 \left( \rh \right) \rightarrow 0$; however, this choice has a less intuitive effect on the estimator's variance. For an efficient design of the MFAB estimator it is necessary to estimate the impact that $\rh$ and $\rl$ have on the correlation and MSE of the estimator. This can be done by leveraging the PCE construction within AB, as discussed in the next section.

\subsection{Leveraging the AB framework to efficiently estimate MFAB properties without re-sampling}
\label{SSEC:PCEmetrics}
We will describe in this section how to estimate bias and correlation between models based on $\rh$ and $\rl$, with minimal computational cost. This knowledge will be fundamental for constructing optimal MFAB estimators.

\subsubsection{Bias quantification}
\label{SSSEC:PCEbias}
We aim to develop bias estimators that require only a small number of pilot samples. The approach is straightforward: first, we estimate the significance of adapted coordinates in contributing to the model's bias. Then, we use a reduced space spanned by the important directions to obtain additional samples and estimate the bias due to truncation. We present two strategies for achieving this goal.

\paragraph{Bias estimator 1: MSE of adapted PCEs.}
The first bias estimator is based on the mean square error (MSE) of the projected polynomial chaos expansion (PCE) onto the adapted space. To obtain this estimator, we first evaluate a set of $N_p$ pilot samples ${Q_H^{(i)}}_{i = 1}^{N_p}$ of the HF model in the original coordinates $\bxiH$. Using Algorithm~\ref{algo1}, we construct the rotation matrix $\AmatH$, and then, without the need for additional samples, we project the pilot samples onto the adapted space to obtain the corresponding adapted variables $\bbetahr = \AmatHr \bxiH$.  
Next, we project $\{Q_H^{(i)}\}_{i = 1}^{N_p}$ to the adapted  PCE 
\begin{equation}
	\label{eq:adapted_PCE_pilot_HF}
	{Q}_{H}(\bbetahr)
	= \sum_{\bm{\gamma} \in \mathcal{J}_{p_H}^{\rh}} c_{H, \bm{\gamma}} \psi_{\bm{\gamma}}(\bbetahr)\,.
\end{equation}
The expansion is the same as Eq~\eqref{eq:3.9}, but specific to the HF. $\mathcal{J}_{p_H}^{\rh}$ represents the sets of PCE multi-indices now for only $\rh$ adapted coordinates. An estimator for the MSE of the PCE can be defined as
\begin{equation}
\label{eq:4-9}
\text{MSE}_{PCE} \approx \frac{1}{N_p} \sum_{i=1}^{N_p} \left( Q_H(\bxiH^{(i)}) - {Q}_{H}(\bbetahr^{(i)}) \right)^2.
\end{equation}
This estimator captures the bias convergence, which will be used as a target for the variance allocation in the estimator, as shown in the numerical results section.

\paragraph{Bias estimator 2: difference in first-order PCEs.} The second estimator uses simpler information and is quantified by the difference of the first-order PCEs in the original space. In Eq~\eqref{eq:3.11}, we explained that the assumption $\bm{\eta}_{H, \neg r} = \bm{0}$ introduces errors in the adapted PCEs. Writing the transformation $\bxiH \mapsto \bbetahr \mapsto \bxiHr$ as $\bxiHr = \AmatHrt \AmatHr \bxiH$, the error is embedded in the difference between $\bxiHr$ and $\bxiH$. Since these two quantities are in the original space, we can use the PCE in this space to quantify the differences without requiring additional model evaluations. Although the high-order PCE in the original space is not available, we have constructed a first-order pilot PCE during the construction of $\AmatH$. Therefore, we can use the difference in the first-order PCEs to quantify the difference.

Suppose the first-order coefficients of the pilot PCE are $\left(c_{H, \bm{e}_1},\ldots, c_{H, \bm{e}_d}\right)$, the first-order components with respect to $\bxiH$ and $\bxiHr$ are $\left(c_{H, \bm{e}_1} \xi_{H, 1}, \ldots, c_{H, \bm{e}_d} \xi_{H, d} \right)$ and $\left( c_{H, \bm{e}_1} \xi_{H, 1}^{r_H}, \ldots, c_{H, \bm{e}_d} \xi_{H, d}^{r_H} \right)$, respectively. The difference between them can be quantified by the $l^2$-norm $\left \lVert \left(c_{H, \bm{e}_1} \xi_{H, 1}, \ldots, c_{H, \bm{e}_d} \xi_{H, d} \right) - \left( c_{H, \bm{e}_1} \xi_{H, 1}^{r_H}, \ldots, c_{H, \bm{e}_d} \xi_{H, d}^{r_H} \right) \right\rVert_2$. 
Similarly to the previous estimator, we only use this quantity to estimate the convergence rather than the MSE's magnitude. 
We introduce the following error estimator
\begin{equation}
	\label{eq:4-17}
	\varpi = \frac{1}{N_p} \sum_{i=1}^{N_p} \left\lVert \bm{w} \odot \bxiHri - \bm{w} \odot \bxiH^{(i)} \right\lVert_2 \,,
\end{equation}
where ``$\odot$'' denotes the element-wise product and $\bm{w}$ indicates the normalized first-order coefficients, i.e., \\$\bm{w} \coloneqq \left(c_{H, \bm{e}_1},\ldots, c_{H, \bm{e}_d}\right) / \lVert \left(c_{H, \bm{e}_1},\ldots, c_{H, \bm{e}_d}\right) \rVert_2$ (obtained from the first row of the rotation matrix $\AmatH$). The estimator $\varpi$ is the mean of the weighted $l^2$-norm of the difference between $\{ \bxiHri \}_{i=1}^{N_p}$ and $\{ \bxiH ^{(i)} \}_{i=1}^{N_p}$. This approach has the promise of more efficiency than the previous one since the first-order coefficients can be obtained with limited model evaluations.

Algorithm~\ref{ALG:bias_estimator} summarizes the numerical procedures designed to obtain both the estimators presented above. After the evaluation of the MSE, a truncation of the adapted space is obtained (according to a user-specified tolerance). It is worth noting that this procedure does not require additional evaluations since the absolute value of the MSE is not computed, but rather the convergence is compared with respect to a threshold level specified as input. 
\begin{algorithm}[htb]
	\caption{Determination of the adapted HF dimension based on the bias estimators}\label{ALG:bias_estimator}
	
	\KwIn{$\{ \bxiH ^{(i)} \}_{i=1}^{N_p}$, $\{Q_H(\bxiH ^{(i)})\}_{i=1}^{N_p}$: pilot input and corresponding HF samples;\newline
		$\AmatH$: rotation matrix of the HF model; \newline
		$\chi$: estimator indicator ($\chi=1$ for estimator 1 and $\chi=2$ for estimator 2); \newline 
		$\vartheta$: threshold  of cumulative difference ratio  ($0<\vartheta<1$).}
	
	\uIf{$\chi=1$ \label{step1}}{
		\For{$r_H = 1, \ldots, d_H$}{
		Compute $\bbetahr^{(i)} = \AmatHr \bxiH^{(i)}$, for $i=1,\ldots, N_p$ ;
		
		Compute the adapted PCE coefficients $c_{H, \bm{\gamma}}$ in ~\eqref{eq:adapted_PCE_pilot_HF} by least square regression;
		
		Evaluate ${Q}_H(\bbetahr^{(i)})$ by the constructed PCE;
		
		Compute the MSE of the PCE surface, $\text{MSE}_{PCE}$, of Eq~\eqref{eq:4-9}.
		}
	}
	\ElseIf{$\chi=2$}{
		Let $\bm{\omega}$ be the first row of $\AmatH$;
		
		\For{$r_H = 1, \ldots, d_H$}{
			Compute $\bxiHri = \AmatHrt \AmatHr \bxiH^{(i)}$, for $i=1,\ldots, N_p$ , and $\varpi$ by Eq~\eqref{eq:4-17}. \label{step10}
		}
	}
	
	Denote the estimator with respect to $\rh$ as $\bm{\theta} = \{\theta_{\rh} \} _{\rh=1} ^{d_H}$ without distinguishing the two estimators; \label{step:estimator}
	
	Compute the bias improvement as the adjacent difference of $\bm{\theta}$ as $\bm{\theta}_{\text{diff}} = \{0, \theta_1-\theta_2, \theta_2-\theta_3, \ldots, \theta_{d_H-1} -  \theta_{d_H}\}$; \label{step:diff_estimator}
	
	Compute the cumulative bias improvement as the cumulative difference $\bm{\theta}_{\text{cdiff}} = \{0, \theta_1-\theta_2, \theta_1-\theta_3, \ldots, \theta_{1} -  \theta_{d_H}\}$; \label{step:cum_diff_estimator}
	
	Compute the cumulative to total difference ratio as 
	\begin{equation}
		\label{eq:cdiff_ratio}
		\bm{\varrho} = \{\varrho_1, \ldots, \varrho_{d_H}\} = \left\{ \theta_{\text{cdiff}, r_H} / (\theta_1 - \theta_{d_H})\right\}_{r_H = 1}^{d_H} = \frac{\bm{\theta}_{\text{cdiff}}}{\theta_1-\theta_{d_H}}\,;
	\end{equation}
	
	Choose $\rh$ as the adapted dimension for HF where $\varrho_{\rh}$ is the first element in $\bm{\varrho}$ such that $\varrho_{\rh} \geq \vartheta$.
	
	\KwOut{$\rh$: the adapted dimension of the HF model.}
\end{algorithm}

\subsubsection{Correlation quantification}
\label{SSSEC:PCEcorrelation}
In this section, we develop a procedure to estimate the variance of the MFAB estimator, which depends on both $\rh$ and $\rl$, as shown in Proposition~\ref{prop:coor}. We extend a simplified approach presented in~\cite{Geraci2018}, which avoids re-evaluating the models by using the PCE in AB.

Similar to Eq~\eqref{eq:adapted_PCE_pilot_HF}, an adapted PCE of LF can be built using the pilot samples as 
\begin{equation}
	\label{eq:adapted_PCE_pilot_LF}
 \begin{split}
 {Q}_L(\bbetalr) &= \sum_{\bm{\gamma} \in \mathcal{J}_{p_L}^{\rl}} c_{L, \bm{\gamma}} \psi_{\bm{\gamma}}(\bbetalr),
 \end{split}
\end{equation}
where $\bbetalr = \AmatLr \bxiLr$, while $\mathcal{J}_{p_L}^{\rl}$ represents the sets of PCE multi-indices corresponding to the truncated space for the LF. The expansion is the same as Eq~\eqref{eq:3.9}, but specific to the LF. The correlation between the HF and LF models can be estimated via the PCE as illustrated in the following proposition.
\begin{proposition}[Correlation between models as function of the number of important directions]
\label{prop:PCE_AB_correlation}
 Let two models be represented by their PCEs, Eqs~\eqref{eq:adapted_PCE_pilot_HF} and \eqref{eq:adapted_PCE_pilot_LF},
their correlation is expressed by
\begin{equation}
 \label{eq:corr_estim}
 \rho\left( {Q}_H(\bbetahr),{Q}_L(\bbetalr) \right) = 
      \frac{ \sum_{ \bm{\gamma}  \in \left(\mathcal{J}_{p_H}^{\rh} \bigcap \mathcal{J}_{p_L}^{\rl} \right) \setminus \bm{e}_0 } c_{H, \bm{\gamma}} c_{L, \bm{\gamma}} }
      { \sqrt{ \left(\sum_{\bm{\gamma} \in \mathcal{J}_{p_H}^{\rh} \setminus \bm{e}_0} c_{H, \bm{\gamma}}^2  \right) 
               \left(\sum_{\bm{\gamma} \in \mathcal{J}_{p_L}^{\rl} \setminus \bm{e}_0} c_{L, \bm{\gamma}}^2  \right)} }.
\end{equation}
\begin{proof}
 The proof is provided in Appendix~\ref{prop:PCE_AB_correlation_proof}.
\end{proof}
\end{proposition}

From Proposition~\ref{prop:PCE_AB_correlation}, the following corollary can be obtained.
\begin{corollary}[The correlation between two PCEs is maximized if each model is represented with only the shared multi-indices] 
 \label{coroll:nonuniform}
	For two PCEs with multi-indices $\mathcal{J}_{p_H}^{\rh}$ and $\mathcal{J}_{p_L}^{\rl}$, their correlation $\rho$ is maximized if both expansions use the shared multi-indices given by $\mathcal{J}_{p_H}^{\rh} \bigcap \mathcal{J}_{p_L}^{\rl}$, that is, with abuse of notation,
	\begin{equation}
	 \argmax_{(\mathcal{J}_{p_H}^{\rh},\mathcal{J}_{p_L}^{\rl})} \rho^2\left( {Q}_H(\mathcal{J}_{p_H}^{\rh}),{Q}_L(\mathcal{J}_{p_L}^{\rl}) \right) = \left(\mathcal{J}_{p_H}^{\rh} \bigcap \mathcal{J}_{p_L}^{\rl},\; \mathcal{J}_{p_H}^{\rh} \bigcap \mathcal{J}_{p_L}^{\rl} \right),
	\end{equation}
    where $\rho^2\left( {Q}_H(\mathcal{J}_{p_H}^{\rh}),{Q}_L(\mathcal{J}_{p_L}^{\rl}) \right)$ is obtained by Eq~\eqref{eq:corr_estim}.
	If we assume that the PCE expansion for the HF model includes all multi-indices from the LF model, we can further write
	$\mathcal{J}_{p_H}^{\rh} \bigcap \mathcal{J}_{p_L}^{\rl} = \mathcal{J}_{p_L}^{\rl}$.
	\begin{proof}
      The proof for the later case is provided in Appendix~\ref{coroll:nonuniform_proof}.
    \end{proof}
\end{corollary}

Corollary~\ref{coroll:nonuniform} demonstrates that increasing the dimension of one model in the adapted space beyond the dimension of the other model reduces the correlation between the HF and LF models. 

We observe that the PCE framework provides a straightforward method to estimate the correlation, as explained in Proposition~\ref{prop:PCE_AB_correlation}. Additionally, the framework enables us to comprehend the benefits of sampling on the shared space over sampling in the original space. This proposition is presented below to illustrate this advantage.

\begin{proposition}[Correlation in the adapted space is greater than correlation in the original space]\label{prop:correl_increase_adapt}
	Consider PCEs of the HF and LF of Eqs~\eqref{eq:adapted_PCE_pilot_HF} and \eqref{eq:adapted_PCE_pilot_LF}, but with polynomial order one, i.e., $p_H = p_L = 1$, and with dimensions $r_H = r_L = d_H = d_L = d$, implying $\bm\eta = \bbeta_s = \bbetahr = \bbetalr$ for the shared adapted variables. If we 
	\begin{itemize}
	 \item[I.] Assume that the first-order PCE coefficients of HF and LF are greater than or equal to zero;
	 \item[II.] The rotation matrices on HF and LF only permute the original variables $\bm{\xi}$ to rearrange them in decreasing order as in $\bm{\eta}$;
	\end{itemize}
	then the squared correlation between the two models obtained by sampling in the important directions is greater than the correlation in the physical variables
	\begin{equation}
		\rho^2\left( {Q}_{H}(\bm{\eta}), {Q}_{L}(\bm{\eta}) \right) \geq \rho^2\left( {Q}_{H}(\bm{\xi}), {Q}_{L}(\bm{\xi}) \right).
	\end{equation}
\begin{proof}
 The proof is provided in Appendix~\ref{prop:correl_increase_adapt_proof}.
\end{proof}
\end{proposition}

This proposition demonstrates that if the variables in both the HF and LF models are ranked by their importance and a shared shape is constructed based on this ranking, their correlation can be increased. Although the construction of the rotation in the proposition is simplified compared to AB, the key assumption of rearranging the variables in descending order of importance is still present. Therefore, this result also applies to AB.

The propositions presented in this section aim to provide intuitions and are valid under specific assumptions. For instance, they do not consider practical scenarios where truncation can introduce rotation errors. Therefore, it is necessary to develop a numerical strategy for selecting the truncation dimensions and incorporating it into the construction of MFAB, as discussed in the next section.

\section{Practical implementation of MFAB}
\label{SEC:MFAB_practical}
In Section~\ref{SEC:ABforMF}, we demonstrated how the correlation between models could be maximized by limiting the number of terms in their AB representations. On the other hand, the MSE of the MFAB estimator depends only on the bias of the AB representation of the HF model, which decreases as $\rh$ approaches $d_H$. Moreover, the optimal truncation for the models needs to be determined from a limited number of available pilot samples in the original coordinates. In this section, we introduce the design of MFAB estimators based on these considerations. We provide two MFAB estimators. The first one balances its variance with the residual bias of the HF model in the truncated AB space, aiming to maximize the correlation among the models by minimizing the shared space as much as possible. The second estimator maintains a full representation of the HF model, which does not introduce any bias, but results in a lower correlation between models. These estimators have different data requirements and can be useful in different scenarios, as discussed later.    

\subsection{Bias-variance balanced MFAB estimator}
The first MFAB estimator aims to maximize the correlation among the HF and LF models by leveraging the smallest possible important directions $\rh$ (see Proposition~\ref{coroll:nonuniform}). As this choice affects the MFAB estimator's MSE (see Proposition~\ref{Prop:MFAB_MSE}), we need to determine the acceptable estimator bias and then use its value as the target variance in the sample allocation step. By relying on the tools presented in the previous section, this approach is computationally straightforward and is summarized in Algorithm~\ref{ALG:biased}, which uses Algorithm~\ref{ALG:MFAB} as a subroutine.

The approach assumes that additional HF and LF model evaluations are allowed, and we refer to this as the \textit{non-legacy dataset} case. To construct the estimator, we start with collecting pilot samples for both models corresponding to the same coordinates in the original space. Algorithm~\ref{algo1} can then be used for the HF and LF models to obtain their rotation matrices $\AmatH$ and $\AmatL$, independently. We use the HF rotation matrix as input (along with other user-defined quantities) for Algorithm~\ref{ALG:bias_estimator} to determine the truncation dimension $\rh$. Then, we perform a grid search on the truncation $\rl$ to evaluate the maximum correlation between the two models. Maximizing the correlation is equivalent to maximizing the use of LF information, which, in turn, corresponds to minimizing the estimator's variance for a fixed cost. The shared space can be defined according to Eq~\eqref{eq:beta_share}, which allows us to evaluate additional samples for both models, if needed.

Since Algorithm~\ref{ALG:biased} balances bias and variance, we must estimate a target variance for the MFAB estimator. The two proposed bias estimators introduced in the previous section only capture the convergence trend, so we also need to estimate the bias associated with $\rh$ in HF. This can be done by evaluating an additional $N_{\rh}$ HF samples in the shared space, as shown in Eq~\eqref{eq:bias_quan}. The MFAB estimator is designed for problems with a small $\rh$, so the additional number of $N_{\rh}$ samples required is also small. Once the shared space has been defined, converging the sample allocation for the MFAB estimator is no different from traditional multilevel/MF estimators, such as MLMC, MFMC, or ACV. This task involves converging on the values for the oversampling ratio $\ratio$ and the number of HF runs $N$, where the total number of LF samples is $\lceil \ratio N \rceil$. The detailed process to compute the MFAB estimator is presented in Algorithm~\ref{ALG:MFAB}, which specifies options for non-legacy datasets and bias-variance balancing.

\begin{algorithm}[htb]
	\caption{MFAB estimator with given adapted HF and LF models}\label{ALG:MFAB}
	
	\KwIn{
		$\{ \bxiH ^{(i)} \}_{i=1}^{N_p}$, $\{Q_H(\bxiH ^{(i)})\}_{i=1}^{N_p}$, 
		$\AmatH$, $\AmatL$, $\rh$, $\rl$; \newline
		\texttt{Legacy\_flag}: using legacy dataset (\texttt{True}) or allow for additional HF runs (\texttt{False}); \newline
		\texttt{Bias\_flag}: select if the bias-variance balanced (\texttt{True}) estimator is desired\newline
		$\sigma^2$: target variance when non-legacy dataset is used.
	}
	
	\uIf{\texttt{Legacy\_flag} == \texttt{True} }
	{
		Project $\bxiH$ to $\bbeta_s$ by Eq~\eqref{eq:beta_share}, map them to $\bxiLr$ by Eq~\eqref{eq:adp2org}, and evaluate $Q_{L}(\bxiLr)$;
		
		Compute $\ratio$ by \eqref{eq:MF_samp_ratio};	
	}
	
	\Else
	{
		Specify the target variance $\sigma^2$;
		
		\uIf{\texttt{Bias\_flag} == \texttt{True}}
		{
			Generate  $N_p$  samples of $\bbeta_s$, map them to $\bxiHr$ and $\bxiLr$ by Eq~\eqref{eq:adp2org}, and evaluate $Q_{H}(\bxiHr)$ and $Q_{L}(\bxiLr)$;
		}
		\Else{
			Generate  $N_p$  samples of $\bxiH$, project them to $\bbeta_s$,  map $\bbeta_s$ to $\bxiLr$, and evaluate $Q_{H}(\bxiH)$ and $Q_{L}(\bxiLr)$;
		}
		\While{True}{
			Compute $\ratio$ by \eqref{eq:MF_samp_ratio} and the required HF samples $N$ by \eqref{eq:required_N_HF};
			
			Compute additional HF samples $\Delta N = N - N_p$;
			
			\uIf{$\Delta N = 0$}{
				\textbf{Break};
			}
			\Else{
				\uIf{\texttt{Bias\_flag} == \texttt{True}}
				{
					Generate $\Delta N$ samples of $\bbeta_s$, map them to $\bxiHr$ and $\bxiLr$, and evaluate $Q_{H}(\bxiHr)$ and $Q_{L}(\bxiLr)$;
				}
				\Else{
					Generate $\Delta N$ samples of $\bxiH$, map to $\bbeta_s$ and then to $\bxiLr$, and evaluate $Q_{H}(\bxiH)$ and $Q_{L}(\bxiLr)$;
				}
				Append the new HF and LF samples to the existing samples and update $N_p = N$;
			}
		}
	}
	
	Generate $\lceil (\ratio-1) N \rceil$ samples of $\bbetalr$, map them to $\bxiLr$, and evaluate $Q_L(\bxiLr)$;
	
	Append LF samples to the existing samples and compute MFAB estimator by \eqref{eq:MFAB_share} and its variance.
	
	\KwOut{$\widehat{Q}^{MFAB}$, MFAB estimator.}
\end{algorithm}

\begin{algorithm}[htb]
	\caption{{Bias-variance balanced MFAB estimator.}}\label{ALG:biased}
	
	\KwIn{$\{ \bxiH ^{(i)} \}_{i=1}^{N_p}$, $\{Q_H(\bxiH ^{(i)})\}_{i=1}^{N_p}$, $\{ \bxiL ^{(i)} \}_{i=1}^{N_p}$, $\{Q_L(\bxiL ^{(i)})\}_{i=1}^{N_p}$; \newline
		$\chi$: bias estimator indicator (choose from $\{1, 2\}$); \newline 
		$\vartheta$: threshold  of cumulative to total difference ratio  ($0<\vartheta<1$); \newline 
		$N_{\rh}$: number of additional samples to quantify the bias of the $\rh$-d adapted HF model.}
		
	Construct rotation matrices of $\AmatH$ and $\AmatL$ by Algorithm \ref{algo1} from pilot samples;
	
	Determine $\rh$  from $\AmatH, \chi, \vartheta$ by Algorithm~\ref{ALG:bias_estimator};

	\SetKwProg{myproc}{Choose the adapted LF dimension with maximized correlation}{}{}
	\myproc{}{
			\For{$r_L$= 1 to $d_L$}{
				Compute the correlation of $r_H$-d adapted HF and $r_L$-d adapted LF by \eqref{eq:corr_estim};
			}
	
		Choose the LF dimension $r_L$ that maximizes the correlation \eqref{eq:corr_estim};
	}

	\SetKwProg{myproc}{Estimate squared bias $\delta^2(\rh)$}{}{}
	\myproc{}{
	Generate addition $N_{\rh}$ samples of $\bbeta_s$, \label{step:bias}
	\begin{equation}
	\label{eq:bias_quan}
    \begin{split}
	 \delta \left( \rh \right) 
	\approx \widehat{\delta}(\rh)
	\coloneqq \frac{1}{N_{\rh}} \sum_{i=1}^{N_{\rh}} Q_H \left( \mathcal{A}_{H,\rh}^\mathrm{T} \bbeta_s^{(i)} \right) - \frac{1}{N_p} \sum_{i=1}^{N_p} Q(\bxiH^{(i)})   \,.
    \end{split}
    \end{equation}
	}
	
	Compute MFAB estimator by Algorithm~\ref{ALG:MFAB} with \texttt{Legacy\_flag}=\texttt{False}, \texttt{Bias\_flag} = \texttt{True}, and target variance $\sigma^2 =  \widehat{\delta}^2(\rh)$.
	
	\KwOut{$\widehat{Q}^{MFAB}$, MFAB estimator.}
	
\end{algorithm}

\subsection{Unbiased MFAB estimator for \emph{legacy} and \emph{non-legacy} high-fidelity data}
\label{SSEC:MFABunbiased}
As shown in the previous section, constructing the bias-variance balanced MFAB estimator always requires evaluating additional HF samples by sampling the shared space $\bbetas$. However, it's not always possible to obtain additional evaluations for the HF model, such as when using a \emph{legacy dataset}. In this case, an MFAB estimator can still be obtained by sampling the LF model to determine the optimal sample allocation, while the number of HF evaluations $N$ remains fixed. This corresponds to using a full (non-truncated) representation of the HF model, while the LF model can still be arbitrarily reduced.

One key difference from the bias-variance balanced strategy is that the estimator is unbiased because the HF model is non-truncated. The shared space spanned by $\bbetas$ can still be constructed and used to maximize correlation. However, the shared samples are not sampled directly but obtained by mapping the available HF realizations to this space through the rotation matrix $\AmatH$. 

A numerical procedure to build this MFAB estimator is presented in Algorithm~\ref{ALG:unbiased}, where the subroutine Algorithm~\ref{ALG:MFAB} is used with the \texttt{Legacy\_flag} set to \texttt{True}. First, we search for the HF important directions that can represent the model accurately by Algorithm~\ref{ALG:bias_estimator}. This step does not require re-evaluating the HF model to determine $\delta^2$, i.e., $\delta^2=0$, given that HF samples are evaluated on the full space. Second, we map the set of HF samples from its original space to its auxiliary space by
\begin{equation}
\label{eq:4-18}
\bbetahr = \AmatHr \bxiH.
\end{equation}
For an assigned number of important directions $\rl$ for the LF, we can assemble the shared space $\bbetas$ by Eq~\eqref{eq:beta_share} and then proceed by sampling the LF given the mapping $\bxiLr = \mathcal{A}_{L,\rl}^\mathrm{T} \bbeta_s$. For the last step of the algorithm, we need to converge the oversampling ratio $\ratio$; by doing so we minimize the variance for the estimator with an assigned number of HF realizations.

The remaining question is how to select the optimal number of important directions for the LF model. Conceptually, this can be done similarly to the previous case (Algorithm~\ref{ALG:biased}) by performing a grid search over $\rl$. However, there is a notable difference: we cannot rely on the correlation expression introduced in Proposition~\ref{prop:PCE_AB_correlation}, as the HF model is not represented by its AB counterpart. Hence, it is necessary here to re-evaluate the LF model for each $1\leq\rl\leq d_L$ and explicitly compute the correlation between the models as
\begin{equation}
	\label{eq:corr_unbiased}
	\begin{split}
		\rho^2 \left( Q_H(\bxiH),\, Q_L(\bxiLr) \right)
		\approx\frac
		{\left( \sum_{i=1}^{N_p} \left( Q_H(\bxi_H^{(i)}) - \widehat{Q}_H(\bxi_H^{(i)})\right) \left( Q_L(\bxiLri) - \widehat{Q}_L(\bxiLri)\right) \right)^2}
		{\sum_{i=1}^{N_p} \left( Q_H(\bxi_H^{(i)}) - \widehat{Q}_H(\bxi_H^{(i)})\right)^2
		\sum_{i=1}^{N_p} \left( Q_L(\bxiLri) - \widehat{Q}_L(\bxiLri)\right) ^2}\,.
	\end{split}
\end{equation}
This process incurs the additional cost of discarding LF evaluations, which, depending on the application and the computational cost of the model, can be more or less feasible. To alleviate the cost of this grid search, we limit it to the space $1\leq \rl \leq \rh$ (knowing that increasing it would not help, see Corollary~\ref{coroll:nonuniform}).

\begin{algorithm}[htb]
	\caption{Unbiased MFAB estimator for legacy and non-legacy HF dataset.}
	\label{ALG:unbiased}
	
	\KwIn{$\{ \bxiH ^{(i)} \}_{i=1}^{N_p}$, $\{Q_H(\bxiH ^{(i)})\}_{i=1}^{N_p}$, $\{ \bxiL ^{(i)} \}_{i=1}^{N_p}$, $\{Q_L(\bxiL ^{(i)})\}_{i=1}^{N_p}$; \newline
		$\chi$: bias estimator indicator (choose from $\{1, 2\}$); \newline 
		$\vartheta$: threshold  of cumulative to total difference ratio  ($0<\vartheta<1$); \newline
		\texttt{Legacy\_flag}: using legacy dataset if \texttt{True} else using non-legacy dataset with target variance.}
	
	Construct rotation matrices of $\AmatH$ and $\AmatL$ by Algorithm \ref{algo1} from pilot samples;
	
	Determine $\rh$  from $\AmatH, \chi, \vartheta$ by Algorithm~\ref{ALG:bias_estimator};

	\SetKwProg{myproc}{Choose the adapted LF dimension with maximized correlation}{}{}
	\myproc{}{

			\For{$r_L$= 1 to $\rh$}{
				Project $\bxiH$ to $\bm{\eta}_{H, \rh}$ by \eqref{eq:4-18}, and assemble $\bbeta_s$ by \eqref{eq:beta_share}; 
				
				Map $\bbeta_s$ to $\bxiLr$ by the second equation in Eq~\eqref{eq:adp2org} and evaluate $Q_L(\bxiLr)$;
				
				Compute the correlation of $\rh$-d projected HF and $r_L$-d adapted LF by \eqref{eq:corr_unbiased};
			}
		
		Choose the LF dimension $r_L$ that maximizes the correlation expressed by Eq~\eqref{eq:corr_unbiased};
	}
	
	Compute MFAB estimator by Algorithm~\ref{ALG:MFAB} with \texttt{Bias\_flag}=\texttt{False}.
	
	\KwOut{$\widehat{Q}^{MFAB}$, MFAB estimator.}
	
\end{algorithm}

Due to the projection error of the HF model, the correlation in this case is smaller than when both models are generating input samples in the reduced space (see Proposition~\ref{prop:PCE_AB_correlation}). Since this contribution would depend on the effect introduced by the truncated $(d_H - \rh)$ variables, it is expected to decay rapidly once the model's response has been captured by the first $\rh$ important variables. 

In principle, the approach with the legacy HF dataset can be easily extended to the non-legacy HF dataset, which is also presented in Algorithms~\ref{ALG:unbiased} and  \ref{ALG:MFAB} by using the \textit{non-legacy dataset} option. The only modification required would happen in the resource allocation step of the Algorithm~\ref{ALG:MFAB}. With the unbiased option, the HF model must be sampled in its original space to obtain the optimal $N$. The resulting algorithm would be an unbiased MFAB estimator with prescribed target variance, which could provide an alternative to Algorithm~\ref{ALG:biased}. We will discuss and compare these two algorithms in the next section.

\section{Numerical examples}
\label{SEC:numres}
In this section, we present several numerical results to demonstrate the features of the MFAB approach. First, we adopt an analytical test problem in Section~\ref{ssec:numres_analytical} to discuss a wide range of results, given the availability of both an exact solution and the low computational cost of the models. Afterward, we illustrate two more challenging computational problems. In Section~\ref{ssec:numres_acoustic}, we consider an acoustic problem inspired by a direct field acoustic test, while in Section~\ref{ssec:numres_fem}, we use a finite element model for a realistic nuclear fuel assembly.

\begin{figure}[htb]
    \centering
    \begin{minipage}{0.5\textwidth}
        \centering
        \includegraphics[width=0.8\textwidth]{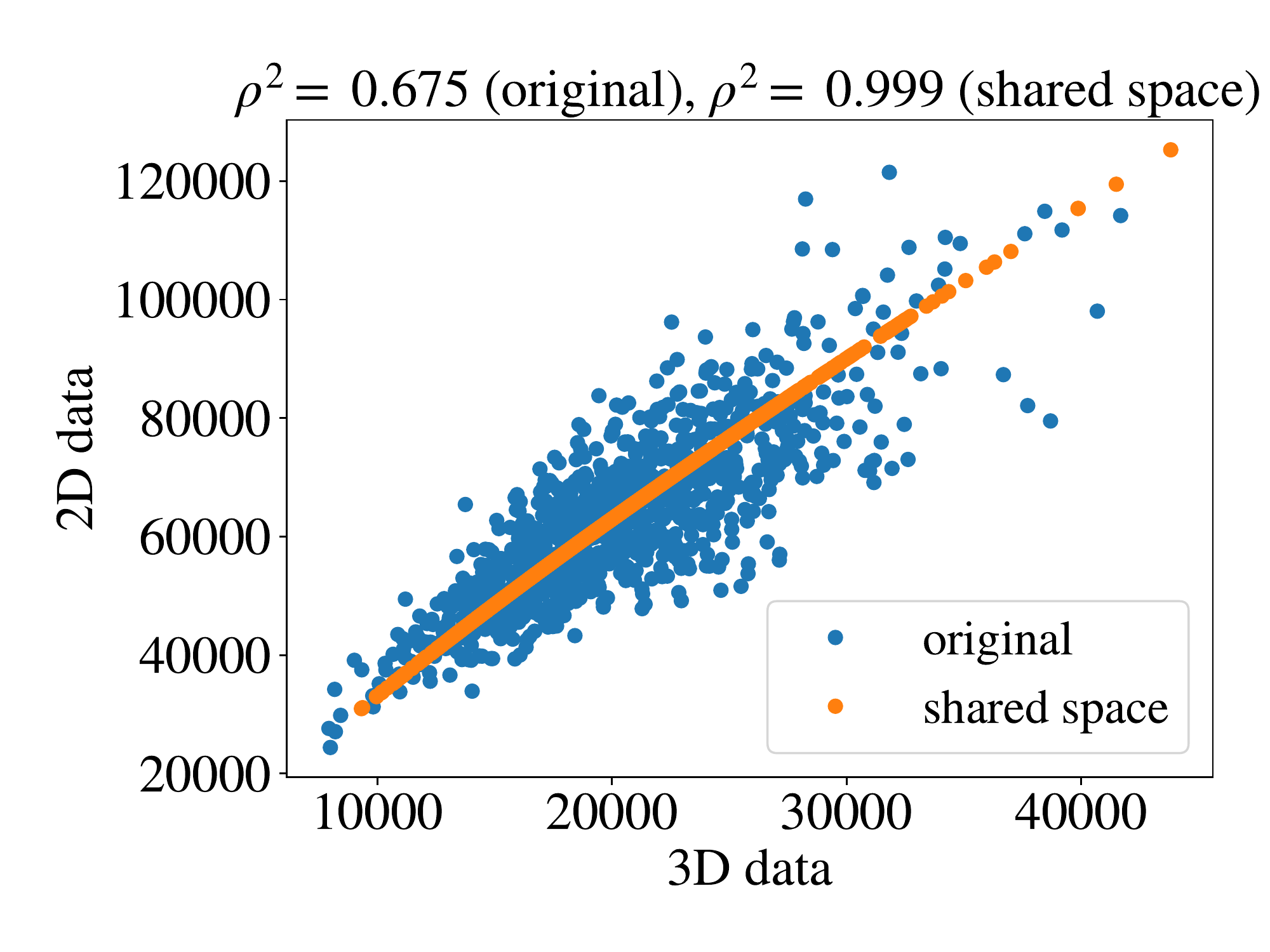}
        \subcaption{Exit pressure $P_e$ [Pa] for the elliptical geometry section}
        \label{fig:scatterplot_3D_2D_ellipse_adp}
    \end{minipage}%
    \begin{minipage}{0.5\textwidth}
        \centering
        \includegraphics[width=0.8\textwidth]{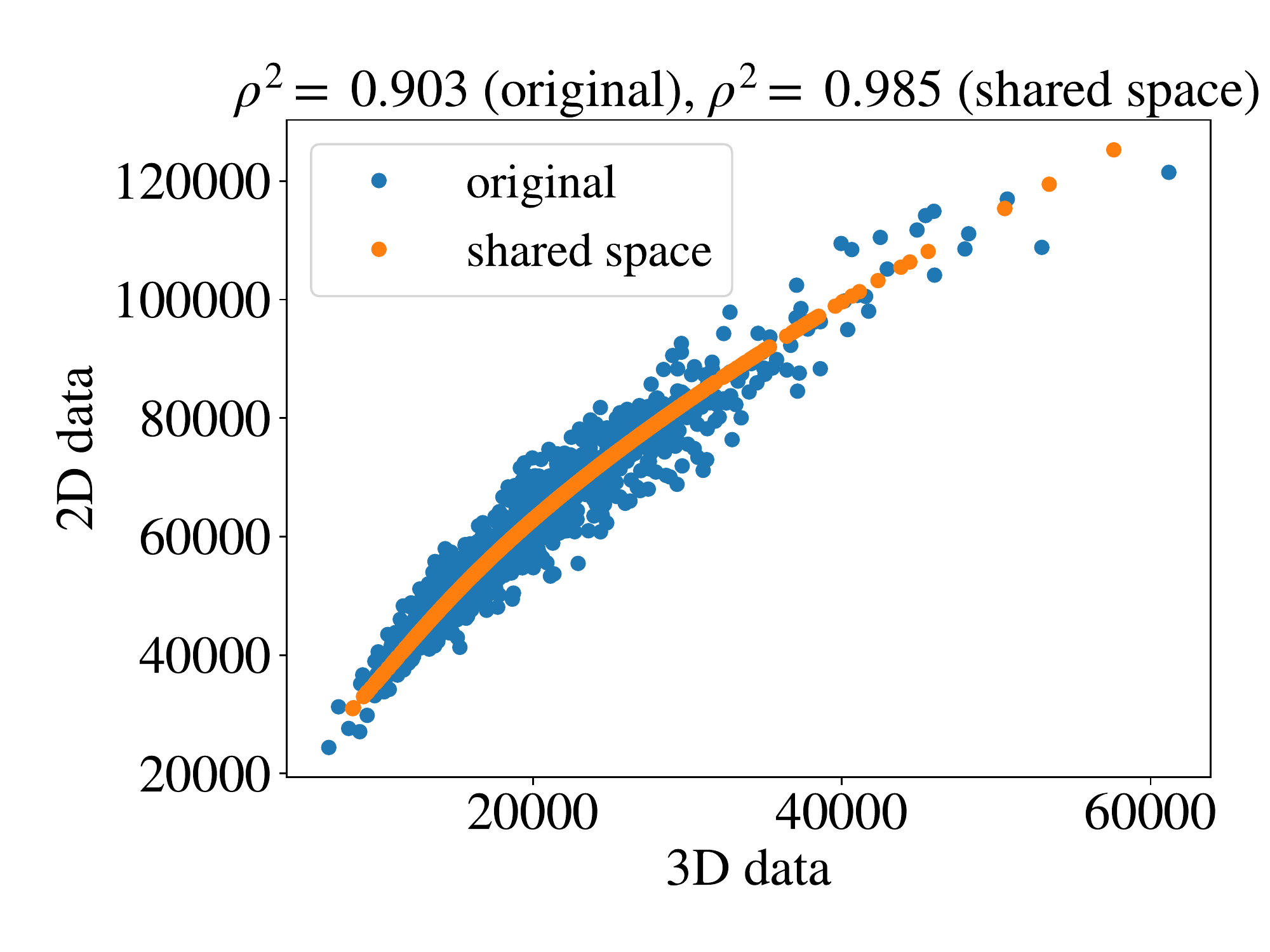}
        \subcaption{Exit pressure $P_e$ [Pa] for the circular geometry section}
        \label{fig:scatterplot_3D_2D_circular_adp}
    \end{minipage}
    \caption{Scatter plot for the nozzle flow in the elliptical (\subref{fig:scatterplot_3D_2D_ellipse_adp}) and circular (\subref{fig:scatterplot_3D_2D_circular_adp}) geometry case. This figure augments Figure~\ref{fig:nozzle_scatter_original} by overlaying the scatter plot of the model sampled in the shared space (orange dots) with the one corresponding to the models sampled in the original space (blue dots). In both cases, the correlation is greatly increased, demonstrating how the variability of the models can be well-captured by resorting to a 1D shared space even if the models are described by different parameters.}
    \label{fig:nozzle_scatter_shared}
\end{figure}

Before considering the numerical examples, let us return to the nozzle example introduced in Section~\ref{SEC:motivation} to demonstrate the increase in correlation in both MF scenarios. The purpose of this simple demonstration is to support our intuition that relying on a shared space to correlate models with dissimilar parametrizations can be understood from the physical understanding of the problem\footnote{similar interpretations for more complex problems, like the ones presented in the following sections, are hampered by the complexity of the underlying physical models}. For the nozzle example, a shared space with a single important direction is sufficient to correlate the elliptical and circular high-fidelity cases, in which the elliptical case has 3 parameters for the HF model and 2 parameters for the LF model. In both cases, the LF model is a 2D approximation with only two parameters: one for the geometry, \textit{in lieu} of the two geometrical parameters of the HF model, and one for the total pressure, which is shared by the models. In Figure~\ref{fig:nozzle_scatter_shared} shows the scatter plots for these two scenarios; this figure augments Figure~\ref{fig:nozzle_scatter_original} with the datapoints obtained by sampling the models at the same shared locations. We note that, in both scenarios, the correlation obtained by sampling the models in the shared space is significantly increased compared to the correlation obtained by sampling the original coordinates. As described in the previous sections, the correlation is the main parameter controlling the variance of the MF estimator (along with the model costs). Therefore, we can easily link the improvement in correlation to the improvement in MFAB efficiency. We will discuss the performance of the MFAB estimator in detail for the numerical examples in the following sections.

\subsection{Analytical test problem}
\label{ssec:numres_analytical}
\begin{figure}[htb]
	\centering
	\begin{subfigure}[t]{0.5\linewidth}
		\centering
		\includegraphics[width=0.8\linewidth]{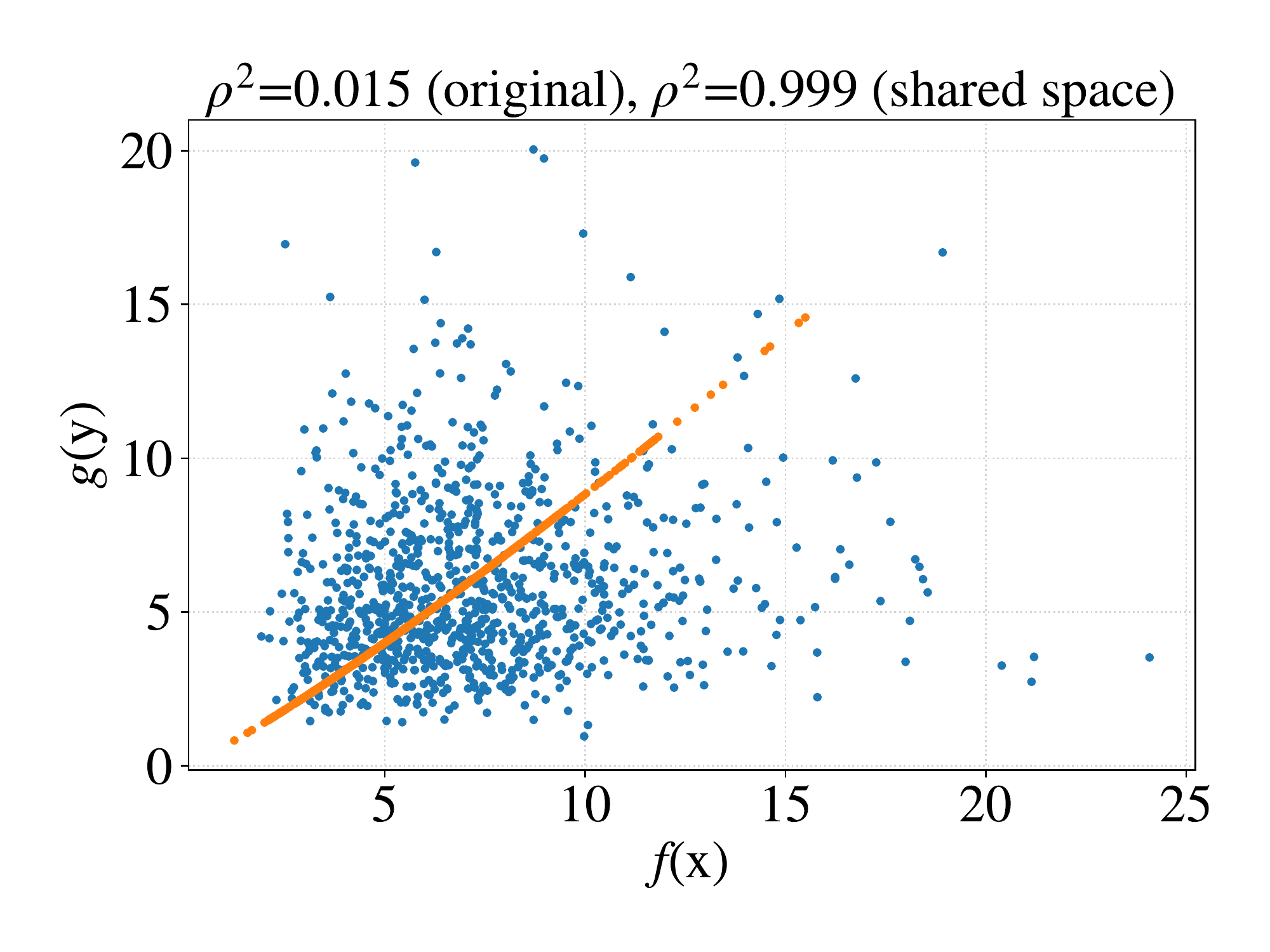}
		\subcaption{Scatter plots}
	\end{subfigure}
    ~\hspace{-5mm}
	\begin{subfigure}[t]{0.5\linewidth}
		\centering
		\includegraphics[width=0.8\linewidth]{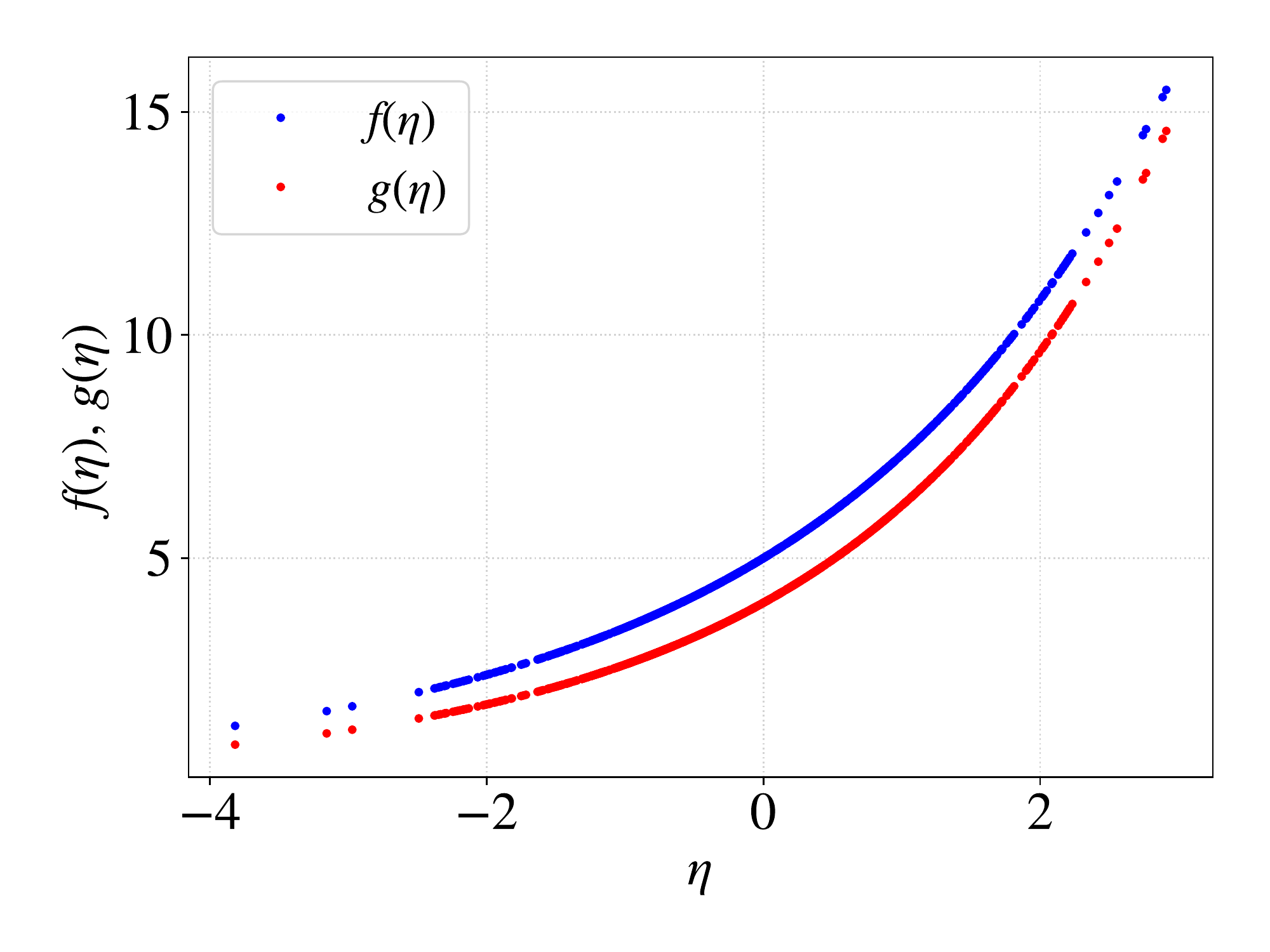}
		\subcaption{Models' response over the first shared variable}
	\end{subfigure}
	\caption{ (a) scatter plots for the models sampled in the original and shared spaces with a single shared variable and (b) their responses over the shared variable.}
	\label{fig:exponential_problem_scatter}
\end{figure}

The first example we consider is an analytical test case, which, by providing closed-form solutions, is helpful to illustrate and verify the algorithm. Consider two models: a HF model $f=f(\bm{x})$, characterized by 10 uncertain parameters, i.e., $\bm{x} \in \mathbb{R}^{10}$, and a LF model $g=g(\bm{y})$ with 8 uncertain inputs, i.e., $\bm{y} \in \mathbb{R}^8$, defined by
\begin{equation}
	\begin{split}
		\label{eq:5.1}
		f(\bm{x}) = &\exp(x_1+0.05x_2) + \exp(0.8 x_3) + \exp(0.8 x_4+0.05x_5+0.05x_6) \\
		&+ \exp(0.8 x_7+0.05x_8) + \exp(0.08x_9 + 0.05x_{10})\,, \\
		g(\bm{y}) = &\exp(0.1y_1+y_2) + \exp( 0.1 y_3 + 0.01 y_4 + 0.8 y_5) + \exp(0.8 y_6 + 0.1y_7) + \exp(0.8 y_8)\,.
	\end{split}
\end{equation}
We assume $x_1, ..., x_{10} \sim \mathcal{N}(0, \,1)$ and  $y_1, ..., y_8 \sim \mathcal{N}(0, \,1)$. To illustrate the characteristics of the models, in Figure~\ref{fig:exponential_problem_scatter} we report the scatter plots (1000 samples) for the models sampled in the original and shared coordinates. Here, we used a single shared variable such that the models' response over this variable can be easily represented (see Figure~\ref{fig:exponential_problem_scatter}(b)). As visible from the plot, the correlation among the models is greatly increased by resorting to the shared variables.

\begin{figure}[htb]
	\centering
	\begin{minipage}{0.48\linewidth}
		\centering
		\includegraphics[width=0.8\linewidth]{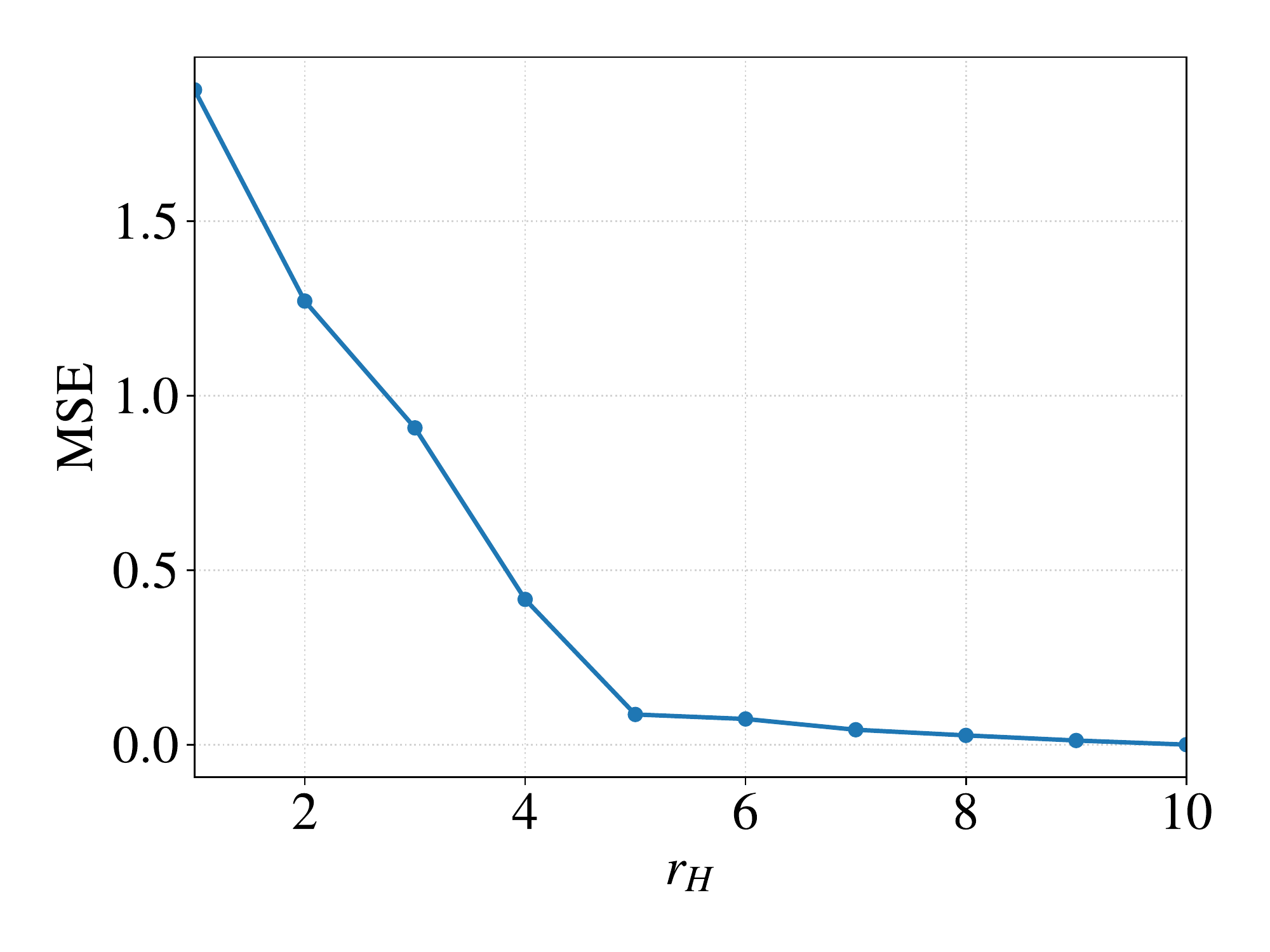}
		\subcaption{MSE of adapted PCEs ($\chi=1$).} 
		\label{fig:5.2}
	\end{minipage}
	%
	%
	\begin{minipage}{0.48\linewidth}
		\centering
		\includegraphics[width=0.8\linewidth]{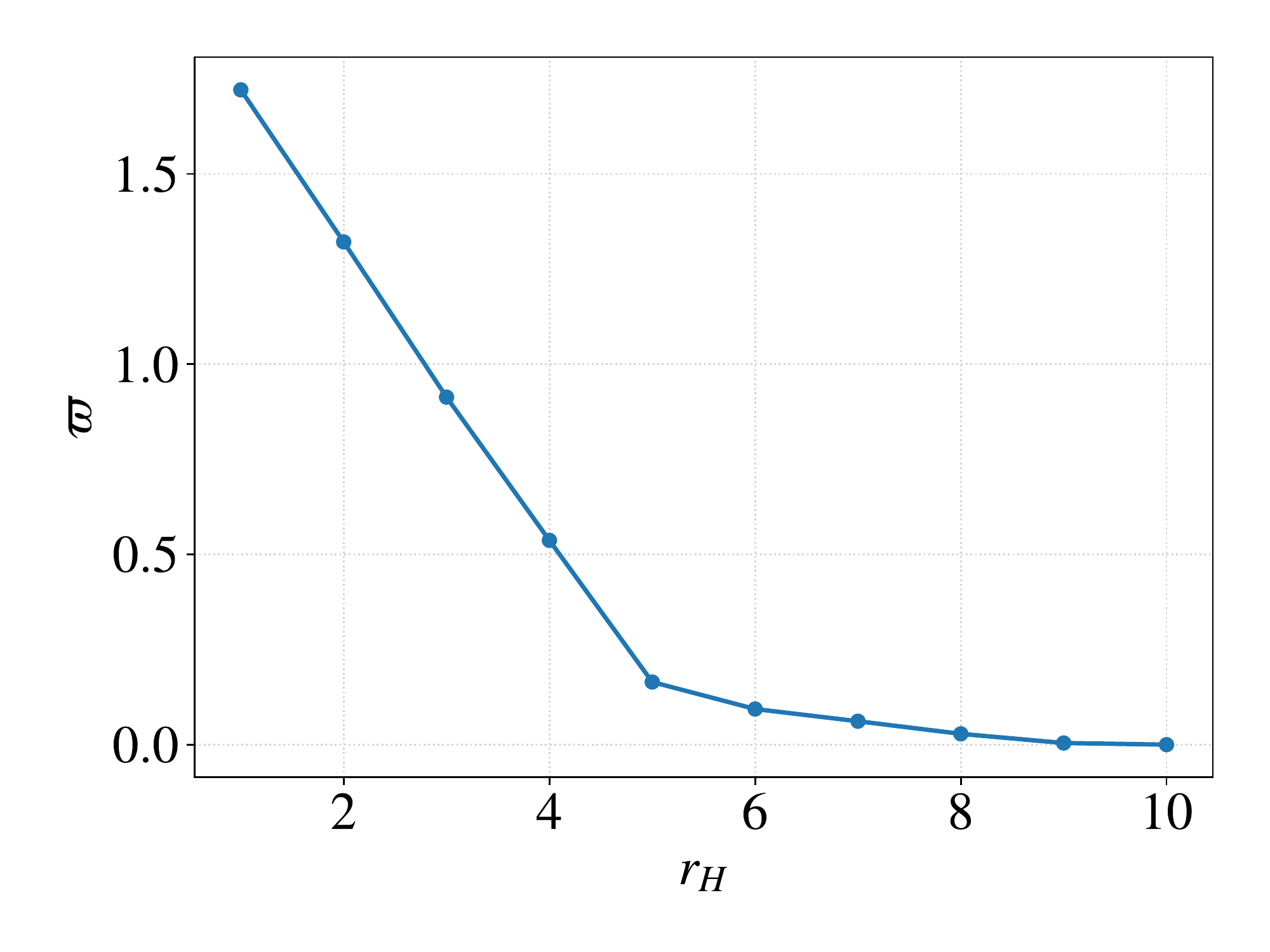}
		\subcaption{Difference in first-order PCEs ($\chi=2$).}
		\label{fig:5.3}
	\end{minipage}
	\caption{Comparison of the bias estimators introduced in Section~\ref{SSSEC:PCEbias} for the analytical test problem.}
	\label{fig:exponential_problem_bias}
\end{figure}

The MFAB algorithms described in Algorithms~\ref{ALG:biased} and \ref{ALG:unbiased} rely on estimating and controlling the bias of the MFAB estimator. In our application, we use $N_p=40$ pilot samples, but we also want to estimate statistics in the full dimensional spaces, where $d_H=10$ and $d_L=8$, so we use a larger dataset of $N'_p = 100$ samples. The bias estimators introduced in Section~\ref{SSSEC:PCEbias} (with $\chi=1, 2$ in Algorithm \ref{ALG:bias_estimator}) are both computed and reported in Figure~\ref{fig:exponential_problem_bias}. From the figure, we can see that using five variables significantly decreases the model's bias, while the remaining five only have a marginal effect.

\begin{figure}[h!]
	\centering
	\begin{minipage}{0.48\linewidth}
		\centering
		\includegraphics[width=0.8\linewidth]{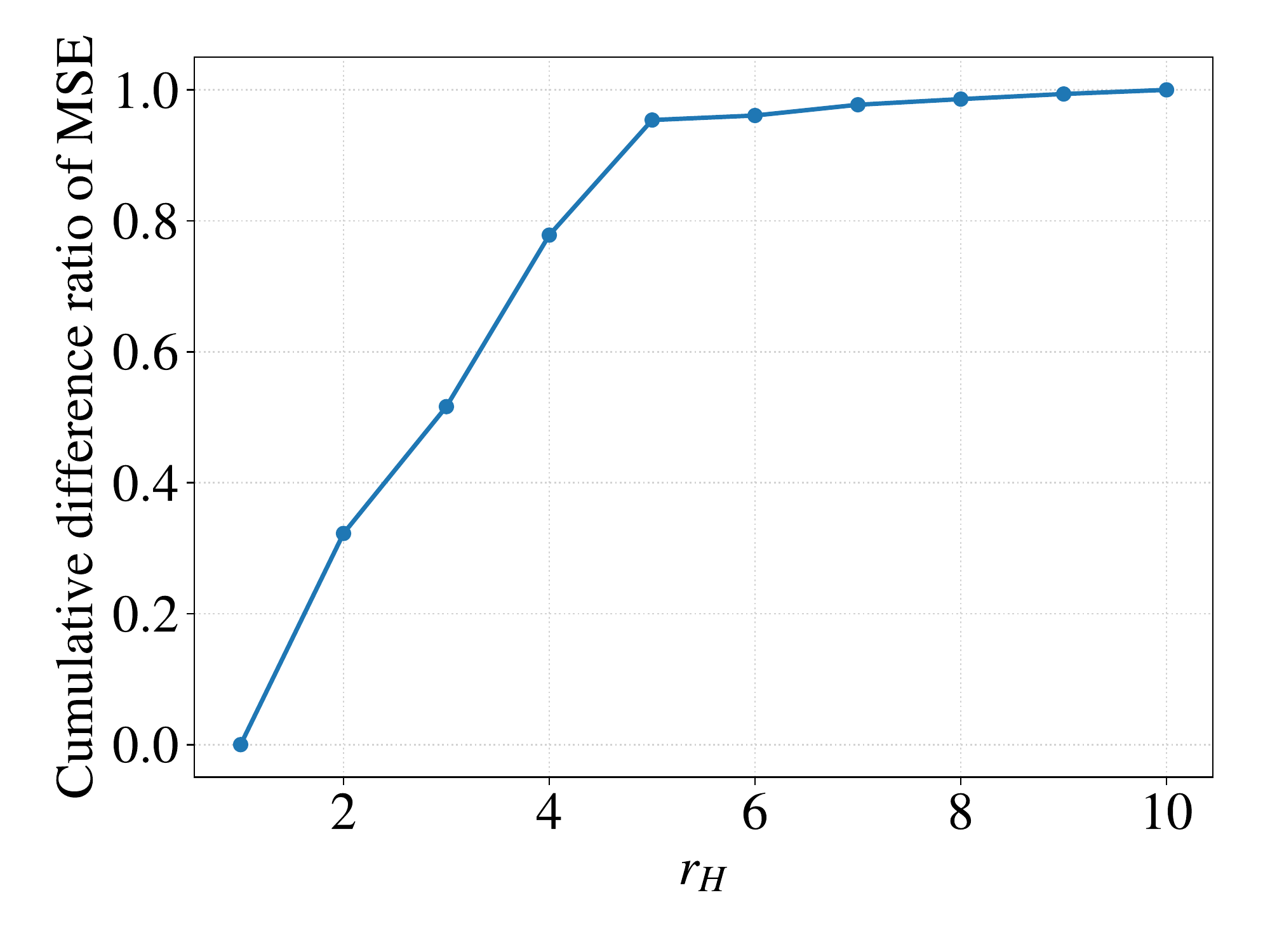}
		\subcaption{MSE of adapted PCEs ($\chi=1$).} 
		\label{fig:5.2-2}
	\end{minipage}
	\begin{minipage}{0.48\linewidth}
		\centering
		\includegraphics[width=0.8\linewidth]{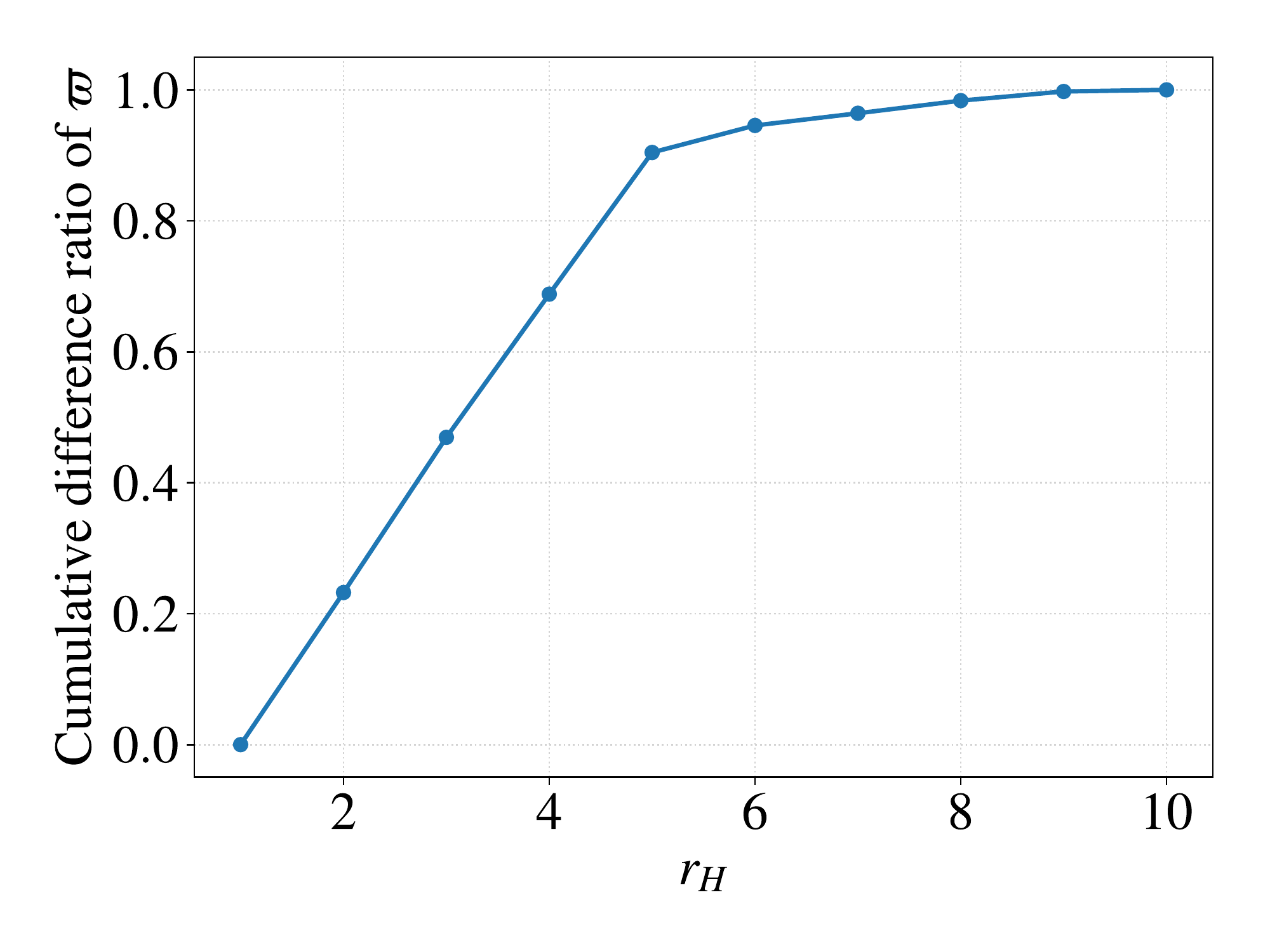}
		\subcaption{Difference in first-order PCEs ($\chi=2$).}
		\label{fig:5.3-2}
	\end{minipage}
    \caption{Adaptive selection of the HF model dimension based on the bias estimators introduced in Section~\ref{SSSEC:PCEbias} and Algorithm~\ref{ALG:bias_estimator} with $\vartheta=1$ for the analytical test problem.}
	\label{fig:exponential_problem_bias_estimators}
\end{figure}

The cumulative to total difference ratio of the two metrics can be computed following Algorithm~\ref{ALG:bias_estimator} with results reported in Figure~\ref{fig:exponential_problem_bias_estimators}. We use $\vartheta=1$ to show all achievable levels as a function of the HF dimension.

\subsubsection{Bias-variance balanced estimator}
\begin{figure}[htb]
	\centering
	\centering
	\includegraphics[width=0.45\linewidth]{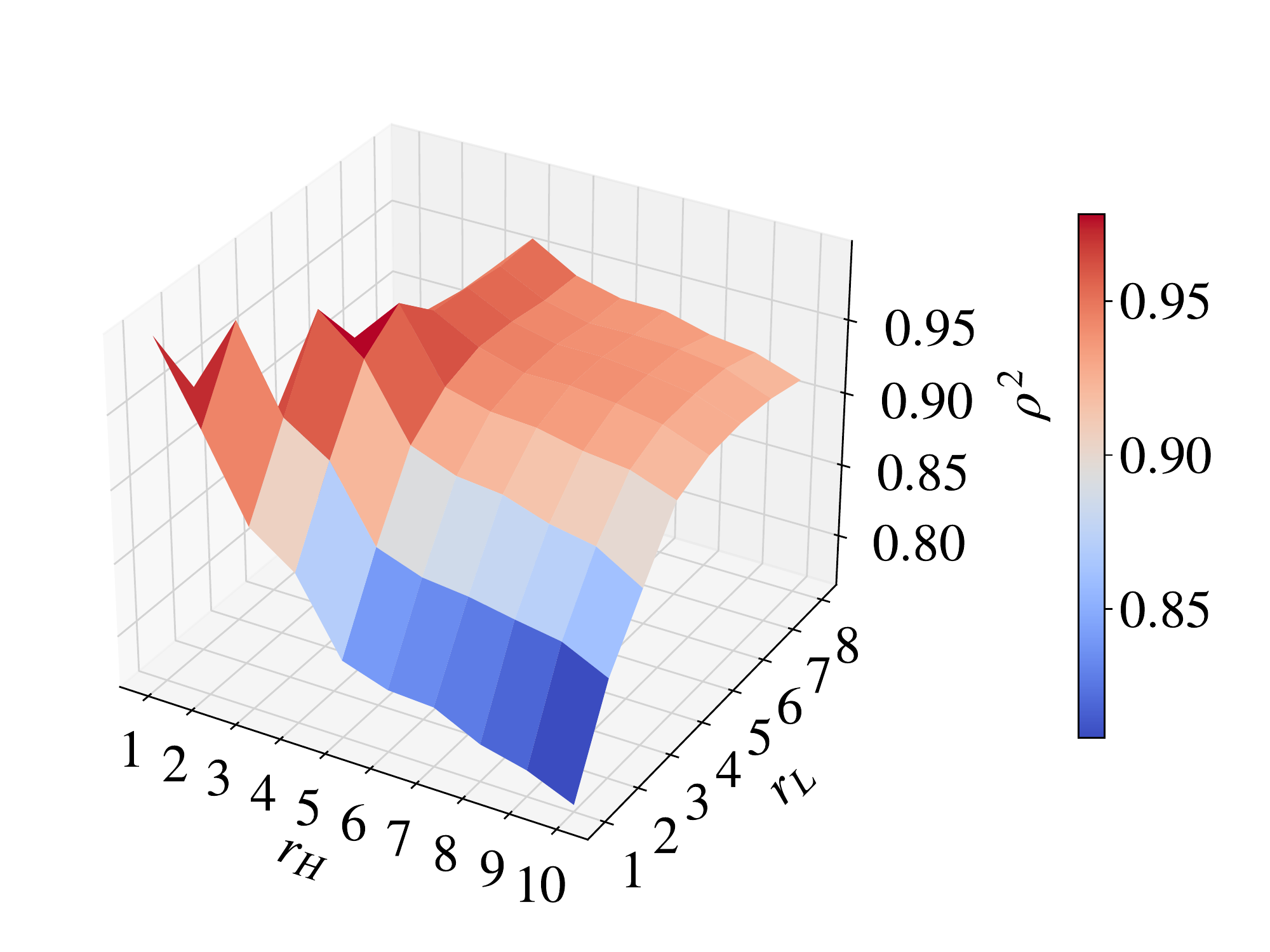}
	\caption{Estimated correlation for the two models in the analytical test problem following Proposition~\ref{prop:PCE_AB_correlation} as function of the truncated dimensions $\rh$ and $\rl$.}
	\label{fig:exponential_problem_est_correlation}
\end{figure}

We start by presenting the bias-variance balanced estimator described in Algorithm~\ref{ALG:biased}. For illustration purposes, in Figure~\ref{fig:exponential_problem_est_correlation}, we report the correlation (Proposition~\ref{prop:PCE_AB_correlation}) among models, for different truncation dimensions $\rh$ and $\rl$, by employing $N'_p=100$ samples.

To estimate the bias of the HF model with user-specified $\rh$ in step \ref{step:bias} of Algorithm~\ref{ALG:biased}, we sample the model on the shared space with additional $N_{\rh}$ samples and evaluate the bias using Eq~\eqref{eq:bias_quan} in Algorithm~\ref{ALG:biased}. The required number of additional HF samples, $N_{\rh}$, is linked to $\rh$. For $\rh=1$, we choose $N_{\rh} = 10$ and estimate the squared bias to be approximately $\widehat{\delta}^2 = 3.1$. For $\rh=3$, we choose $N_{\rh} = 20$ and obtain $\widehat\delta^2 = 0.5$, and for $\rh=5$, we choose $N_{\rh} = 30$ and obtain $\widehat\delta^2 = 0.04$. Once the HF dimension is fixed, the LF dimension is chosen to maximize correlation, which can be determined through a grid search as shown in Figure~\ref{fig:exponential_problem_est_correlation}.

Once the dimensions of adapted HF and adapted LF are fixed, one can specify the desired level of variance of the MF estimator and run the optimization problem (Algorithm~\ref{ALG:MFAB} with \textit{non-legacy dataset} and bias-variance balanced options) to find parameters $N$, $\ratio$, $\alpha$ by Eq~\eqref{eq:MFAB_share}. Here, to balance the two contributions of the MSE, we require $\sigma^2= \widehat\delta^2$. 

\begin{figure}[h!]
	\centering
	\begin{subfigure}[t]{0.48\linewidth}
		\centering
		\includegraphics[width=0.8\linewidth]{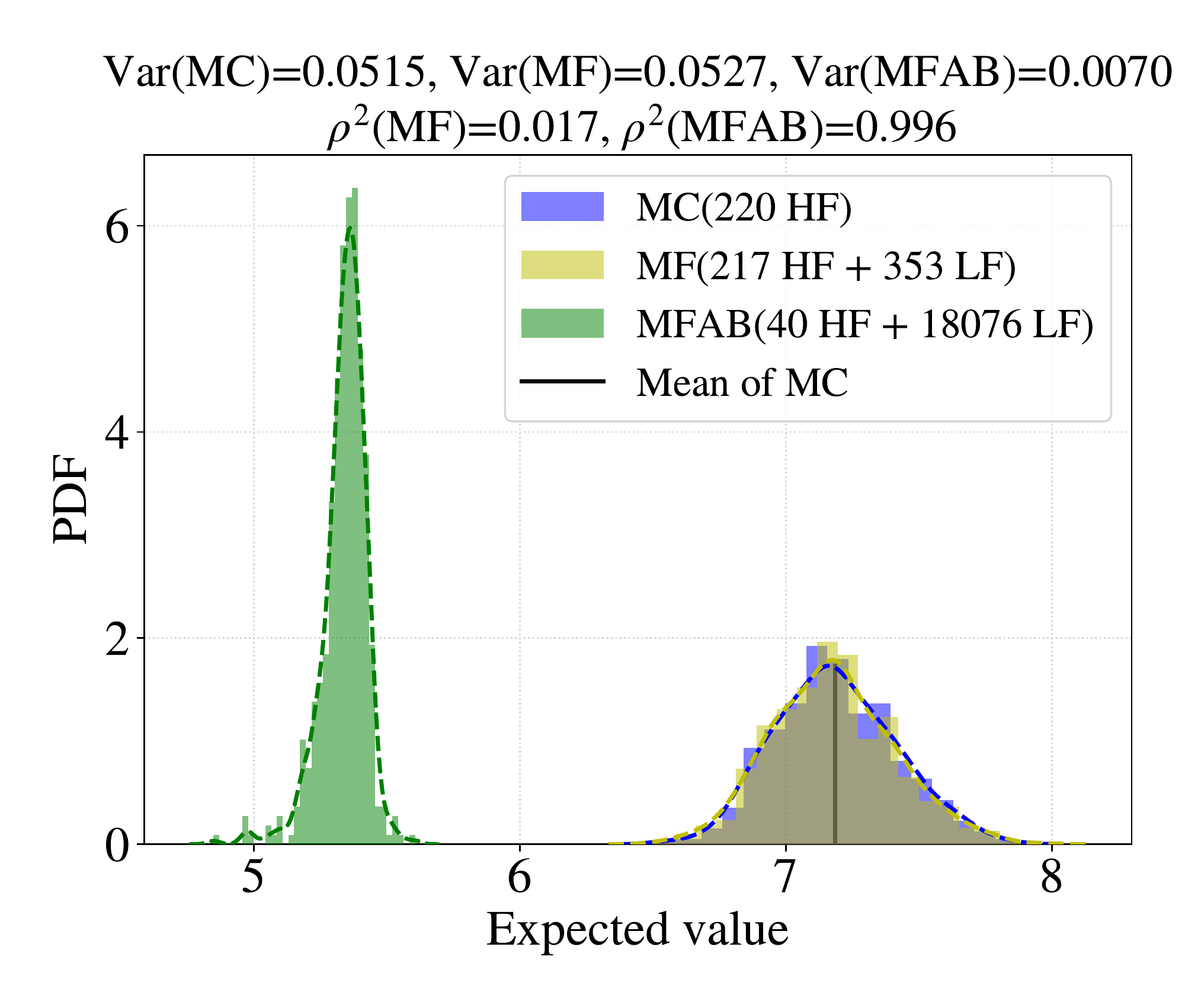}
		\caption{$\rh=1$ and $\rl=1$}\label{fig:5.5}
	\end{subfigure}
	\begin{subfigure}[t]{0.48\linewidth}
		\centering
		\includegraphics[width=0.8\linewidth]{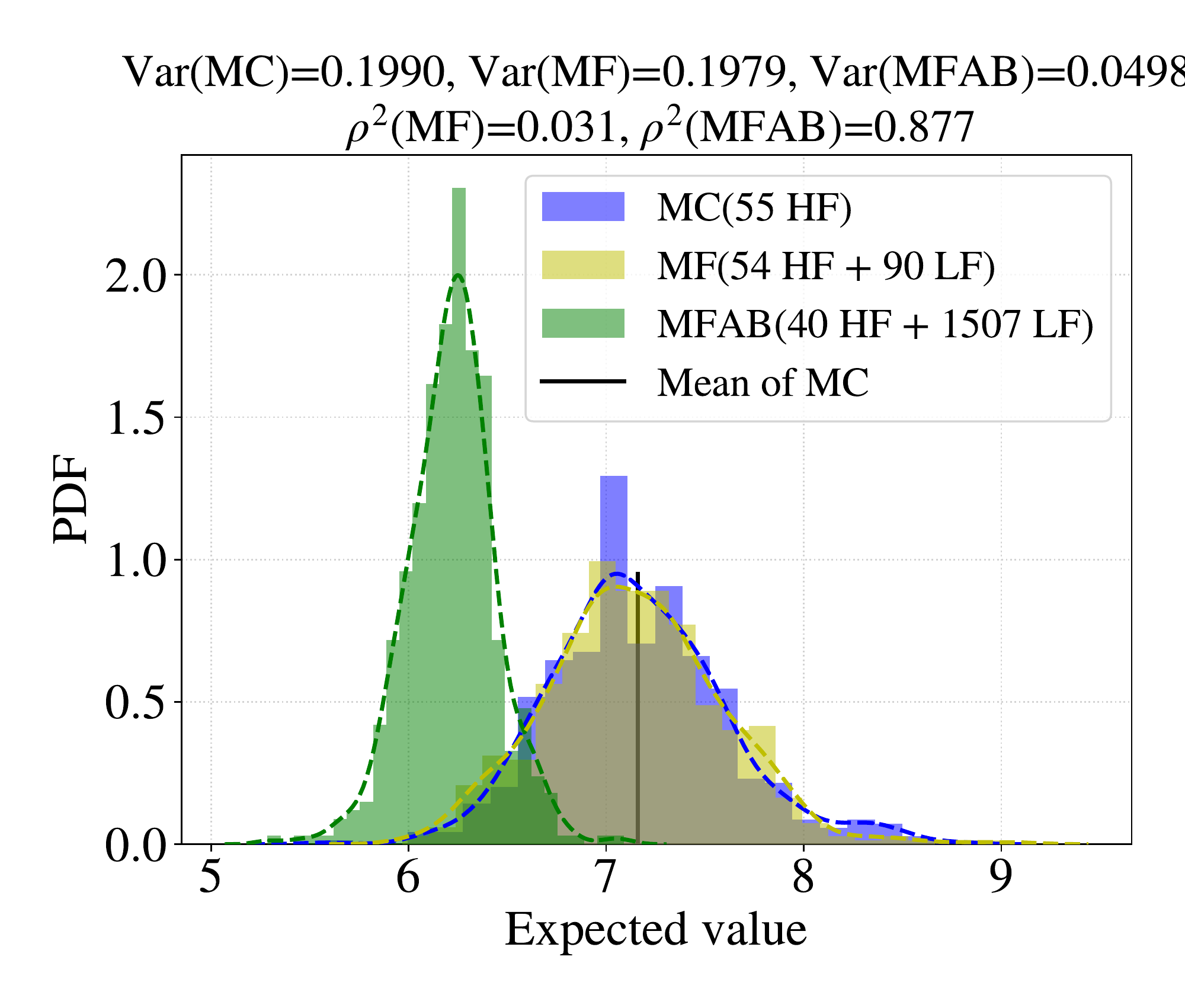}
		\caption{$\rh=3$ and $\rl=3$
		} 
		\label{fig:5.6}
	\end{subfigure}
    \vspace{3mm}
    \begin{subfigure}[t]{0.48\linewidth}
		\centering
		\includegraphics[width=0.8\linewidth]{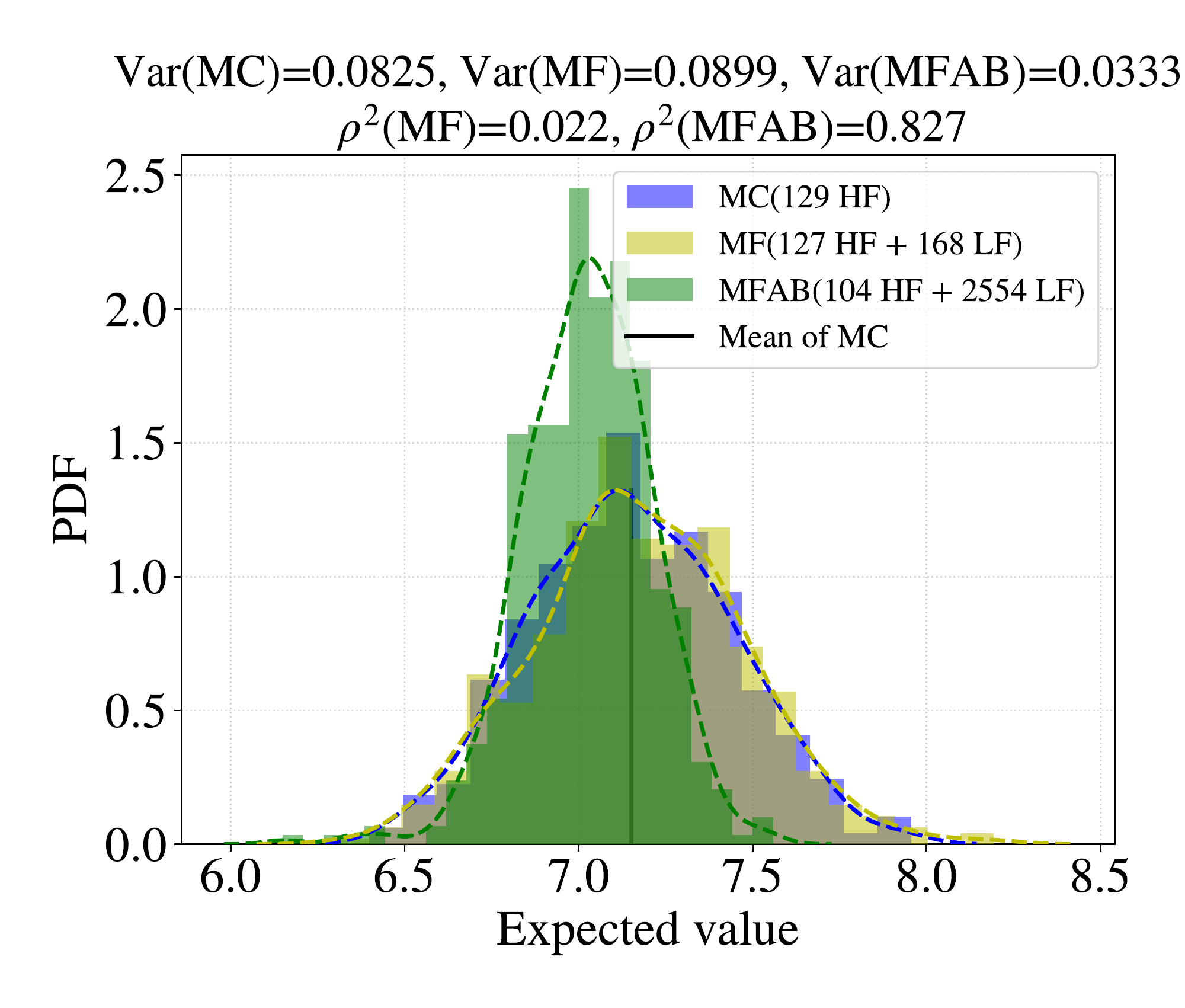}
		\caption{$\rh=5$ and $\rl=5$
		} 
		\label{fig:5.7}
	\end{subfigure}
	\begin{subfigure}[t]{0.48\linewidth}
		\centering
		\includegraphics[width=0.8\linewidth]{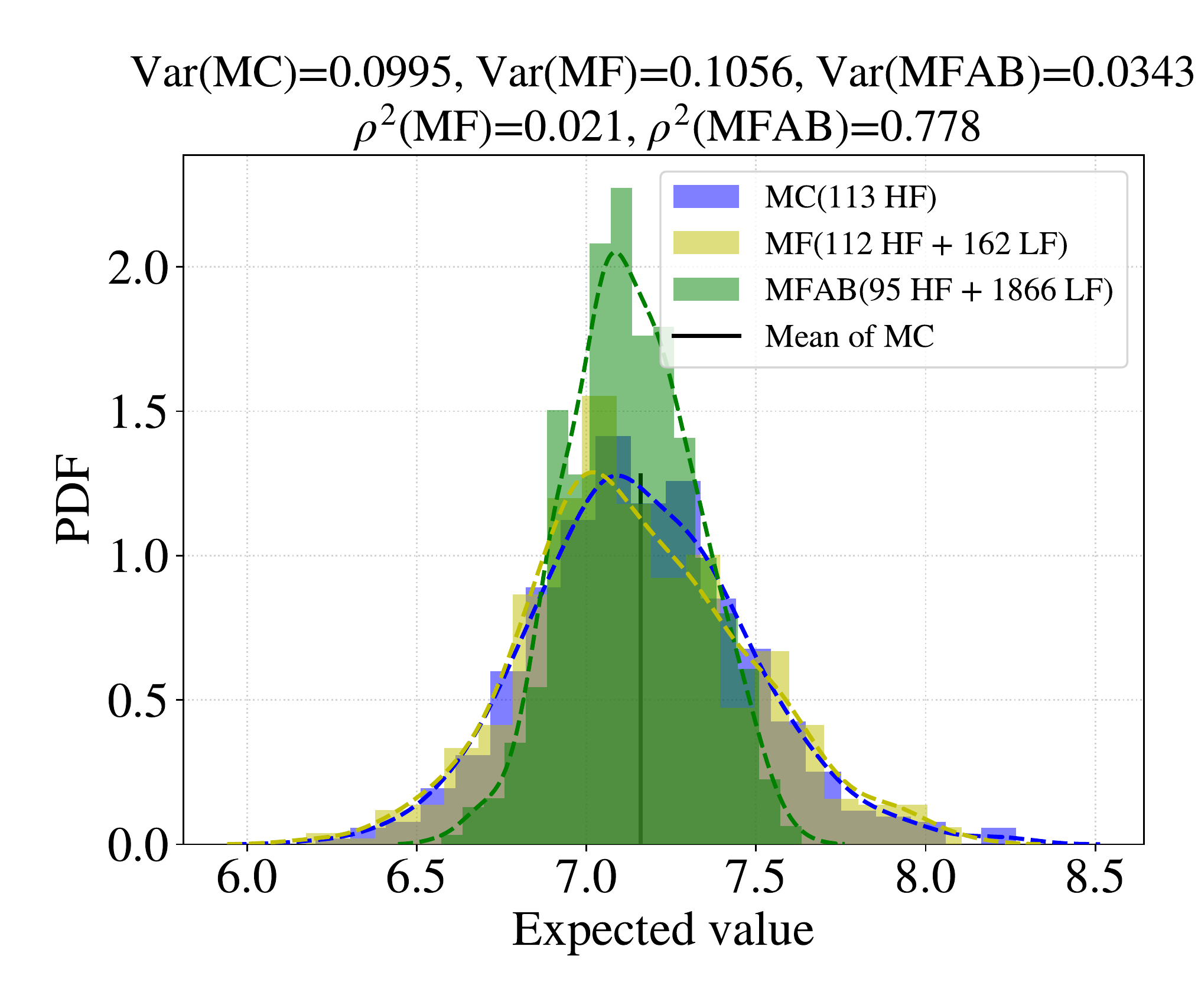}
		\caption{$\rh=6$ and $\rl=5$
		} 
		\label{fig:5.7-2}
	\end{subfigure}
	\caption{Probability density functions for 500 realizations of the MFAB, MF, and MC estimators for the analytical test problem with different choices of $\rh$ and $\rl$. Increasing the number of dimensions $\rh$ contributes to reducing the bias, as predicted by the theory, while the number of LF dimensions $\rl$ is determined to maximize the correlation between the models.}
\end{figure}

To demonstrate the performance of the MFAB estimator, we conduct multiple repetitions based on a pilot set containing 40 samples. We run 500 repetitions of the estimators to obtain their probability density functions (PDFs), as shown in Figure~\ref{fig:5.5}. 
We use the MFAB estimator cost to compute the equivalent single-fidelity MC and MF estimators, both of which are obtained by sampling the models in their original coordinates. The mean value of the MC estimator is represented by a vertical solid line in Figure~\ref{fig:5.5}. 

The squared correlation between HF and LF for the MFAB estimator has significantly increased to 0.996 from about 0.017. As a result, the MFAB estimator has reduced the variance from 0.0515 to 0.007 compared to the MC estimator. With high correlation between HF and LF, the variance of the MFAB estimator is already less than the target variance, $\widehat{\delta}^2 = 3.1$, with only 40 HF samples. However, it is also observed that, although the variance of the MFAB estimator has decreased significantly, it has a substantial bias due to the model reduction of the HF model. In this case, the MSE can be further reduced by increasing $\rh$. For instance, for $\rh=3$, an optimal value of $\rl=3$ can be selected, as shown in Figure~\ref{fig:exponential_problem_est_correlation}. The 500 repetitions for this case are reported in Figure~\ref{fig:5.6} for the three estimators. The correlation of LF and HF in MFAB has dropped to 0.877 compared to the previous case; however, it is still a substantial increase from 0.031. Moreover, the optimal MFAB estimator can achieve a variance of 0.0498, which is again a significant improvement from the variance of the MC estimator of 0.199. As expected, the bias of the MFAB estimator has decreased compared to the case of $\rh=1$.

Similar results can be obtained for $\rh=5$ and $\rl=5$, where $\widehat{\delta}^2 = 0.04$, as shown in Figure~\ref{fig:5.7}. The correlation of HF and LF in the MFAB estimator decreases with respect to $(\rh, \rl) = (3, 3)$; however, it is still much larger than that in the MF estimator obtained by sampling the models in their original space. The optimal estimator to obtain an estimator with a variance less than 0.04 has a variance of 0.0333, substantially less than the MC and classical MF estimator with the same cost. Moreover, although the correlation of HF and LF has dropped slightly, the bias of the MFAB estimator in this case is much smaller than in the previous one. A virtually unbiased estimator can be obtained by increasing $\rh$ even more, e.g., $\rh=6$. The results are reported in Figure~\ref{fig:5.7-2}; in this case, the optimal correlation is obtained with $\rl=5$.

\subsubsection{Unbiased MFAB estimator}
In this case, we still assume non-legacy data. Algorithm~\ref{ALG:unbiased} can be used to construct the MFAB estimator with \texttt{bias\_flag=False} in the subroutine Algorithm~\ref{ALG:MFAB}.

\begin{figure}[htb]
	\centering
	\begin{subfigure}{0.48\linewidth}
		\centering
		\includegraphics[width=0.9\linewidth]{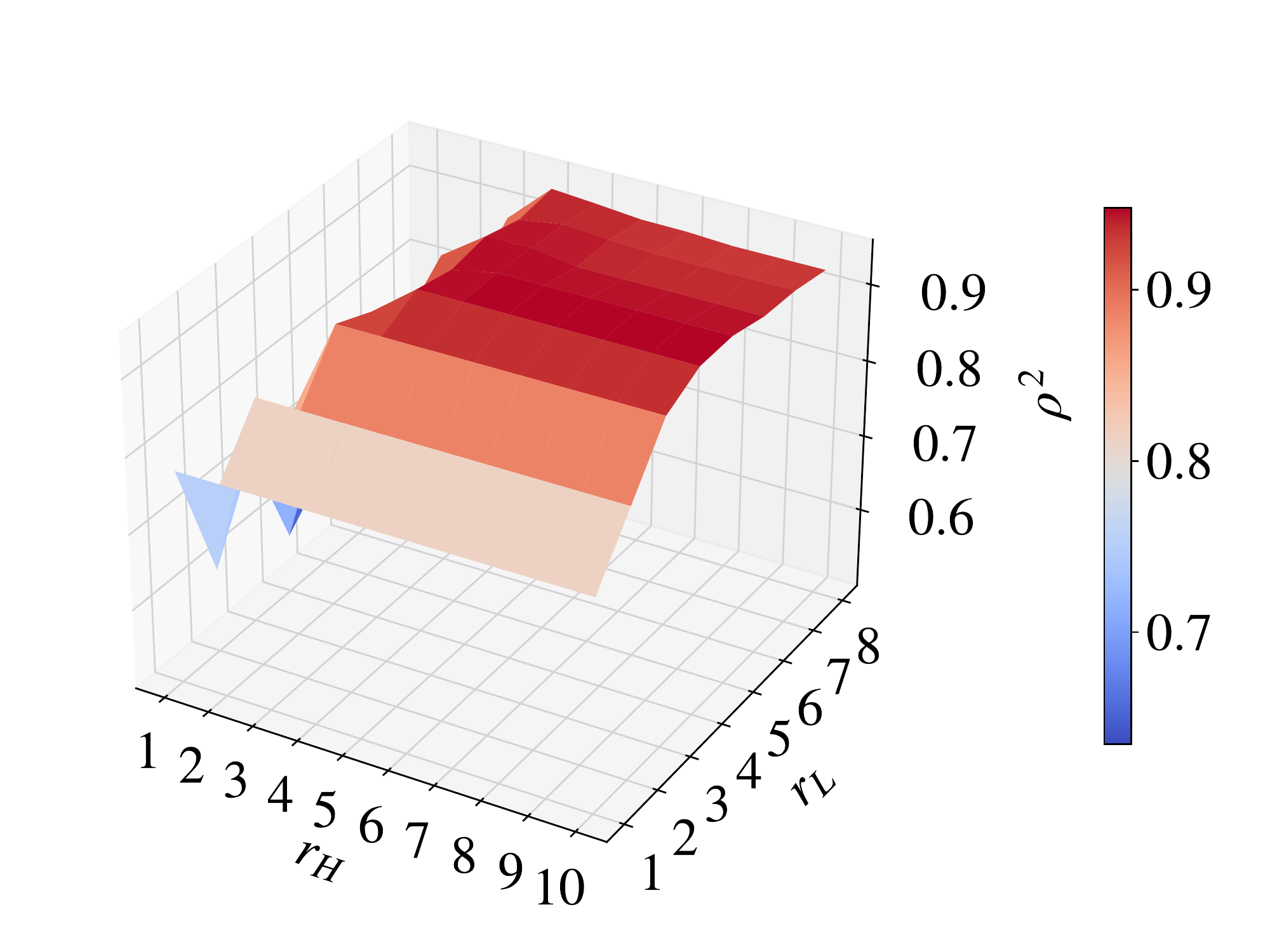}
		\subcaption{}
		\label{fig:exponential_problem_est_correlation2}
	\end{subfigure}
	\begin{subfigure}{0.48\linewidth}
		\centering
		\includegraphics[width=.8\linewidth]{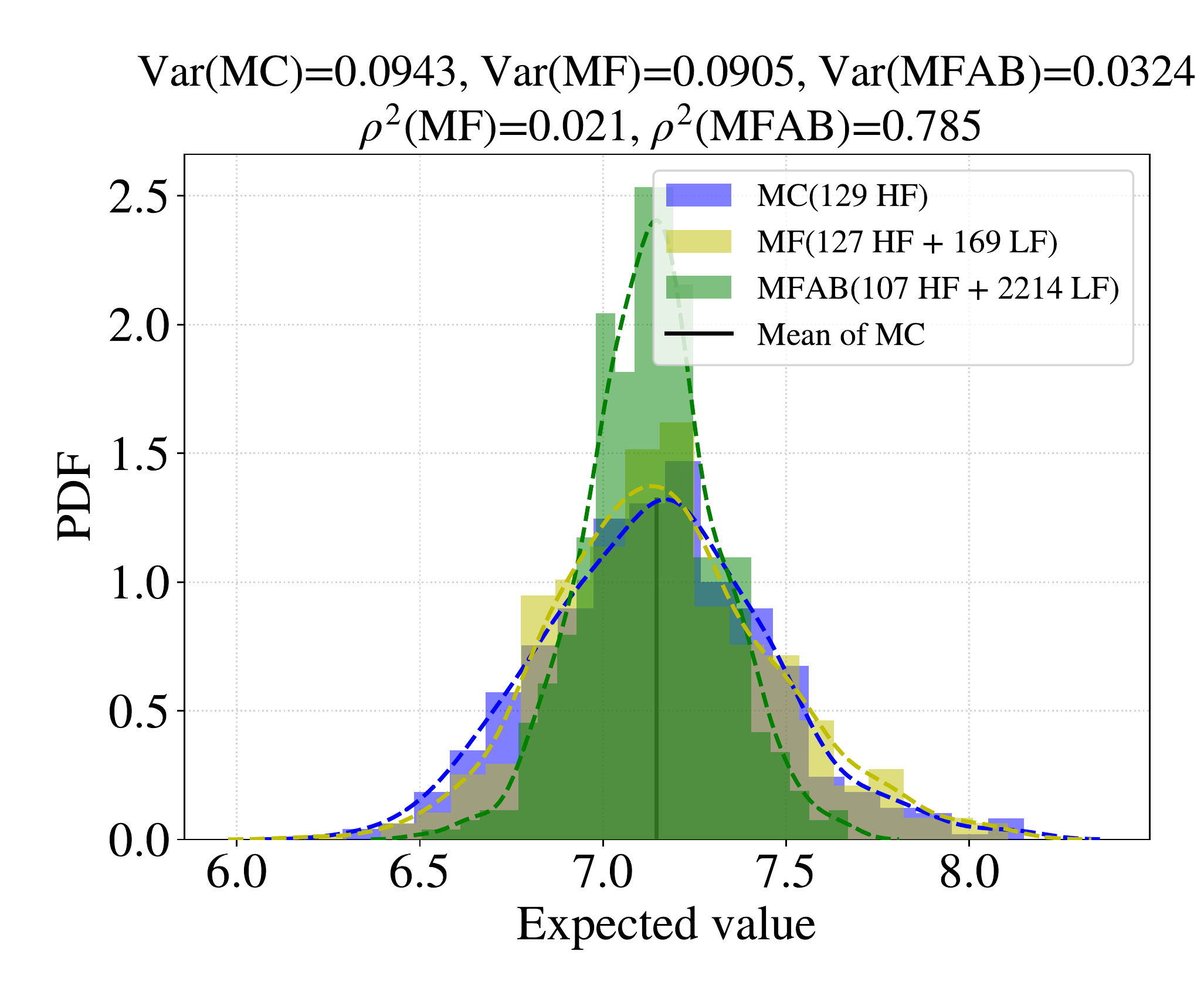}
		\subcaption{}
		\label{fig:5.9}
	\end{subfigure}
	\caption{(a) Estimated correlation for the two models in the analytical test problem with the unbiased strategy, and (b) probability density functions for 500 realizations of the MFAB, MF, and MC estimators, where the unbiased strategy is used in the MFAB estimator with $r_H = 5$ and $r_L = 5$.}
\end{figure}

The first two steps of Algorithm~\ref{ALG:unbiased} are identical to those in Algorithm~\ref{ALG:biased}. The adapted HF model is converged using $\rh = 5$. In the next step of Algorithm~\ref{ALG:unbiased}, we compute the correlation between HF and LF using Eq~\eqref{eq:corr_unbiased}, with LF re-evaluated on the shared space for $\rl = 1, \ldots, \rh$. The results are shown in Figure~\ref{fig:exponential_problem_est_correlation2}, where we expand the correlation calculation to full dimensions for both HF and LF for demonstration purposes only. 
When $\rh=5$, the value of $\rl=5$ maximizes their correlation. However, $\rh \times N_p$ evaluations of the LF model are required, which is the additional cost of the unbiased strategy. In this case, the correlation cannot be estimated without re-running the LF model.

We can specify a desired level of variance and compute the optimal MF estimator by finding the parameters $N$, $\ratio$, $\alpha$ that satisfy Eq~\eqref{eq:MFAB_share}. To compare the results with those obtained with the biased strategy in the previous section, we choose $\sigma^2 = 0.04$. We calculate three estimators 500 times, and their PDFs are presented in Figure~\ref{fig:5.9}. The corresponding (equivalent total cost) MC and MF estimators are also reported for comparison.
We observe that the correlation between HF and LF for the MFAB estimator is 0.785, which is a significant improvement from the correlation of 0.021 in the original space. As a consequence, the variance of the MFAB estimator is reduced to 0.0324 from the value of 0.0943 for the MC estimator, which represents a gain of approximately 65\%. As in the previous case, the MF estimator, which samples the models in their original coordinates, cannot achieve better results than MC since the correlation between them is too low.

\subsection{A direct field acoustic testing application}
\label{ssec:numres_acoustic}
\begin{figure}[htb]
  \centering
  \begin{subfigure}{0.48\linewidth}
    \centering
	\includegraphics[width=0.75\linewidth]{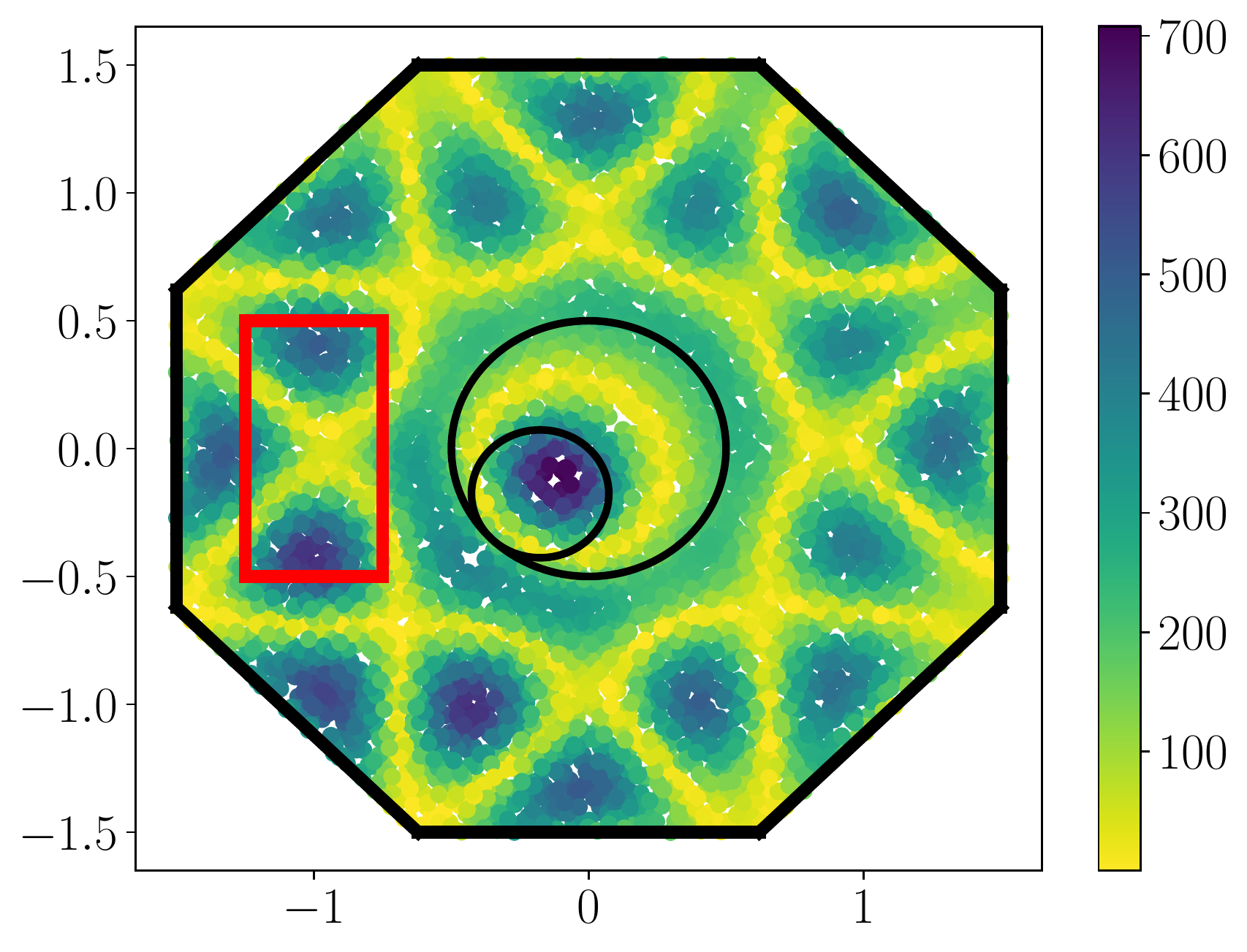}
        \subcaption{}
        \end{subfigure}
        \begin{subfigure}{0.48\linewidth}
          \centering
          \includegraphics[width=.8\linewidth]{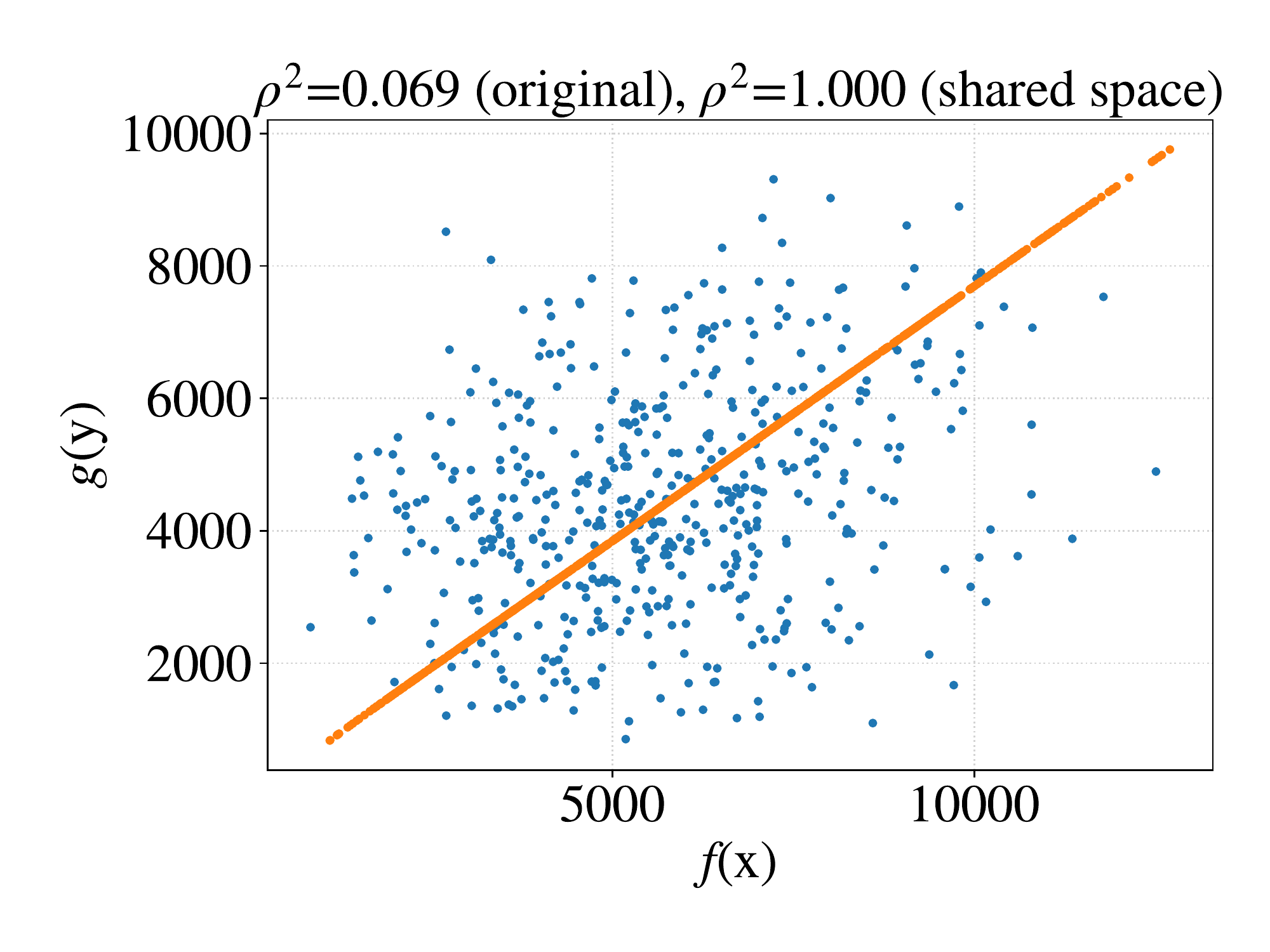}
          \subcaption{}
        \end{subfigure}
        \caption{(a) Example pressure plot of the speaker cabinet. The QoI is the average pressure inside the red rectangle. (b) Scatter plot of HF and LF in the original and shared spaces with a single shared variable of the acoustic application.}
	\label{fig:5.10}
\end{figure}

We will now introduce a MF benchmark case, which is included in the PyApprox software \cite{pyapprox} and is inspired by a direct field acoustic test described in~\cite{Stasiunas2016}. The problem involves a circular scatterer with a circular air inclusion placed in the center of an octagonal domain, where each side is a speaker with its cabinet (see Figure~\ref{fig:5.10}(a)). The acoustic pressure $u$ in the octagonal domain $D$ is governed by the real Helmholtz equation under the assumptions of the scatterer being a dense fluid and the absence of impedance for the speakers' cabinets. For a fixed angular frequency $\omega = 2 \pi \nu$, the equation is given by
\begin{equation}
 \begin{split}
  \Delta u + \kappa^2 u &= 0 \quad \mathrm{in} \quad D \\
  \dfrac{\partial u}{\partial n} &= \rho_0 \omega \sum_{j=1}^s h_j \chi_j \quad \mathrm{on} \quad \partial D,
 \end{split}
\end{equation}
where $\kappa=\omega/c$ is the wavenumber, $c$ is the local speed of sound, $\rho_0$ is the fluid density, and $\chi_j: \partial D\rightarrow {0,1}$ is the characteristic function of the $j$th speaker, which oscillates with velocity $h_j \cos{\left(\omega t\right)}$. The QoI in this problem is the average sound pressure, which is measured in the red rectangle area induced by eight loudspeakers located at the center of each side of an octagonal speaker cabinet, as shown in Figure~\ref{fig:5.10}(a). Our goal is to estimate the expected value of this QoI.

In this problem, the pressure is obtained by activating the speakers, whose amplitudes follow a uniform distribution. The scatterer is considered to be made of aluminum, for which the sound speed varies as $\mathcal{U}(6000,6600)$ m/s in aluminum, and the density of air is distributed as $\mathcal{U}(0.8,1.6)$ Kg/$m^3$. The liner frequency is $\nu=400$Hz. The complete list of random variables for this problem is summarized in Table~\ref{tab:uq_acoustic}.
 \begin{table}[htb]
 \centering
 \begin{tabular}{lll}
 \hline\hline
 Uncertain Parameter                                    &    Distribution           &  Units    \\ \hline  
Amplitude oscillation for Speakers 1 and 2 ($\theta_1$)     & $\mathcal{U}[1,9]$        &  $m/s$    \\
Amplitude oscillation for Speakers 3 and 4 ($\theta_2$)     & $\mathcal{U}[1,9]$        &  $m/s$    \\
Sound speed of aluminum ($\theta_3$)                    & $\mathcal{U}[6000,6600]$  &  $m/s$    \\
Air density ($\theta_4$)                                & $\mathcal{U}[0.8,1.6]$    &  $Kg/m^3$ \\
Amplitude oscillation for Speakers 5 and 6 ($\theta_5$)     & $\mathcal{U}[1,9]$        &  $m/s$    \\
Amplitude oscillation for Speakers 7 and 8 ($\theta_6$)     & $\mathcal{U}[1,9]$        &  $m/s$    \\ \hline \hline  
 \end{tabular}  
\caption{Uncertain parameters and distribution for the Acoustic problem. Speakers are ordered counterclockwise, with the first speaker located on the right vertical edge of the octagon.} 
\label{tab:uq_acoustic}
\end{table}

The Helmholtz equation is solved using a finite element (FE) method, which involves linear superposition of basis functions. The HF model has a dimension of 6 and is denoted as $f(\bm{x})$, where $\bm{x} = [\theta_1, ..., \theta_6]^T$, and the mesh size for the FE,  denoting the number of element in one side of the speaker cabinet,  is 21. The LF model, denoted as $g(\bm{y})$, has a dimension of 4 and includes only the first four activated speakers, with their distributions being the same as in the HF model. The sound speed in aluminum and atmospheric pressure are assumed to follow the same distributions as in the HF model. The mesh size for the LF model is 11, which is coarser than the HF model, resulting in a cost of only 10\% of the HF model. Figure~\ref{fig:5.10}(b) presents scatter plots (500 samples) for the models sampled in the original and shared coordinates. For demonstration purposes, a single shared variable is used. The scatter plots show that the correlation among the models significantly increases by using shared variables.

\begin{figure}[htb]
	\begin{subfigure}{0.48\linewidth}
		\centering
		\includegraphics[width=0.8\linewidth]{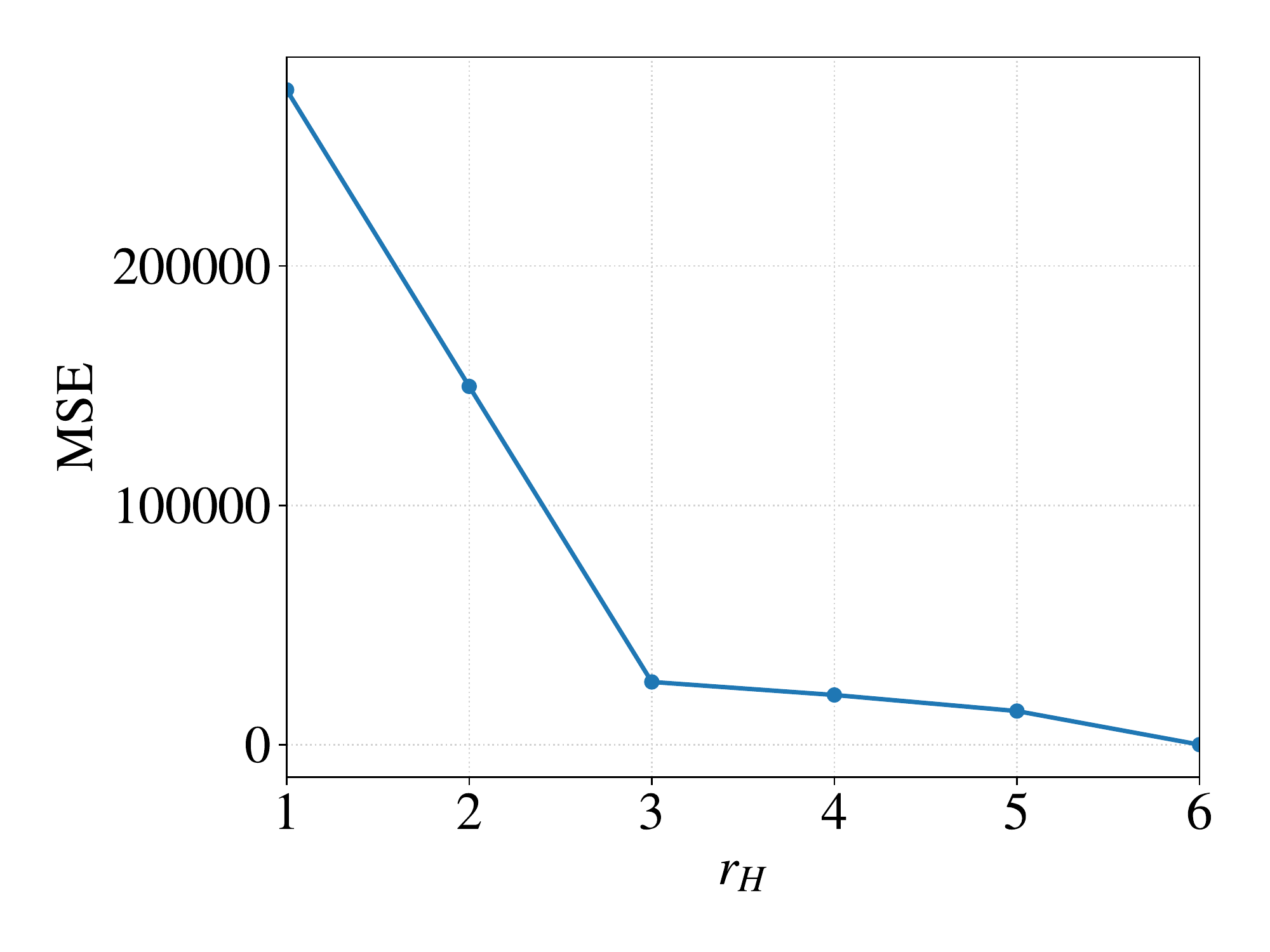}
		\subcaption{MSE of adapted PCEs ($\chi=1$).}
	\end{subfigure}
	\begin{subfigure}{0.48\linewidth}
		\centering
		\includegraphics[width=0.8\linewidth]{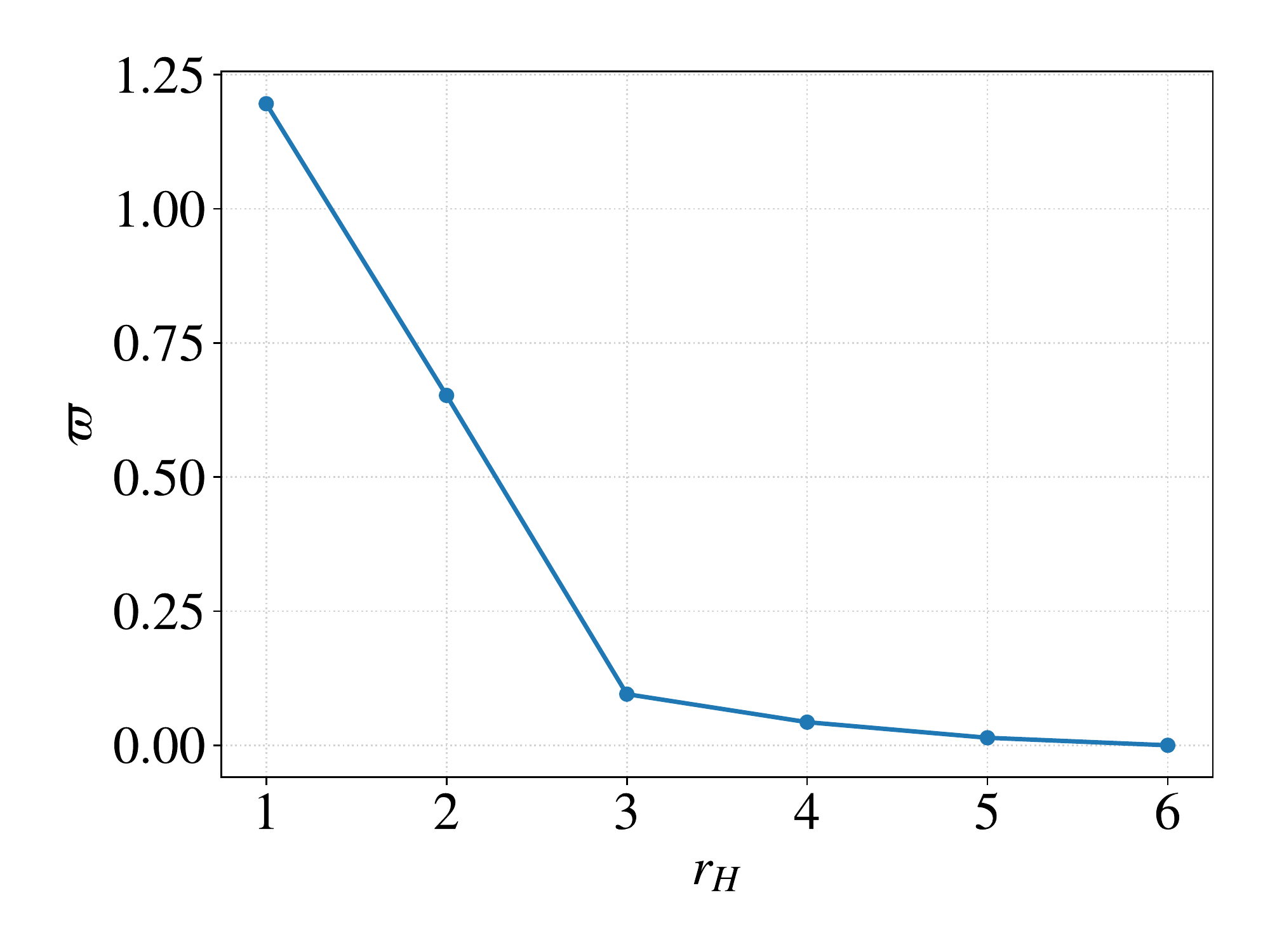}
		\subcaption{Difference in the first-order PCEs ($\chi=2$).}
	\end{subfigure}
	\caption{Comparison of the bias estimators introduced in Section~\ref{SSSEC:PCEbias} for the acoustic application.}
	\label{fig:5.13}
\end{figure}

\begin{figure}[h!]
	\begin{subfigure}{0.48\linewidth}
		\centering
		\includegraphics[width=0.8\linewidth]{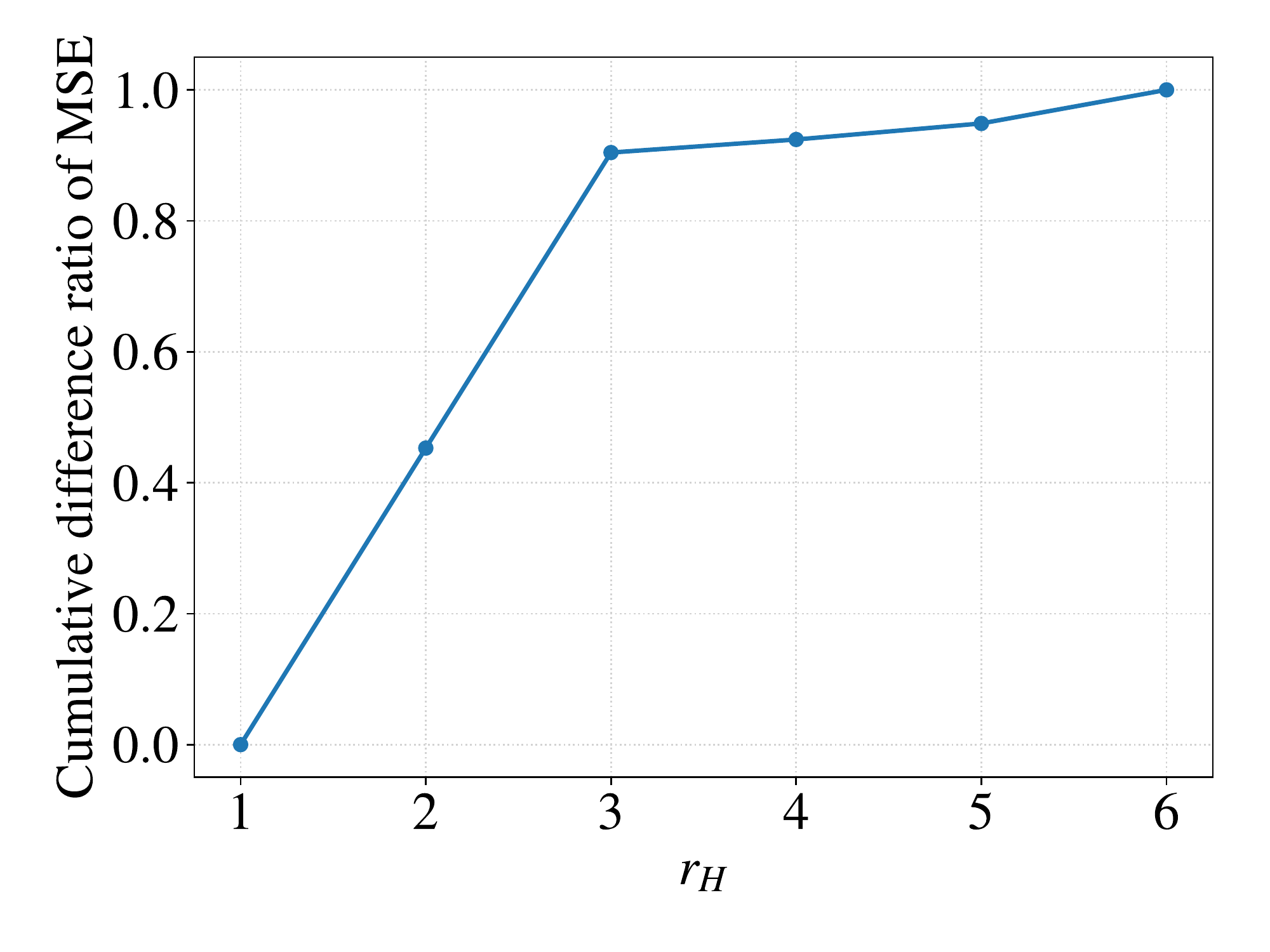}
		\subcaption{MSE of adapted PCEs ($\chi=1$).}
	\end{subfigure}
	\begin{subfigure}{0.48\linewidth}
		\centering
		\includegraphics[width=0.8\linewidth]{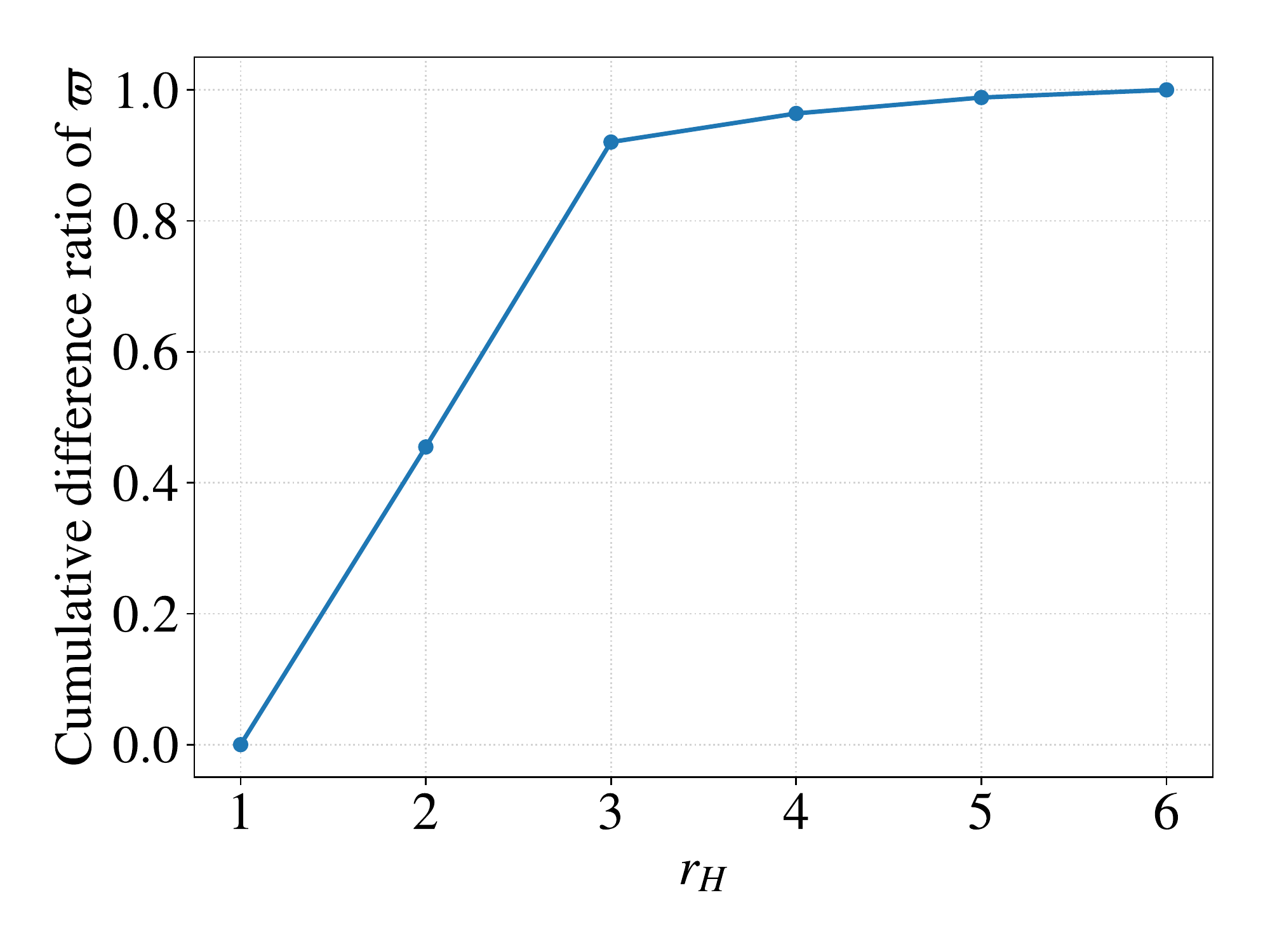}
		\subcaption{Difference in first-order PCEs ($\chi=2$).}
	\end{subfigure}
	\caption{Adaptive selection of the HF model dimension based on the bias estimators introduced in Section~\ref{SSSEC:PCEbias} and Algorithm~\ref{ALG:bias_estimator} with $\vartheta=1$ for the acoustic application. }
	\label{fig:5.13-2}
\end{figure}

To implement Algorithms~\ref{ALG:biased} and \ref{ALG:unbiased}, the first two steps involve estimating and controlling the estimator bias. We use $N_p = 40$ pilot samples in this application. However, for demonstration purposes, we employ a larger dataset of $N'_p=100$ samples to explore the full space for quantities like bias and correlation, as we did in the previous section. Algorithm~\ref{ALG:bias_estimator} with $\chi=1, 2$ is used to compute the two bias metrics, which are presented in Figure~\ref{fig:5.13}.
We notice a rapid decrease in bias as the dimension of the adapted HF increases from 1 to 3, after which the bias remains relatively stable, indicating that the adapted HF will converge to the full-dimensional model with a dimension of 3. Furthermore, we compute the cumulative-to-total difference ratio of the two bias metrics using Algorithm~\ref{ALG:bias_estimator}, and the results are shown in Figure~\ref{fig:5.13-2} for all achievable levels as a function of the HF dimension with $\vartheta=1$.

\subsubsection{Bias-variance balanced estimator}
\begin{figure}[htb]
		\centering
		\includegraphics[width=0.45\linewidth]{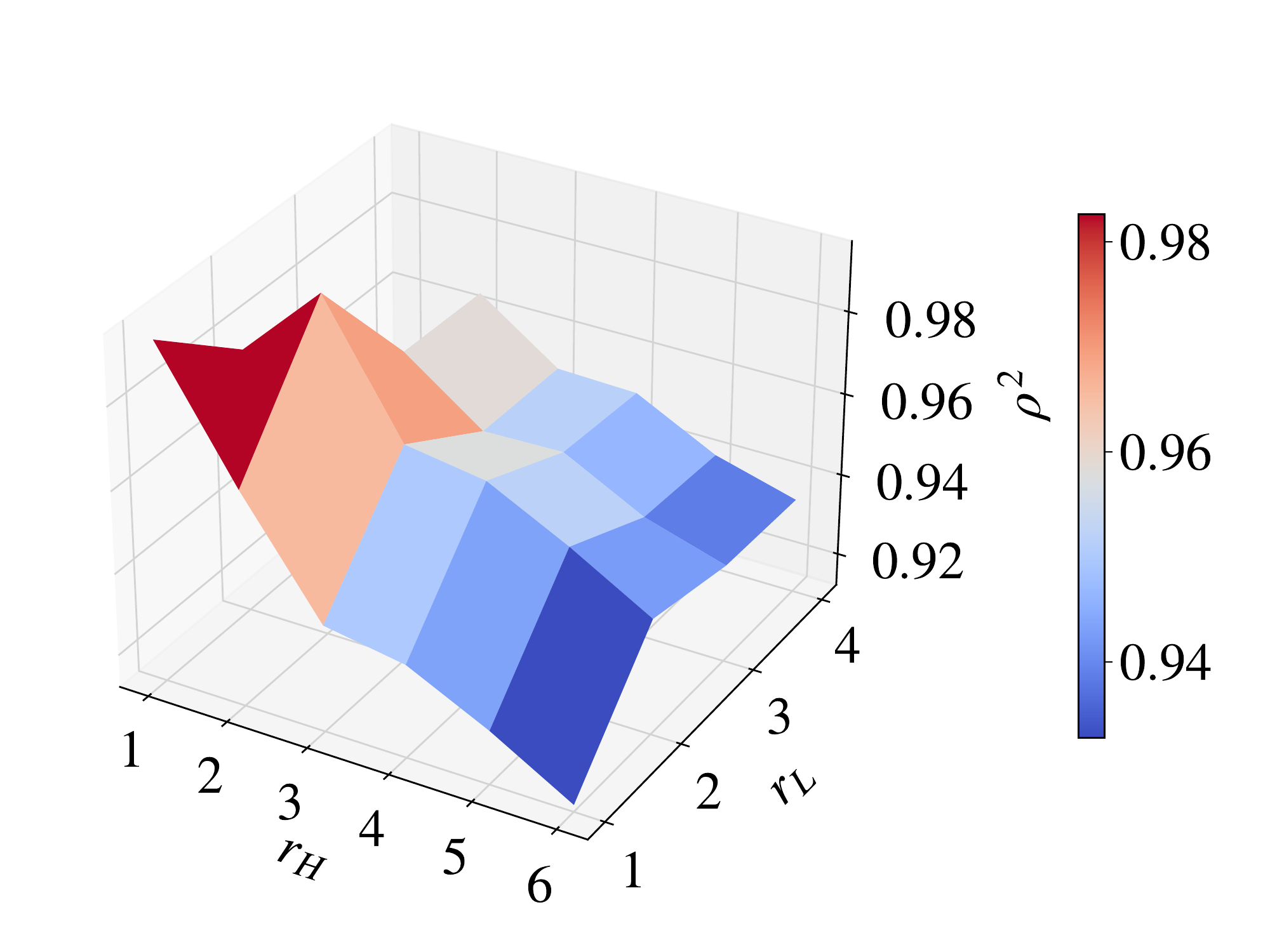}
		\caption{Estimated correlation for the two models in the acoustic application following Proposition~\ref{prop:PCE_AB_correlation} as function of the truncated dimensions $\rh$ and $\rl$. }
		\label{fig:5.15}
\end{figure}

The correlations estimated for different truncation dimensions $\rh$ and $\rl$ are presented in Figure~\ref{fig:5.15}.
\begin{figure}[htb]
	\centering
	\begin{subfigure}[t]{0.33\linewidth}
		\centering
		\includegraphics[width=\linewidth]{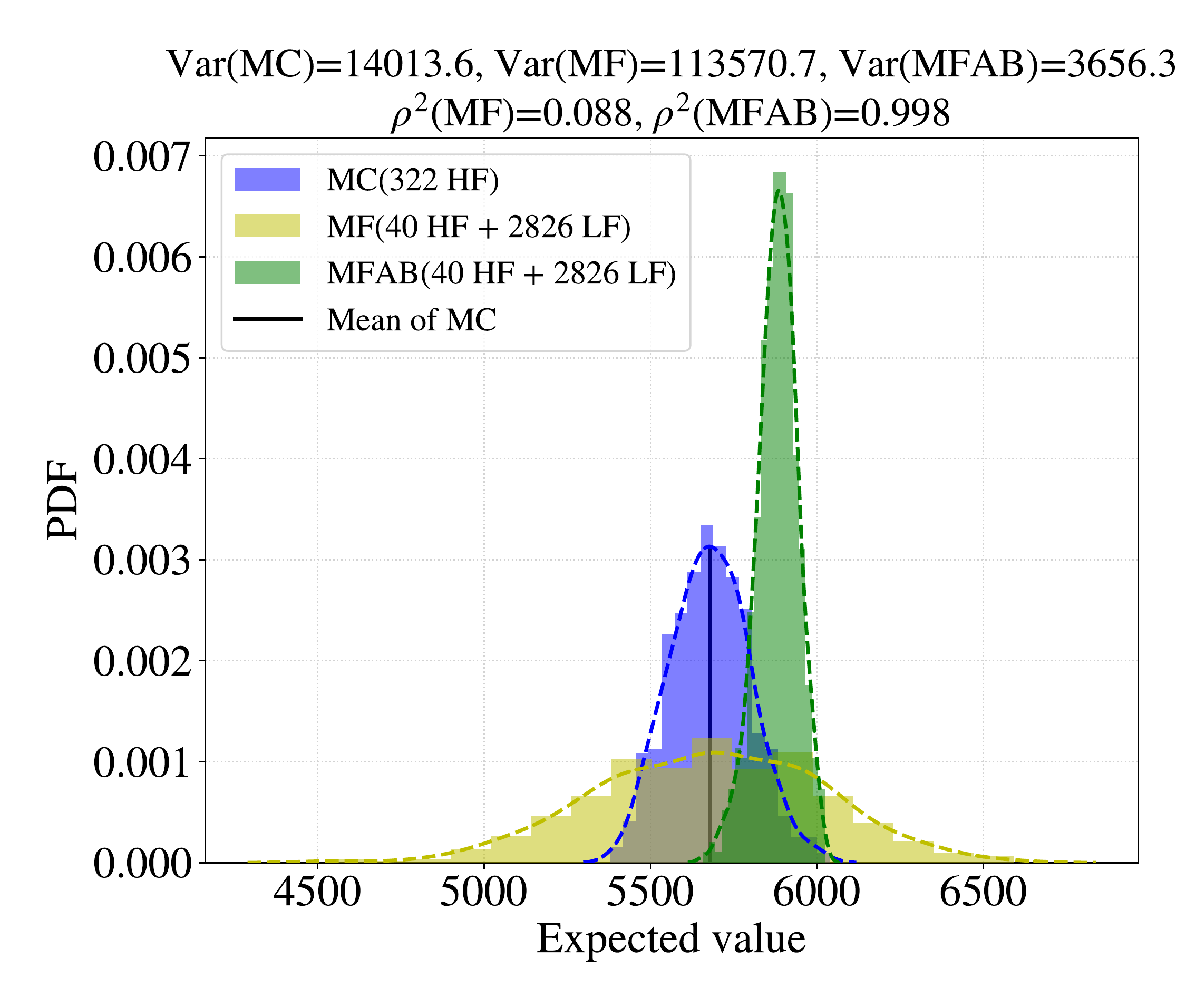}
		\caption{$\rh=1$ and $\rl=1$}\label{fig:5.16}
\end{subfigure}
\begin{subfigure}[t]{0.33\linewidth}
	\centering
	\includegraphics[width=\linewidth]{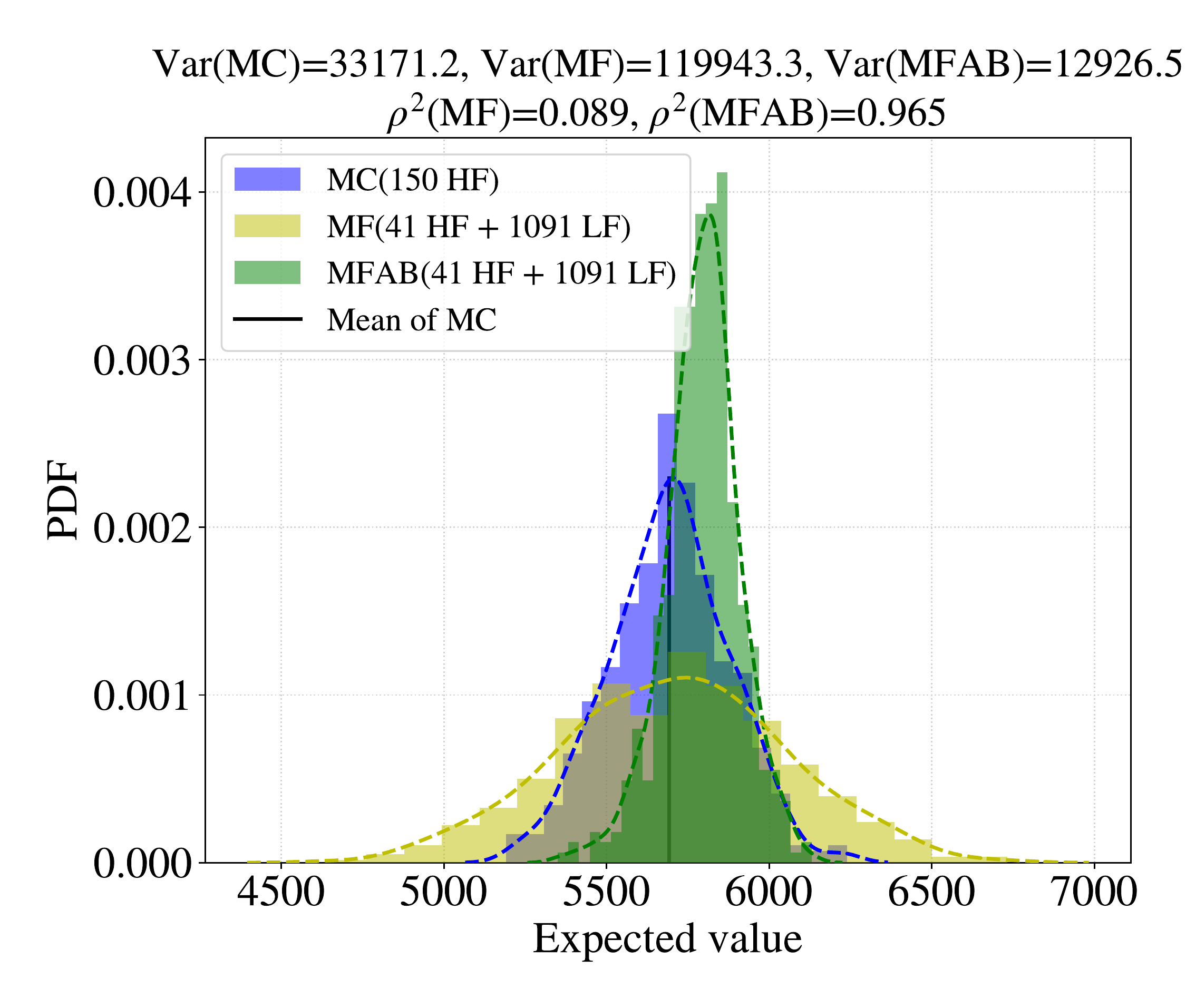}
	\caption{$\rh=2$ and $\rl=2$
	} 
	\label{fig:5.17}
\end{subfigure}
\begin{subfigure}[t]{0.33\linewidth}
	\centering
	\includegraphics[width=\linewidth]{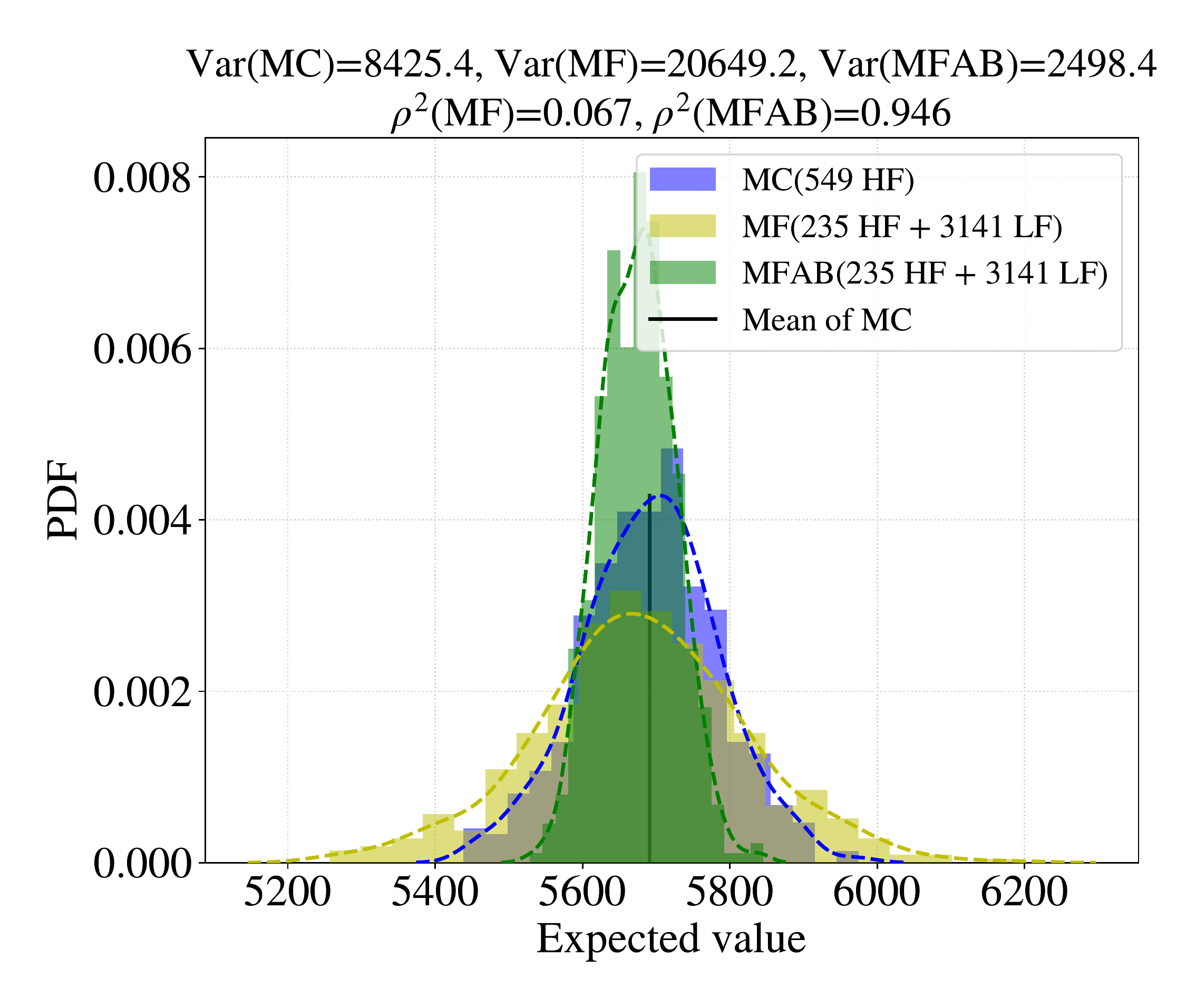}
	\caption{$\rh=3$ and $\rl=2$
	} 
	\label{fig:5.18}
\end{subfigure}
\caption{Probability density functions for 500 realizations of the MFAB, MF, and MC estimators for the acoustic application with different choices of $\rh$ and $\rl$. Increasing the number of dimensions $\rh$ contributes to reducing the bias, as predicted by the theory, while the number of LF dimensions $\rl$ is determined to maximize the correlation between the models.}
\end{figure}
The bias is estimated using an additional $N_{\rh}$ samples. For $\rh=1$, we choose $N_{\rh}=10$, and the estimated squared bias is $\widehat{\delta}^2 = 29864.0$. For $\rh=2$, we choose $N_{\rh}=20$, and obtain $\widehat{\delta}^2 = 20201.4$. For $\rh=3$, we choose $N_{\rh}=30$, and obtain $\widehat{\delta}^2 = 2496.3$. As with the previous example, we consider different truncations. For $\rh=\rl=1$, $\sigma^2= \widehat{\delta}^2 = 29864.0$. 
Figure~\ref{fig:5.16} shows the results for the three estimators (MC, MF, and MFAB) for 500 repetitions obtained with $N_p=40$.
From Figure~\ref{fig:5.16}, we can see that compared to the MF estimator, the squared correlation between HF and LF of the MFAB estimator has increased from 0.088 to 0.998. As a result, the MFAB estimator has a much-reduced variance compared to the MC estimator, and the former is about 26\% of the latter. Additionally, due to the high correlation between HF and LF, the variance of the MFAB estimator is already less than the target variance, $\widehat{\delta}^2 = 29864.0$, with 40 HF samples. However, we can also see a significant bias introduced in the estimator. A bias reduction can be obtained for $\rh=2$, for which the optimal truncation $\rl=2$ is obtained; see Figure~\ref{fig:5.15}. For this case, the 500 repetitions for the estimators are reported in Figure~\ref{fig:5.17}.
Compared to the previous case, the squared correlation has dropped to 0.965, but it is still a significant improvement from sampling in the original coordinates, which only achieves 0.089. Therefore, the variance of the MFAB estimator is only 39\% of the MC estimator. The MFAB estimator's bias has also decreased from the previous case.
A virtually unbiased estimator can be obtained using $\rh = 3$ and $\rl = 2$, for which a target variance of $\widehat{\delta}^2 = 2496.3$ is obtained. The 500 repetitions for the three estimators are reported in Figure~\ref{fig:5.18}. The squared correlation between HF and LF has decreased slightly from 0.965 to 0.946 compared to the case when $(\rh, \rl) = (2, 2)$, but the variance of the MFAB estimator is only 30\% of its MC counterpart. As by design, the bias of the MFAB estimator is almost negligible.

\subsubsection{Unbiased MFAB estimator}
\begin{figure}[htb]
	\centering
	\begin{subfigure}{0.48\linewidth}
		\centering
		\includegraphics[width=.9\linewidth]{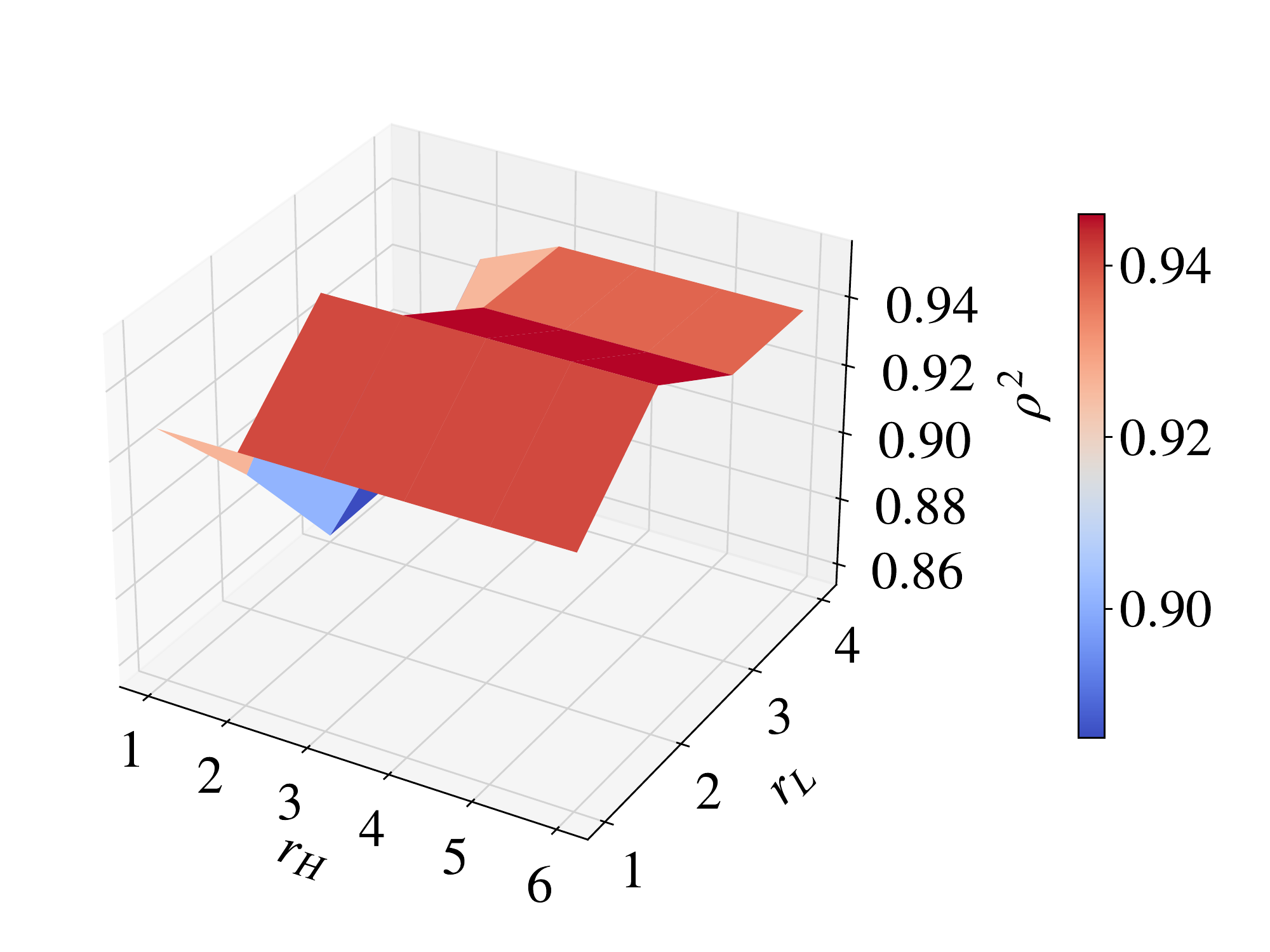}
		\subcaption{}
		\label{fig:5.19}
	\end{subfigure}
	\begin{subfigure}{0.48\linewidth}
		\centering
		\includegraphics[width=.8\linewidth]{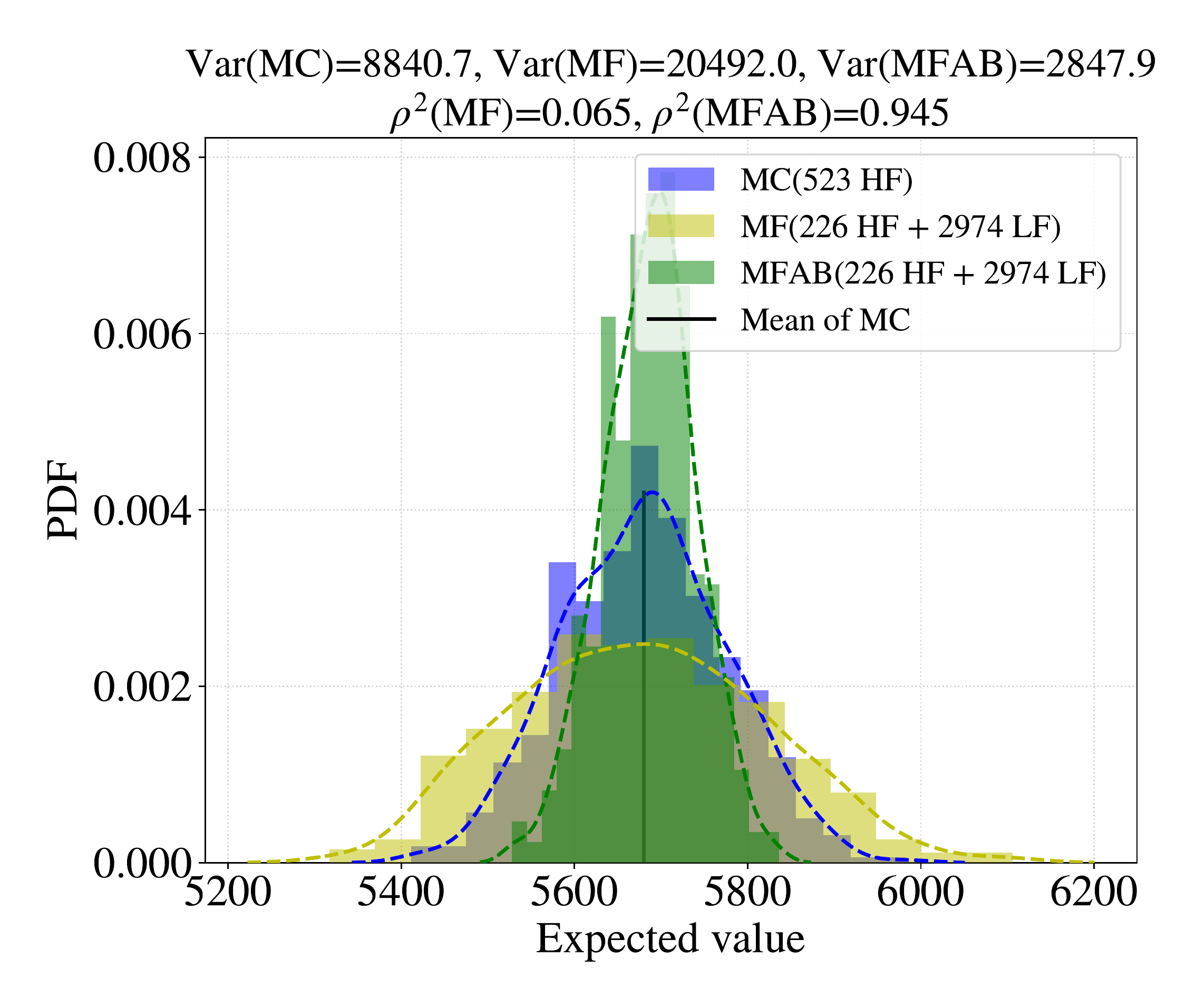}
		\subcaption{}
		\label{fig:5.20}
	\end{subfigure}
	\caption{(a) Estimated correlation for the two models in the acoustic application with unbiased strategy, and (b) probability density functions for 500 realizations of the MFAB, MF, and MC estimators, where the unbiased strategy is used in the MFAB estimator with $r_H = 3$ and $r_L = 2$.}
\end{figure}

In this application, we can still afford to re-run the HF model and, as a consequence, we assume non-legacy data options in the algorithm. Figure~\ref{fig:5.19} illustrates the correlations for the complete range of choices for $\rh$ and $\rl$.
We observe that the maximum correlation is obtained when $\rh=3$ and $\rl=2$. At this step, $\rh \times N_p$ LF evaluations are performed, which is the additional cost of the unbiased strategy.
To ensure fair comparison, we use the same target variance as in the previous case, i.e., $\sigma^2 = 2496.3$. Figure~\ref{fig:5.20} presents the 500 repetitions for the three estimators obtained using 40 pilot samples. The squared correlation has increased from 0.065 to 0.945, leading to a substantial decrease in the variance of the MFAB estimator, which is only 32\% of the MC estimator. Moreover, as expected, the MFAB estimator is unbiased. In this application, the oversampling ratio $\ratio$ for the MF estimator is assumed to be the same as the MFAB estimator since the ratio $\ratio$ computed by the pilot samples is less than 1, implying that leveraging the LF model in the MF estimator has no gain when sampling the original parameters space. Since we are also forcing the estimators to have the same cost, the number of HF samples in the MF estimator is less than in the MC estimator. Since the additional LF samples in the MF estimator does not help to improve the performance, the MF estimator has a greater variance than the MC estimator.

\subsection{Nuclear spent fuel assembly model}
\label{ssec:numres_fem}
\begin{figure}[htb]
	\centering
	\begin{subfigure}[t]{0.48\linewidth}
		\centering
		\includegraphics[width=0.4\linewidth]{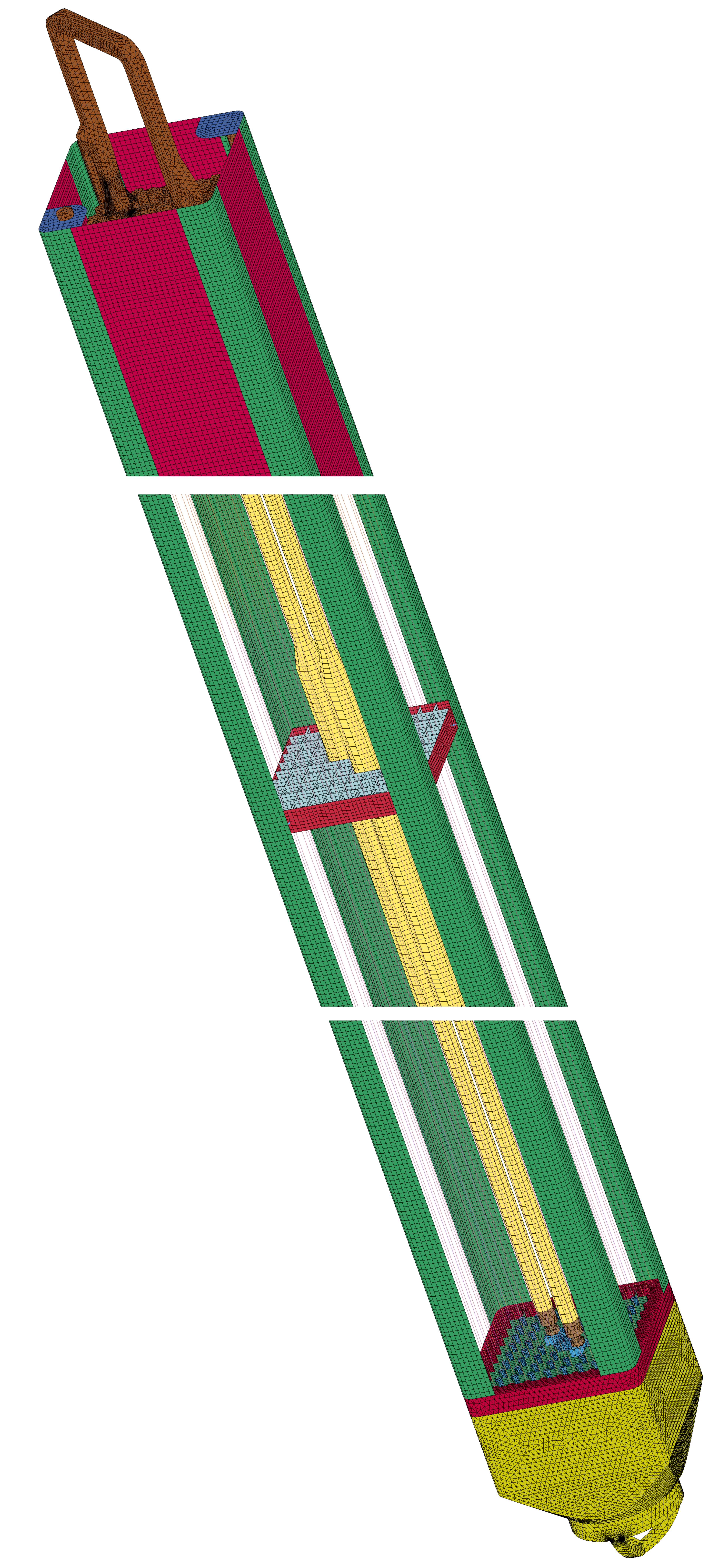}
		\caption{Detailed fuel assembly}
		\label{fig:fa_profile}
	\end{subfigure}
	\hspace{0.5cm}
	\begin{subfigure}[t]{0.4\linewidth}
		\centering
		\includegraphics[width=0.55\linewidth]{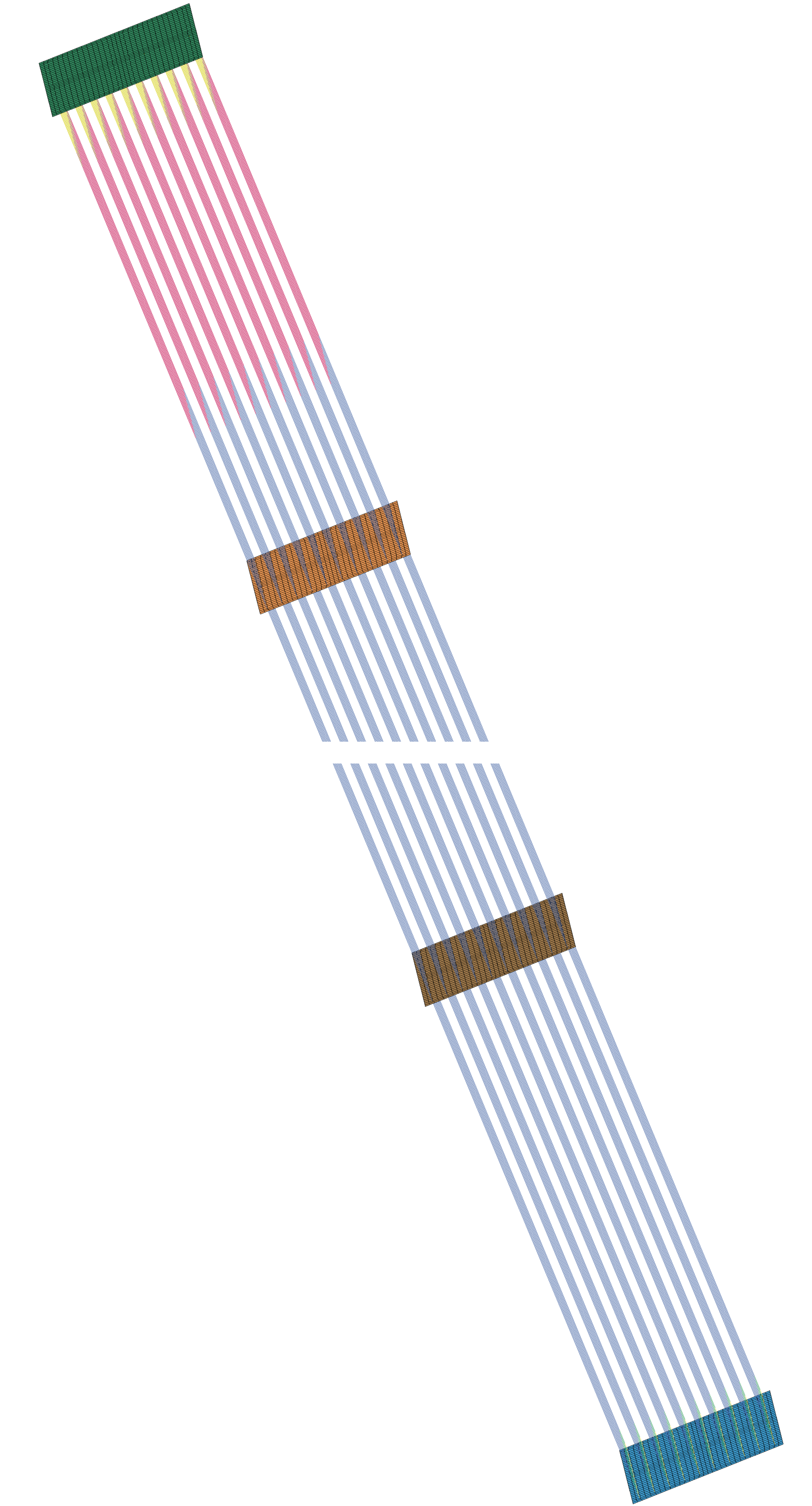}
		\caption{Simplified fuel assembly}
		\label{fig:fa_simplified}
	\end{subfigure}
	\caption{Finite element model of the detailed (\subref{fig:fa_profile}) / simplified (\subref{fig:fa_simplified}) fuel assembly: view of the top (top), the middle span (middle), and the bottom (bottom) without the channel thin panel.}
\end{figure}

We will next demonstrate the application of the developed methodologies to a nuclear fuel assembly model. The fuel assembly model is based on the General Electric GE14 design~\cite{ge14} and is utilized in boiling water reactors. The model is a slender structure with dimensions of 4517.7 mm in length and 140.2 mm in width. The fuel rods are arranged in a $10\times 10$ array, held together by eight uniformly spaced spacer grids, and the upper and lower tie plates. Two water rods are inserted after removing eight fuel rods. These components are placed in a slender squared tube known as the channel. In the finite element (FE) model, linear springs are used to model the connections between the fuel rods and spacer grids, as well as between the spacer grids and the channel. The channel is connected to the lower tie plate by finger springs and is attached to a post of the upper tie plate. The FE model assumes that the channel is clamped to the lower tie plate and the post of the upper tie plate. Figure \ref{fig:fa_profile} illustrates local views of the FE model.

\begin{figure}[htb]
	\centering
	\begin{subfigure}[t]{0.48\linewidth}
		\centering
		\includegraphics[width=.8\linewidth]{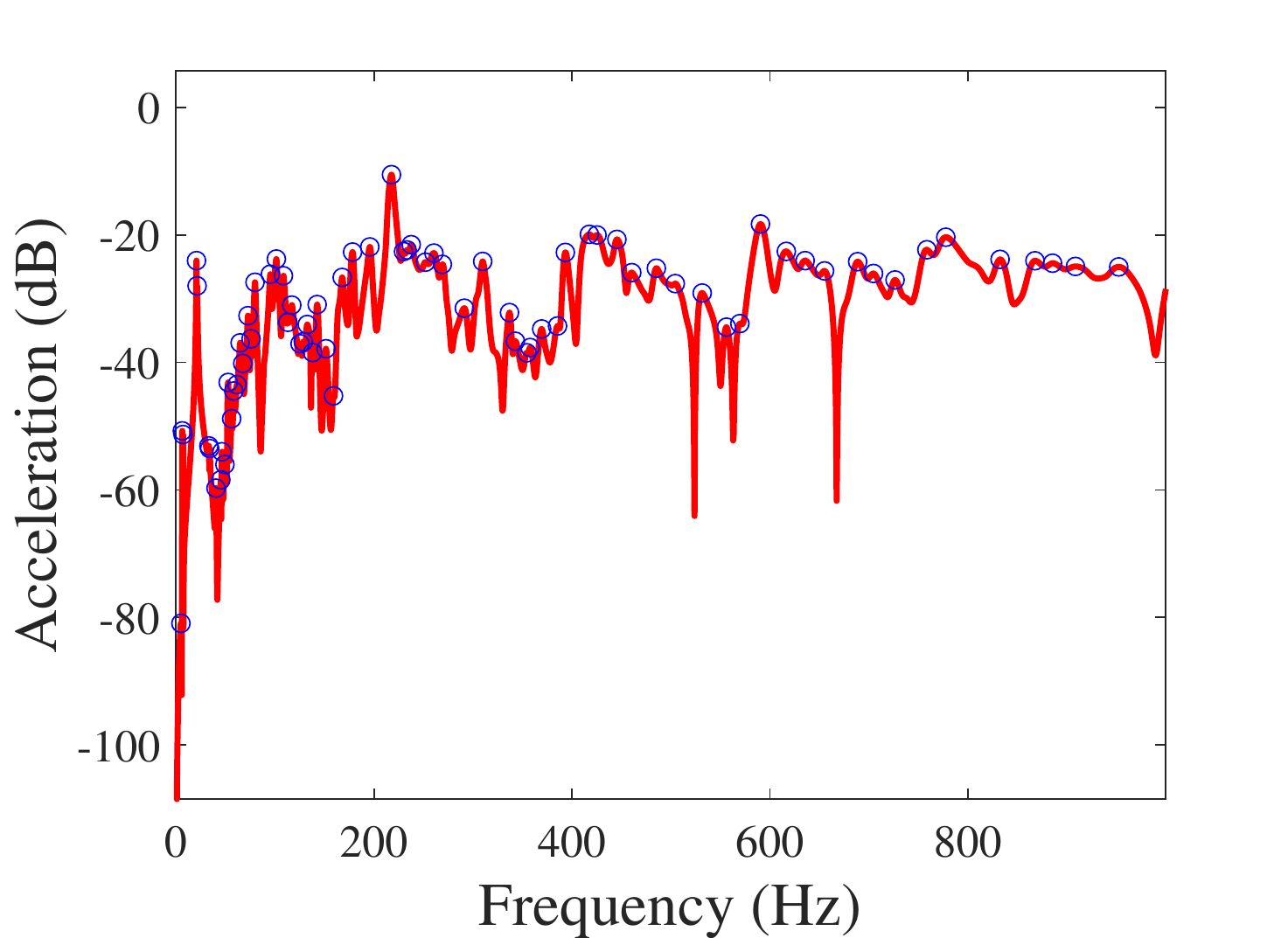}
		\caption{Detailed fuel assembly model.}
		\label{fig:frf_fa}
	\end{subfigure}
	\begin{subfigure}[t]{0.48\linewidth}
		\centering
		\includegraphics[width=.8\linewidth]{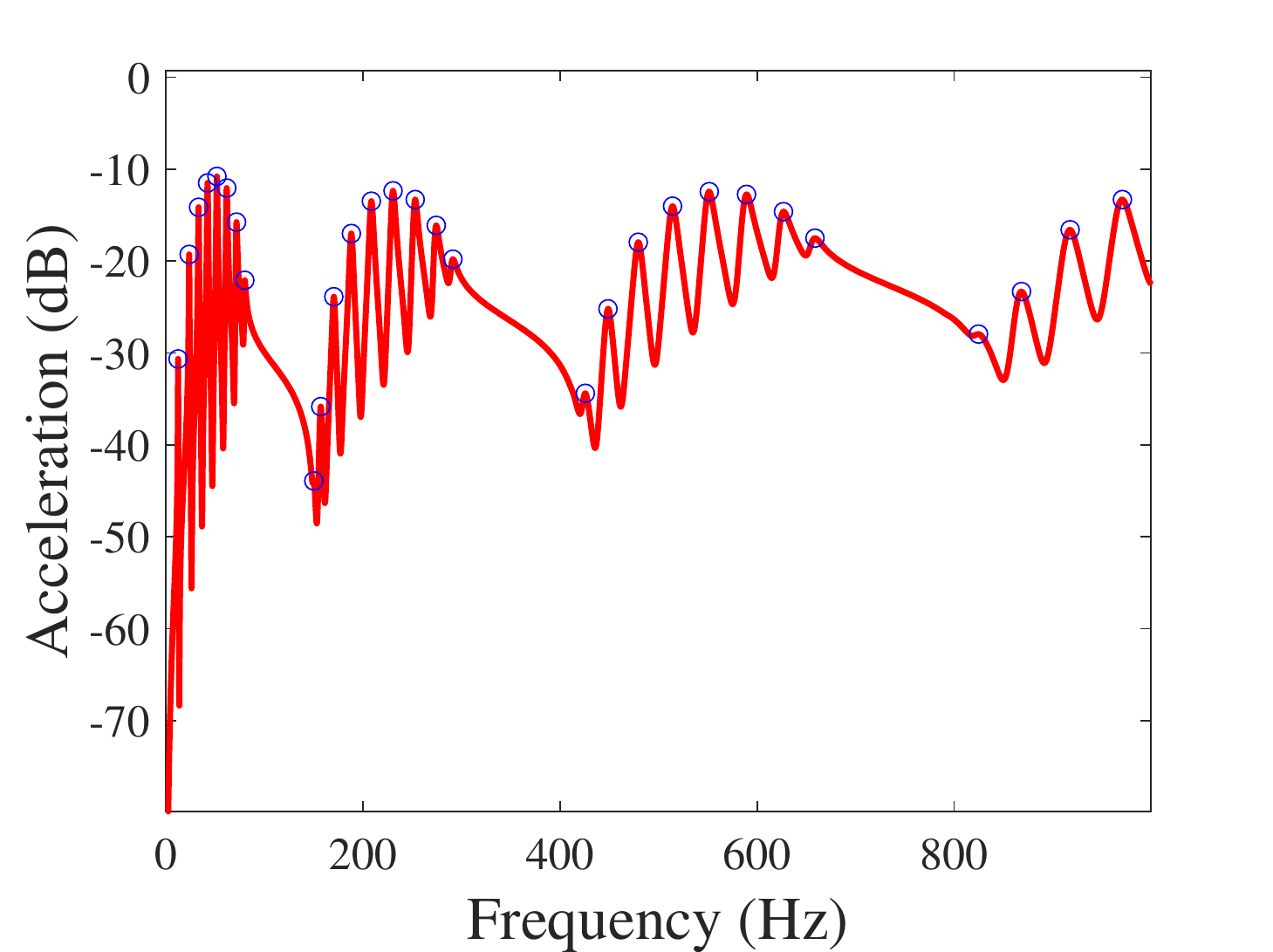}
		\caption{Simplified fuel assembly model.}
		\label{fig:frf_fa_simp}
	\end{subfigure}
	\caption{Frequency response function on mean values for the detailed fuel assembly model~(\subref{fig:frf_fa}) and simplified~(\subref{fig:frf_fa_simp}) model.}
\end{figure}

The FE model consists of $N = 1,912,506$ degrees-of-freedom (DOFs). The analysis focuses on vibration in the frequency range $\mathcal{B}=2\pi\times ]0, 1000]$ rad/s. The lower tie plate's tip node is fully constrained, and the normal translation DOFs of two nodes near the top for each of the four channel faces are also constrained. The QoI of interest is the frequency response function (FRF) of acceleration excited at a node on one face of the lower tie plate and observed at a node on the adjacent face, both in the normal direction. The excitation and observation points are located in the red region of the lower tie plate shown in Figure \ref{fig:fa_profile}. The QoI is the average value of the eight largest peaks in the frequency range $\mathcal{B}$. Accurate estimations can be obtained by using the Craig-Bampton sub-structuring method described in \cite{ezvan2020multiscale}, where all the modes in the frequency band $\mathcal{B}_c=2\pi\times]0, 1200]$ rad/s are considered. The modes in $2\pi\times [1000, 1200]$ rad/s are included to consider the spread contribution to the response within 1000 Hz. To reduce the computational cost, the shift-invert Lanczos method is used to solve the eigenvalue problem \cite{ezvan2020multiscale, ezvan2021dominant}. The model can be solved by 50 computer nodes (16-core Intel(R) Xeon(R) CPU E5-2640 v3 @ 2.60 GHz) in about 5 minutes. The connections between different structural levels, rod-to-grid, and grid-channel, are critical to computing the QoI and are intrinsically uncertain due to the modeling process. Therefore, these two types of connections are randomized by Beta distributions with mean values of 0.15 kN/mm and 1.5 kN/mm, respectively, and coefficients of variation of 20\%. The connections associated with a spacer grid share the same random germ. Therefore, the problem has eight random variables. Choosing the parameters to be their mean values and solving the dynamic problem yields the FRF, as shown in Figure \ref{fig:frf_fa}.
In the figure the peaks of the FRF are marked by blue circles. %

A much simpler LF model has also been constructed, which is a $10\times 10$ array of fuel rods held by eight spacer grids, an upper tie plate, and a lower tie plate. The upper and lower tie plates have been simplified as transverse shell elements, and the spacer grids have also been modeled using transverse shell elements, as shown in the FE model in Figure \ref{fig:fa_simplified}. The simplified model does not include any water rods or channel structure, resulting in $N = 368,892$ DOFs. The upper and lower tie plates are constrained to simulate the same boundary conditions as the HF model. In this model, the FRF of acceleration is excited at a node on one side of the first spacer grid from the bottom and observed at a node on the adjacent side, both along the normal direction. Although the computational cost of the LF model by mode superposition has already been significantly reduced from the HF model, we further reduce the cost by applying the global reduced-order model (ROM) introduced in \cite{ezvan2019dominant}. The cost to obtain the FRF is about 3 minutes using one computer node. Therefore, the HF and LF models have a cost ratio of about 80.

The simplified fuel assembly model employs shared nodes to connect the fuel rods to the spacer grids. In addition, we attempt to introduce randomness by modeling the Young's modulus of the spacer grids as Beta distributions with a coefficient of variation of 20\% and a mean value of 70.0 kN/mm$^2$. Each spacer grid contains one random variable, resulting in a UQ dimension of 8. By setting the parameters to their mean values and solving the dynamic problem using global ROM, we obtain the FRF, which is shown in Figure \ref{fig:frf_fa_simp}. The FRF of the simplified fuel assembly differs significantly from that of the detailed fuel assembly in terms of the shape and magnitude of the response. This is not unexpected, given that the simplified version is much simpler in several aspects and that the excitation and observation locations differ from those of the detailed model. The differences in modeling result in significant differences in the models' parameterizations.

\begin{figure}[htb]
	\centering
	\begin{subfigure}{0.48\linewidth}
		\centering
		\includegraphics[width=.8\linewidth]{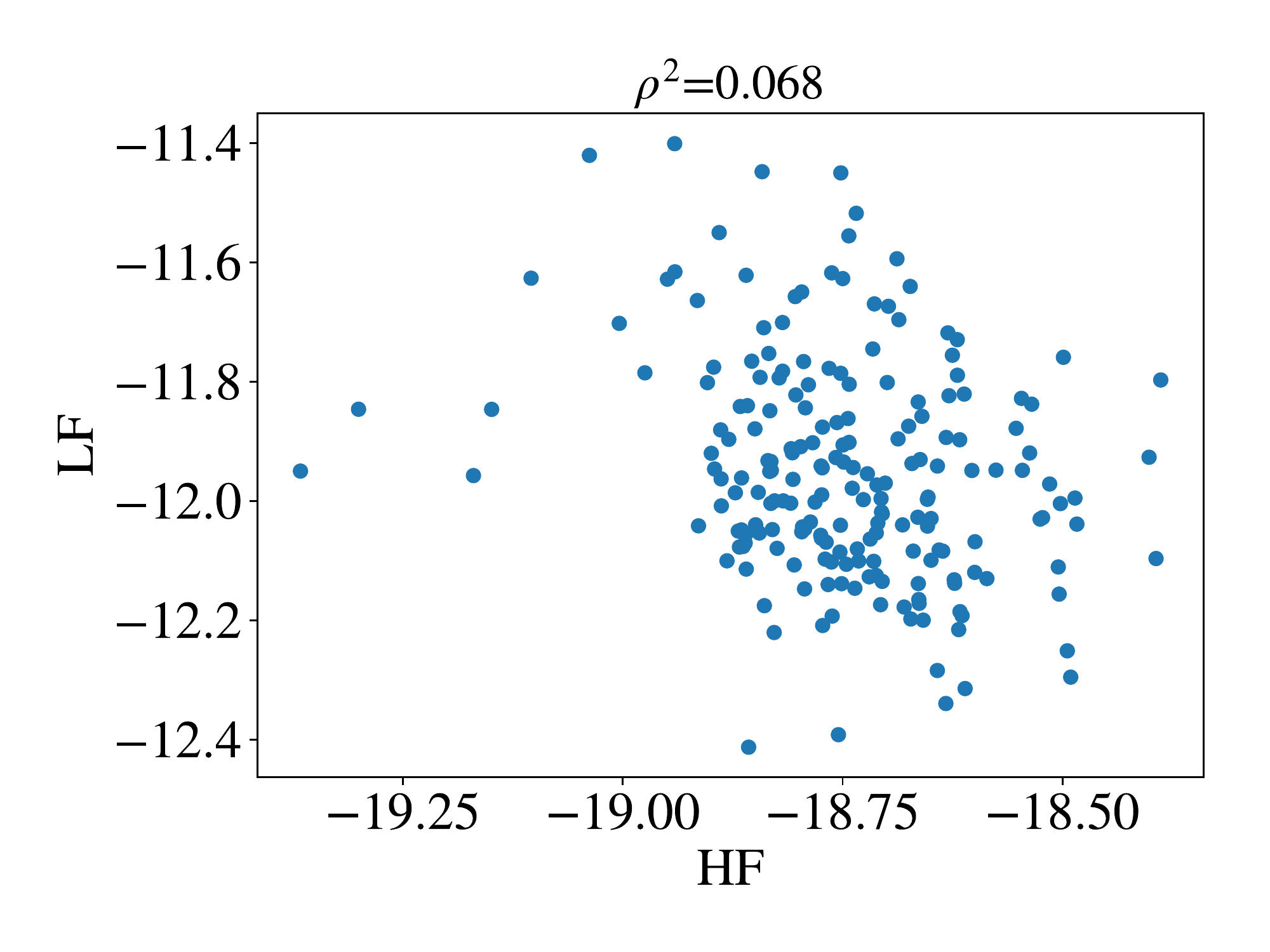}
		\subcaption{}
		\label{fig:scatter_org_FA}
	\end{subfigure}
	\begin{subfigure}{0.48\linewidth}
		\centering
		\includegraphics[width=.8\linewidth]{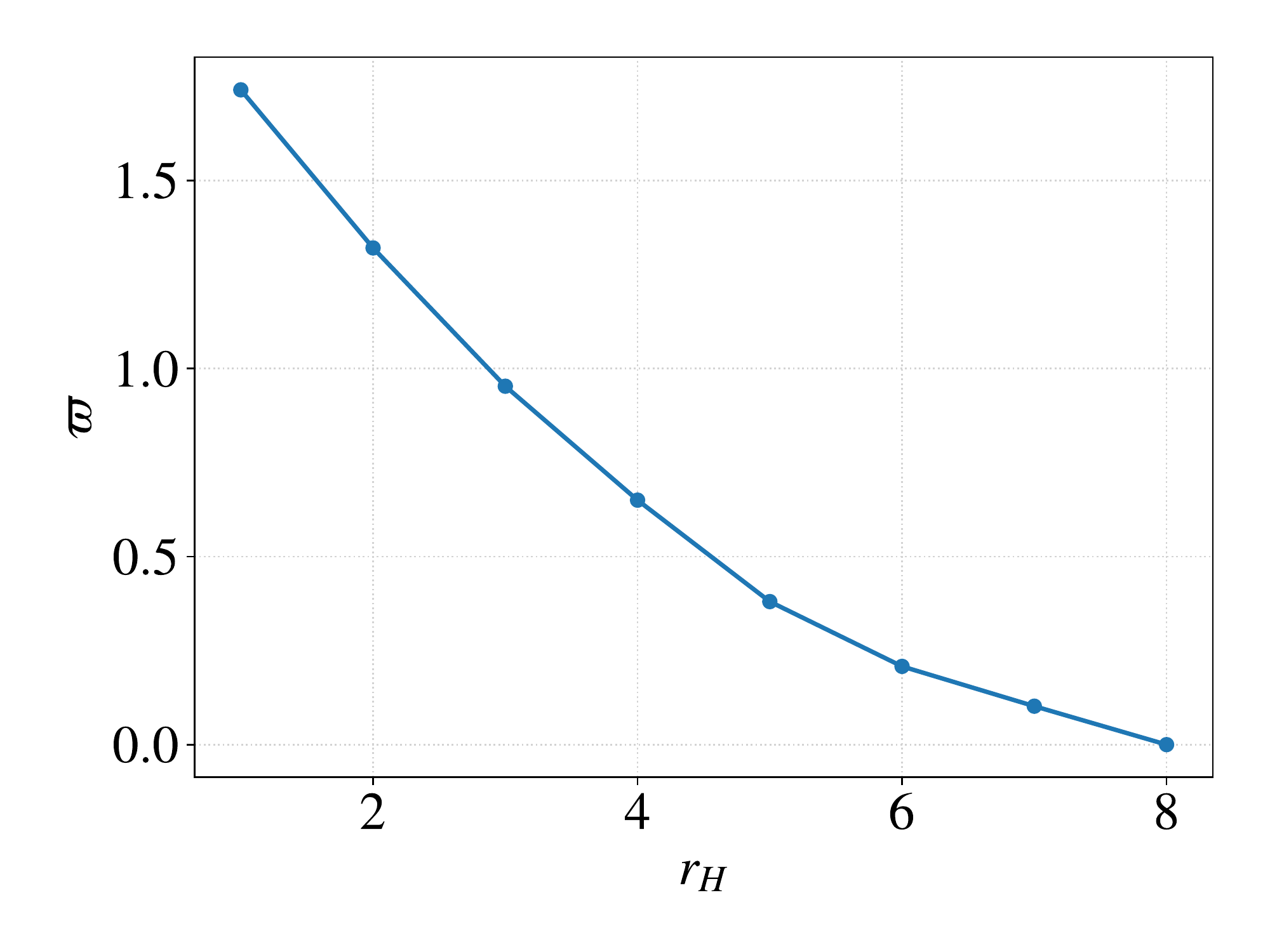}
		\subcaption{}
		\label{fig:bias2_fa}
	\end{subfigure}
	\caption{(a) Scatter plot of the HF and LF 	models on the original coordinates, and (b) bias estimator based on first order information for the fuel assembly application ($\chi=2$).}
\end{figure}

We will now use the MFAB method to estimate the expected value of the QoI and compare it with the MC and MF methods. Since the HF model is expensive, it is unrealistic to obtain repeated realizations of the estimators and then compare them. Therefore, we will use the unbiased approach with the \textit{legacy dataset} options. For this application, we will generate $N_p=200$ pilot samples of the HF model, which will be reused to obtain different estimators for comparison. Although the HF samples are fixed, we can generate as many LF samples. The squared correlation of the HF and LF models on the original coordinates for the pilot samples is 0.068, as shown in Figure~\ref{fig:scatter_org_FA}. The two models are almost uncorrelated.
Thus, the gain of leveraging LF models is limited when using the MF estimator, which samples the models in their original space. Following the first two steps of Algorithm~\ref{ALG:unbiased}, we can compute the bias estimators of the HF model. Here, we will use the estimator based on first-order information for presentation, as shown in Figure \ref{fig:bias2_fa}.
There is no clear convergence trend, implying that all adapted directions are important and cannot be reduced, making correlation enhancement more challenging. We highlight that this feature reflects that the problem response cannot be satisfactorily represented on a lower dimensional manifold. However, this is an intrinsic property of this application and does not depend on the proposed strategy.

\begin{figure}[htb]
	\centering
	\begin{subfigure}{0.48\linewidth}
		\centering
		\includegraphics[width=.9\linewidth]{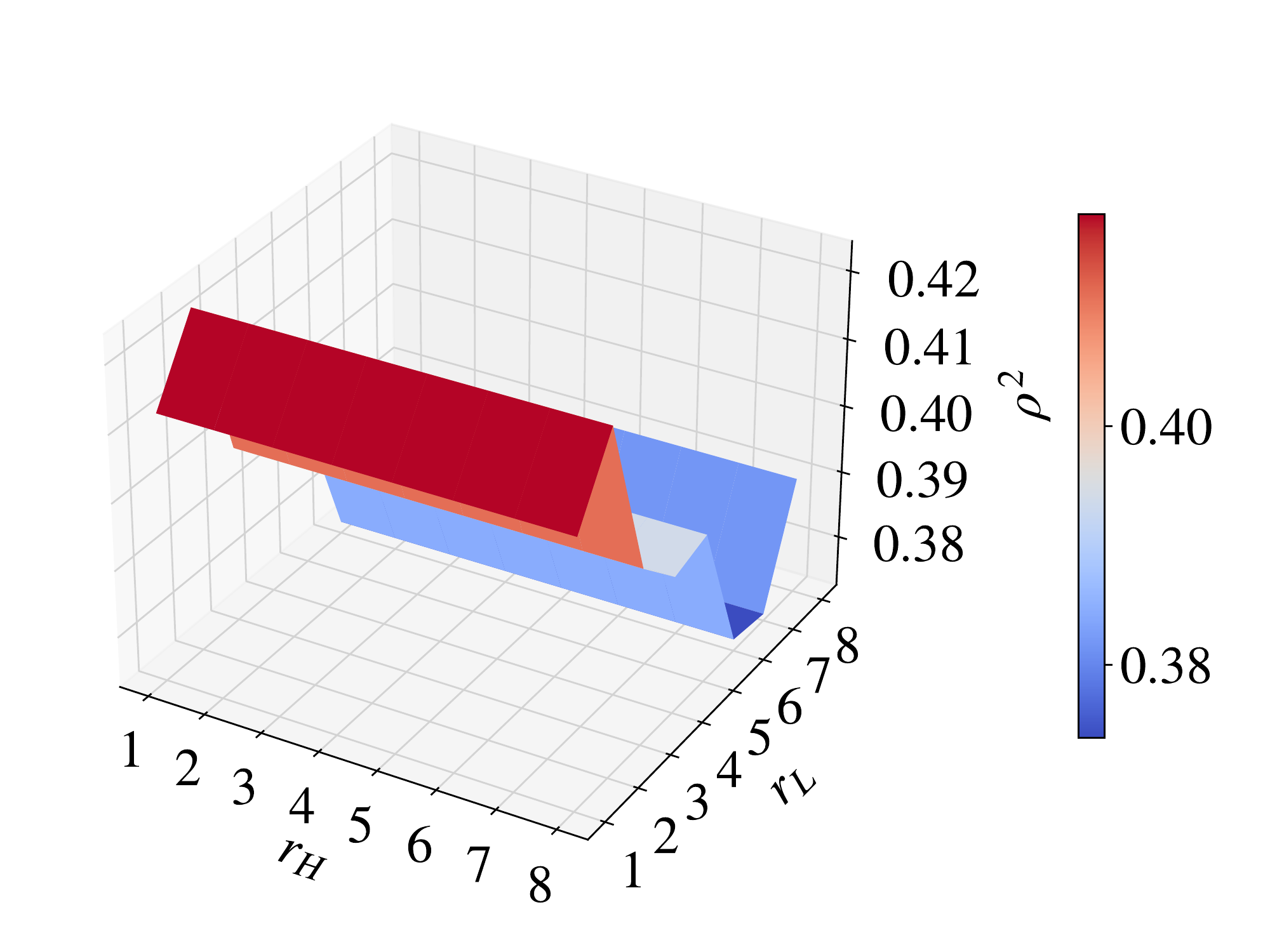}
		\subcaption{}
		\label{fig:correlation_fa}
	\end{subfigure}
	\begin{subfigure}{0.48\linewidth}
		\centering
		\includegraphics[width=.8\linewidth]{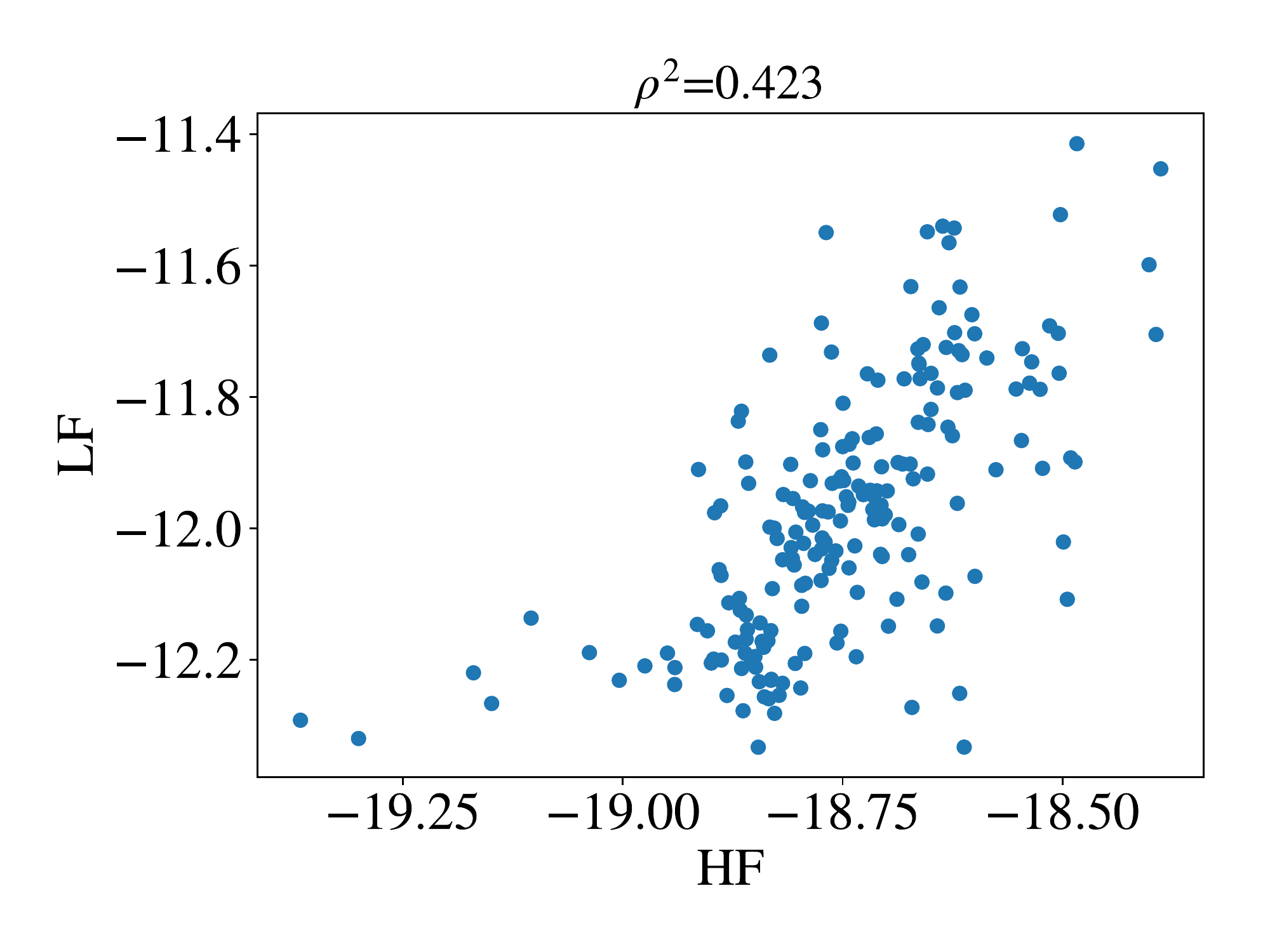}
		\subcaption{}
		\label{fig:scatter_fa_rerun_lf}
	\end{subfigure}
	\caption{(a) Estimated correlation for the two models in the fuel assembly application with the unbiased strategy, and (b) scatter plot of the pilot HF and LF samples after rerun the LF model in the shared space.}
\end{figure}

Next, we need to estimate the correlations for the unbiased estimator using Eq.\eqref{eq:corr_unbiased}. The results for various dimensions are shown in Figure\ref{fig:correlation_fa}. 
When $\rh=\rl=2$, we observe that the squared correlation is 0.423, which is the highest among all dimensions. Although the value is lower than the previous two examples, it still represents a significant increase from 0.068. Subsequently, with $\rh=\rl=2$, we project the pilot HF samples onto the 2-dimensional adapted space, map these samples to the original coordinates of the LF model, and reevaluate the LF model on these samples. The scatter plot of the pilot HF samples and the new pilot LF samples is presented in Figure~\ref{fig:scatter_fa_rerun_lf}. Once again, we observe an increase in the correlation between HF and LF.

\begin{figure}[htb]
	\centering
	\includegraphics[width=0.45\linewidth]{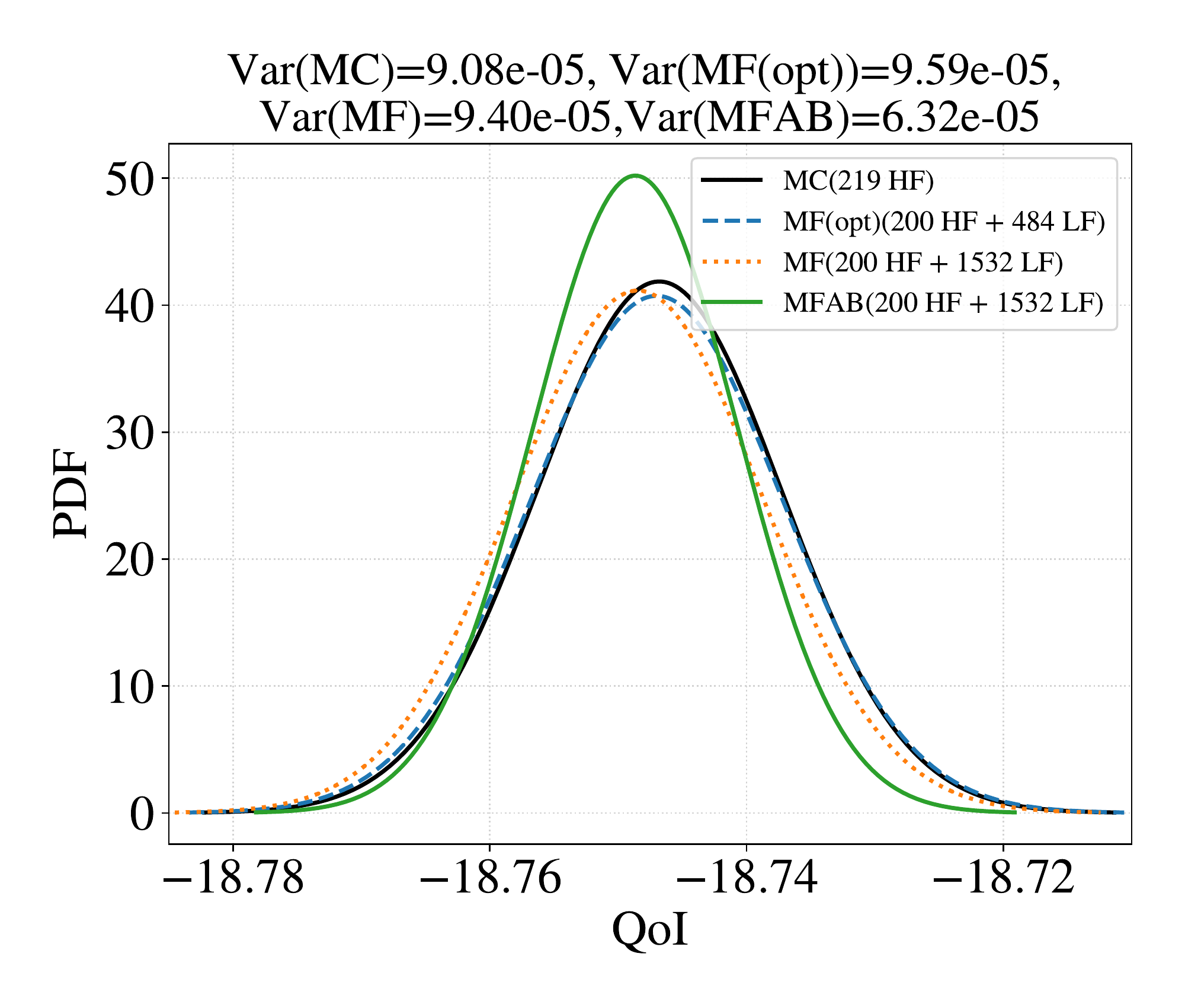}
	\caption{Scatter plot of the pilot HF and LF samples after rerun the LF model for the fuel assembly application.}
	\label{fig:pdf_comparison_fa}
\end{figure}

Following Algorithm~\ref{ALG:MFAB}, we estimate the oversampling ratio $\ratio$ of the LF model to be $\ratio=7.66$, and compute the number of independent LF samples as $\lceil (\ratio-1)N_p \rceil = 1332$. Once the unbiased sampling strategy generates these 1332 independent LF samples, we use them to compute the MFAB estimator and its variance. Unlike the previous two applications, where multiple estimators could be computed, in this application, only one estimator is computed using the \textit{legacy dataset} option. To provide a graphical representation of the estimators' performance, we plot a normal distribution with the estimator mean and variance in Figure~\ref{fig:pdf_comparison_fa}.
We also compare the MFAB estimator with the MC estimator, the optimal MF estimator, and the MF estimator with the same HF and LF samples as the MFAB estimator. For the MC estimator, we first use the 200 pilot HF samples to estimate the mean and variance, and then scale the variance to match the cost of the MFAB estimator. The optimal MF estimator is obtained using the pilot samples sampled in the original coordinates. Due to the low correlation in the original space, the oversampling ratio is calculated as $\ratio=2.42$, which is much smaller than that of the MFAB estimator. Figure~\ref{fig:pdf_comparison_fa} presents a comparison of these estimators.
The variances of the two MF estimators are close to that of the MC estimator, while the variance of the MFAB estimator is 30\% lower than that of the MC estimator. Comparing the two MF estimators, we observe that increasing the number of LF samples barely changes the MF estimator's performance since the correlation between the models in the original spaces is negligible.

The fuel assembly application presents significant challenges because the LF model is simplified to such an extent that its physics differs dramatically from the HF model. Additionally, as shown in Figure~\ref{fig:bias2_fa}, the HF model cannot be accurately represented on a lower-dimensional space, which further complicates the challenge of correlation enhancement. Generally, as the number of required dimensions for accurate representation in the adapted space increases, the correlation enhancement decreases. However, the MFAB approach can still increase the correlation between the models, thereby improving the performance of the MF estimator.

\section{Concluding remarks}
\label{SEC:conclusions}
Multifidelity uncertainty quantification (MF UQ) approaches play a crucial role in deploying UQ analyses for realistic scientific and engineering problems. However, the effectiveness of the various methods developed in the literature is often greatly hindered by the presence of models with dissimilar parameterization. For example, a subset of the parameters may not be shared among models, which reduces the correlation between models by introducing an independent source of variability. Therefore, the dissimilar parameterization poses a significant challenge to MF UQ approaches since their performance is highly dependent on model correlations.

In this work, we addressed this challenge by relying on dimension reduction strategies to identify a shared set of parameters that does not correspond to the original model parameterization but can be obtained from it. We extended the previous work~\cite{Geraci2018,Geraci2018b} by integrating the Adaptive Basis (AB)~\cite{Tipireddy2014,zeng2021accelerated} dimension reduction strategy with a MF estimator corresponding to the MF Monte Carlo estimator with a single low-fidelity model. AB allows for easy and efficient identification of lower-dimensional manifolds in the parameter space that can be used to sample the models and improve their correlations. We presented a general discussion about the design of such estimators and introduced several strategies to support the practical construction of these estimators from available pilot samples. As a result, we proposed two novel strategies to embed AB into MF estimators. One strategy leads to bias-variance balanced MF estimators, enabling the flexibility of controlling bias and variance. In contrast, the other strategy leads to unbiased estimators with variance control. Thanks to AB, both strategies have enhanced correlation among the models and consequently improve the performance and applicability of the MFUQ approaches.

Finally, we presented an array of numerical test cases to illustrate the features of the methods in different practical scenarios, such as in the presence of legacy and non-legacy high-fidelity datasets. For all numerical tests presented, the novel estimator was able to make use of low-fidelity models that would otherwise be too poorly correlated to be exploited within existing MF methods. Our current work focuses on integrating this strategy with multiple low-fidelity models based on the Approximate Control Variate~\cite{Gorodetsky_GEJ_JCP_2020} method. Moreover, we are focusing on using this approach in the context of MF surrogate construction, such as MF networks~\cite{Gorodetsky_JGE_IJUQ_2020}. Preliminary results for this latter case are presented in~\cite{Zeng2023}.

\section*{Acknowledgements}
The authors were partially supported by the Laboratory Directed Research Development (LDRD) program at Sandia National Laboratories and DOE SciDAC FASTMath institute. Sandia National Laboratories is a multi-mission laboratory managed and operated by National Technology and Engineering Solutions of Sandia, LLC., a wholly owned subsidiary of Honeywell International, Inc., for the U.S. Department of Energy's National Nuclear Security Administration under contract DE-NA-0003525. The views expressed in the article do not necessarily represent the views of the U.S. Department of Energy or the United States Government.

\bibliographystyle{plain}
\bibliography{biblio}

\appendix

\begin{appendices}

	\section{Proof of Proposition~\ref{prop:coor}}
	\label{prop:coor_proof}
	\begin{proof}
		Given the assumptions, $\rs = \mathrm{max}(\rh,\rl) = \rh=\rl$. It follows that 
		\begin{equation*}
			\bbeta_s = \bbetahr = \bbetalr\,.
		\end{equation*}
		Then, from Eq~\eqref{eq:adp2org}, $\bxiLr = \AmatLrt \bbetalr = \AmatLrt \bbeta_s$, implying $\bbetahr = \bbeta_s =  \AmatLr \bxiLr$.
		It follows that
		\begin{equation*}
			\bxiHr =	\AmatHrt \bbetahr
			=  \AmatHrt \AmatLr \bxiLr \,.
		\end{equation*}
		Similarly, we can prove the second equation of ~\eqref{eq:mapH2L}.
	\end{proof}

	\section{Proof of Proposition \ref{Prop:MFAB_MSE}}
	\label{Prop:MFAB_MSE_proof}
	\begin{proof}
		\begin{equation*}
			\begin{split}
				\text{MSE}\left(\widehat{Q}^{\text{MFAB}}\right) &= \EE{ \left( \widehat{Q}^{MFAB} - \EE{Q_{H}} \right)^2 } \\
				&= \EE{ \left( \widehat{Q}^{MFAB} - \EE{\widehat{Q}^{MFAB}} + \EE{\widehat{Q}^{MFAB}} - \EE{Q_{H}} \right)^2 } \\
				&= \EE{ \left( \widehat{Q}^{MFAB} - \EE{\widehat{Q}^{MFAB}} \right)^2 } 
				+ \left( \EE{\widehat{Q}^{MFAB}} - \EE{Q_{H}} \right)^2 \\
				&= \Var{ \widehat{Q}^{MFAB} } + \left( \EE{Q_H \left( \mathcal{A}_{H,\rh}^\mathrm{T} \bbeta_s \right)} - \EE{ Q_{H}(\bxiH) } \right)^2.
			\end{split}
		\end{equation*}
		And, by Eq~\eqref{eq:var_MF}
		\begin{equation*}
			\Var{ \widehat{Q}^{MFAB} } = \Var{ \hQh^{\rh}(\rh,\ubbetas) } 
			\left( 1 - \frac{\ratio-1}{\ratio} \rho^2_{r_L, r_H} \right)\,.
		\end{equation*}
	\end{proof}

	\section{Proof of Proposition \ref{prop:PCE_AB_correlation}}
	\label{prop:PCE_AB_correlation_proof}
	\begin{proof}
		The expected values for the two expansions are
		\begin{equation*}
			\begin{split}
				\EE{{Q}_H(\bbetahr)} 
				&= \mathbb{E} \begin{bmatrix}
					\sum_{\bm{\gamma} \in \mathcal{J}_{p_H}^{\rh}} c_{H, \bm{\gamma}} \psi_{\bm{\gamma}}(\bbetahr)
				\end{bmatrix}
				= c_{H, \bm{e}_0} + \sum\nolimits_{\bm{\gamma} \in \mathcal{J}_{p_H}^{\rh} \setminus \bm{e}_0} c_{H, \bm{\gamma}} \EE{\psi_{\bm{\gamma}}(\bbetahr)}
				= c_{H, \bm{e}_0} \\
				\EE{{Q}_L(\bbetalr)} 
				&= \mathbb{E} \begin{bmatrix}
					\sum_{\bm{\gamma} \in \mathcal{J}_{p_L}^{\rl}} c_{L, \bm{\gamma}} \psi_{\bm{\gamma}}(\bbetalr)
				\end{bmatrix}
				= c_{L, \bm{e}_0} + \sum\nolimits_{\bm{\gamma} \in \mathcal{J}_{p_L}^{\rl} \setminus \bm{e}_0} c_{L, \bm{\gamma}} \EE{\psi_{\bm{\gamma}}(\bbetalr)} = c_{L, \bm{e}_0}\,,
			\end{split}
		\end{equation*}
		by properties of Hermite polynomial chaos. In the above equation, $c_{H, \bm{e}_0}$ and $c_{L, \bm{e}_0}$ are zero order coefficients of the two models. It follows that
		\begin{equation*}
			\label{eq:cov_pce}
			\begin{split}
				\Cov{{Q}_H(\bbetahr)}{{Q}_L(\bbetalr)} 
				&= \EE{\left({Q}_H(\bbetahr) - \EE{{Q}_H(\bbetahr)}\right) \left({Q}_L(\bbetalr) - \EE{{Q}_L(\bbetalr)}\right)} \\
				&=\EE{ \begin{pmatrix}
						\sum_{\bm{\gamma} \in \mathcal{J}_{p_H}^{\rh} \setminus \bm{e}_0} c_{H, \bm{\gamma}} \psi_{\bm{\gamma}}(\bbetahr)
					\end{pmatrix} 
					\begin{pmatrix}
						\sum_{\bm{\gamma} \in \mathcal{J}_{p_L}^{\rl} \setminus \bm{e}_0} c_{L, \bm{\gamma}} \psi_{\bm{\gamma}}(\bbetalr)
				\end{pmatrix} }\,.
			\end{split}
		\end{equation*}
		Since $\bbetahr$ and $\bbetalr$ are standard Gaussian variables, they share the same probability measure. Then, by the orthonormality of Hermite polynomial chaos, the terms in the above equation are nonzero only when the multi-indices $\bm{\gamma}$  are identical for both expansions. Then, by noting $\langle\psi_{\bm{\gamma}},  \psi_{\bm{\gamma}} \rangle = 1$, it follows that
		\begin{equation*}
			\label{eq:cov_pce2}
			\Cov{{Q}_H(\bbetahr)}{{Q}_L(\bbetalr)} = \sum\nolimits_{ \bm{\gamma}  \in \left(\mathcal{J}_{p_H}^{\rh} \bigcap \mathcal{J}_{p_L}^{\rl} \right) \setminus \bm{e}_0 } c_{H, \bm{\gamma}} c_{L, \bm{\gamma}} \,.
		\end{equation*}
		By the orthonormality of Hermite polynomial chaos, we can also get 
		\begin{equation*}
			\begin{split}
				\Var{{Q}_H(\bbetahr)} 
				&= \EE{ \left({Q}_H(\bbetahr) - \EE{{Q}_H(\bbetahr)}\right)^2}
				= \EE{ \begin{pmatrix}
						\sum\limits_{\bm{\gamma} \in \mathcal{J}_{p_H}^{\rh} \setminus \bm{e}_0} c_{H, \bm{\gamma}} \psi_{\bm{\gamma}}(\bbetahr)
					\end{pmatrix} ^2}
				= \sum_{\bm{\gamma} \in \mathcal{J}_{p_H}^{\rh} \setminus \bm{e}_0} c_{H, \bm{\gamma}}^2 \\
				\Var{{Q}_L(\bbetalr)} 
				&= \EE{ \left({Q}_L(\bbetalr) - \EE{{Q}_L(\bbetalr)}\right)^2}
				= \EE{\begin{pmatrix}
						\sum\limits_{\bm{\gamma} \in \mathcal{J}_{p_L}^{\rl} \setminus \bm{e}_0} c_{L, \bm{\gamma}} \psi_{\bm{\gamma}}(\bbetalr)
					\end{pmatrix} ^2}
				= \sum_{\bm{\gamma} \in \mathcal{J}_{p_L}^{\rl} \setminus \bm{e}_0} c_{L, \bm{\gamma}}^2 \\
			\end{split}
		\end{equation*}
		Then, by definition, the correlation of ${Q}_H(\bbetahr)$ and ${Q}_L(\bbetalr)$ has the expression of ~\eqref{eq:corr_estim}.
	\end{proof}

	\section{Proof of Corollary \ref{coroll:nonuniform}}
	\label{coroll:nonuniform_proof}
	\begin{proof}
		By the definition of correlation we can write
		\begin{equation*}
			\label{eq:nonuniform1}
			\begin{split}
				\rho^2 &= \frac{\left(\Cov{{Q}_H}{{Q}_L} \right)^2} { \Var{{Q}_H} \Var{{Q}_L} }
				= \frac{ \begin{pmatrix}
						\sum_{  \bm{\gamma} \in \mathcal{J}_{p_L}^{\rl} \setminus \bm{e}_0 } c_{H, \bm{\gamma}} c_{L, \bm{\gamma}} \EE{\psi_{\bm{\gamma}}^2}
					\end{pmatrix}^2 }
				{ \begin{pmatrix}
						\sum_{{{\bm{\gamma}} \in \mathcal{J}_{p_H}^{\rh} \setminus \bm{e}_0 }} c_{H, \bm{\gamma}}^2 \EE{\psi_{\bm{\gamma}}^2}
					\end{pmatrix} 
					\begin{pmatrix}
						\sum_{{{\bm{\gamma}} \in \mathcal{J}_{p_L}^{\rl} \setminus \bm{e}_0 }} c_{L, \bm{\gamma}}^2 \EE{\psi_{\bm{\gamma}}^2}
					\end{pmatrix}
				} \\
				&= \frac{ \begin{pmatrix}
						\sum_{{{\bm{\gamma}} \in \mathcal{J}_{p_L}^{\rl} \setminus \bm{e}_0 }} c_{H, \bm{\gamma}} c_{L, \bm{\gamma}} \EE{\psi_{\bm{\gamma}}^2}
					\end{pmatrix} ^2 }
				{ \begin{pmatrix}
						\sum_{{{\bm{\gamma}} \in \mathcal{J}_{p_L}^{\rl} \setminus \bm{e}_0 }} c_{H, \bm{\gamma}}^2 \EE{\psi_{\bm{\gamma}}^2} 
						+ \sum_{{{\bm{\gamma}} \in \mathcal{J}_\Delta \setminus \bm{e}_0 }} c_{H, \bm{\gamma}}^2 \EE{\psi_{\bm{\gamma}}^2}
					\end{pmatrix} 
					\begin{pmatrix}
						\sum_{{{\bm{\gamma}} \in \mathcal{J}_{p_L}^{\rl} \setminus \bm{e}_0 }} c_{L, \bm{\gamma}}^2 \EE{\psi_{\bm{\gamma}}^2}
					\end{pmatrix}
				} \\   
				&\leq \frac{ \begin{pmatrix}
						\sum_{{{\bm{\gamma}} \in \mathcal{J}_{p_L}^{\rl} \setminus \bm{e}_0 }} c_{H, \bm{\gamma}} c_{L, \bm{\gamma}} \EE{\psi_{\bm{\gamma}}^2} 
					\end{pmatrix} ^2 }
				{ \begin{pmatrix}
						\sum_{{{\bm{\gamma}} \in \mathcal{J}_{p_L}^{\rl} \setminus \bm{e}_0 }} c_{H, \bm{\gamma}}^2 \EE{\psi_{\bm{\gamma}}^2} 
					\end{pmatrix}
					\begin{pmatrix}
						\sum_{{{\bm{\gamma}} \in \mathcal{J}_{p_L}^{\rl} \setminus \bm{e}_0 }} c_{L, \bm{\gamma}}^2 \EE{\psi_{\bm{\gamma}}^2}
					\end{pmatrix}
				} = \rho^2_{\mathcal{J}_{p_L}^{\rl}}\\            
			\end{split}
		\end{equation*}
		where $\mathcal{J}_\Delta = \mathcal{J}_{p_H}^{\rh} \setminus \mathcal{J}_{p_L}^{\rl}$. The last inequality is obtained because all terms corresponding to multi-indices belonging to the set $\mathcal{J}_\Delta$ are contributing to the HF variance only, thus decreasing the correlation squared.
	\end{proof}

	\section{Proof of Proposition \ref{prop:correl_increase_adapt}}
	\label{prop:correl_increase_adapt_proof}
	\begin{proof}
		For simplified notation, we write the two first-order PCEs in the adapted spaces as
		\begin{equation*}
			{Q}_{H}(\bbeta) = c_{H,0}+\sum_{k=1}^d c_{H,k} \eta_{k}\,,\qquad
			{Q}_{L}(\bbeta) = c_{L,0}+\sum_{k=1}^d c_{L,k} \eta_{k}\,.
		\end{equation*}
		By assumption, we have $c_{H,k} \geq 0$ and $c_{L,k} \geq 0$, for $k=1, \cdots, d$, and 
		\begin{equation*}
			c_{H,1} \geq \cdots \geq c_{H,d} \quad \text{and} \quad c_{L,1} \geq \cdots \geq c_{L,d}\,.
		\end{equation*}
		The correlations in the adapted space and the original space can be written as
		\begin{equation*}
			\begin{split}
				\rho\left( {Q}_{H}(\bm{\eta}), {Q}_{L}(\bm{\eta}) \right) &= \frac{\sum_{k=1}^{d}c_{H,k} c_{L,k}}{\sqrt{(\sum_{k=1}^{d}c_{H,k}^2) (\sum_{k=1}^{d} c_{L,k}^2)}}\,, \\
				\rho\left( {Q}_{H}(\bm{\xi}), {Q}_{L}(\bm{\xi}) \right) &= \frac{\sum_{k=1}^{d} c_{H,\sigma(k)} c_{L,k}}{\sqrt{(\sum_{k=1}^{d}c_{H,k}^2) (\sum_{k=1}^{d} c_{L,k}^2)}}\,,
			\end{split}
		\end{equation*}
		respectively, where $\sigma_{(1)}, \cdots, \sigma_{(d)}$ is a permutation such that $c_{H, \sigma_{(k)}}$ and $c_{L,k}$ are PCE coefficients associated with a same basis in the original space (i.e. $\bm{\xi}$ space). The rearrangement inequality in mathematics states that
		\begin{equation*}
			x_{n} y_{1}+\cdots+x_{1} y_{n} \leq x_{\sigma(1)} y_{1}+\cdots+x_{\sigma(n)} y_{n} \leq x_{1} y_{1}+\cdots+x_{n} y_{n}
		\end{equation*}
		for all choices of real numbers
		\begin{equation*}
			x_{1} \leq \cdots \leq x_{n} \quad \text { and } \quad y_{1} \leq \cdots \leq y_{n}
		\end{equation*}
		and any permutation
		\begin{equation*}
			x_{\sigma(1)}, \ldots, x_{\sigma(n)}
		\end{equation*}
		of $x_{1}, \ldots, x_{n}$.
		Apply the rearrangement inequality to the correlations we immediately get 
		\begin{equation*}
			\rho\left( {Q}_{H}(\bm{\eta}), {Q}_{L}(\bm{\eta}) \right) \geq \rho\left( {Q}_{H}(\bm{\xi}), {Q}_{L}(\bm{\xi}) \right).
		\end{equation*}
		Since we also assume that the PCE coefficients are greater than or equal to 0,
		\begin{equation*}
			\rho^2 \left( {Q}_{H}(\bm{\eta}), {Q}_{L}(\bm{\eta}) \right) \geq \rho^2 \left( {Q}_{H}(\bm{\xi}), {Q}_{L}(\bm{\xi}) \right).
		\end{equation*}
	\end{proof}

\end{appendices}

\end{document}